\documentstyle[11pt]{article}
\input amssym.def
\input amssym.tex

\def\goth{\frak}
\def\double{\Bbb}

\def\cc{{\double C}}

\def\rr{{\double R}}
\def\zz{{\double Z}}

\def\ep{{\cal E}}
\def\ppp{{\cal P}}
\def\sss{{\cal S}}
\def\www{{\cal W}}
\def\enep{{\hbox{End}(\ep)}}

\def\dep{{\cal D}(\ep)}

\def\da{{\partial}_{\!\mbox{\tiny A}}}
\def\db{{\partial}_{\!\mbox{\tiny B}}}
\def\dD{{\partial}_{\mbox{\tiny D}}}
\def\ddd{{/\!\!\!\partial}}
\def\DDD{{/\!\!\!\!D}}
\def\dda{{/\!\!\!\partial}_{\!\!\mbox{\tiny A}}}
\def\ddb{{/\!\!\!\partial}_{\!\!\mbox{\tiny B}}}
\def\ddD{{/\!\!\!\partial}_{\!\mbox{\tiny D}}}

\def\naep{{\nabla^{\mbox{\tiny$\ep$}}}}

\def\ot{\otimes}
\def\op{\oplus}

\def\mapright#1{\smash{\mathop{\longrightarrow}\limits^{#1}}}

\def\bb{\begin{eqnarray}}
\def\ee{\end{eqnarray}}

\newtheorem{definition}{Definition}[section]
\newtheorem{lemma}{Lemma}[section]
\newtheorem{satz}{Theorem}[section]
\newtheorem{korollar}[satz]{Corollary}
\newtheorem{proposition}{Proposition}[section]

\begin{document}

%\hsize 15truecm
%\vsize 20truecm
\font\twelve=cmbx10 at 13pt
\font\eightrm=cmr8
\def\petit{\def\rm{\fam0\eightrm}}
\baselineskip 15pt

\begin{titlepage}
\title{Real ${\rm Z}\hspace{-.25cm}{\rm Z}_2-$Bi-Gradings, Majorana Modules and\\
the Standard Model Action}
\author{J\"urgen Tolksdorf\thanks{email: juergen.tolksdorf@mis.mpg.de}\\
Max-Planck Institute for Mathematics in the Sciences\\Leipzig, Germany}
\date{December 13, 2009}

\maketitle

\begin{abstract}
The action functional of the Standard Model of particle physics is intimately related to a
specific class of first order differential operators called Dirac operators of Pauli type
(``Pauli-Dirac operators''). The aim of this article is to carefully analyze the geometrical structure of this class of Dirac operators on the basis of real Dirac operators of simple
type. On the basis of simple type Dirac operators, it is shown how the Standard Model action (STM action) may be viewed as generalizing the Einstein-Hilbert action in a similar way the
Einstein-Hilbert action is generalized by a cosmological constant. Furthermore, we demonstrate how the geometrical scheme presented allows to naturally incorporate also Majorana mass terms within the Standard Model. For reasons of consistency these Majorana mass terms are shown to dynamically contribute to the Einstein-Hilbert action by a ``true'' cosmological constant. Due to its specific form, this cosmological constant can be very  small. Nonetheless, this
cosmological constant may provide a significant contribution to dark matter/energy. In the geometrical description presented this possibility arises from a subtle interplay between Dirac and Majorana masses.

\end{abstract}

\vspace{3.5cm}

\begin{tabbing}
{\bf Keywords:}
Dirac Type Differential Operators, (Real) Clifford Modules,\\
General Relativity, Gauge Theories, Majorana Masses, Cosmological Constant
\end{tabbing}

\vspace{0.15cm}

\noindent
{\bf MSC:} 53C05, 53C07, 70S05, 70S15, 83C05\\
{\bf PACS:} 02.40.Hw, 02.40.Ma, 04.20.-q, 14.80.Bn

\end{titlepage}

\section{Introduction}
The dynamical description of fermions and bosons is usually based upon different geometrical
schemes. The fermionic actions are always defined in terms of Dirac type operators. In contrast, the gravitational and the Yang-Mills functionals are defined in terms of the respective curvatures associated with connections. Accordingly, the following ``ambiguity''
in the definition of the fermionic action is not taken into account. Let
$\psi\in{\goth Sec}(M,\ep)$ be a section of the Hermitian Clifford module
\bb
\label{cliff mod}
(\ep,\gamma_{\!\mbox{\tiny$\ep$}})\twoheadrightarrow(M,g_{\mbox{\tiny M}})
\ee
over an oriented (semi-)Riemannian manifold of even dimension and arbitrary signature. Also,
let $\,\DDD_{\!\!\mbox{\tiny$\ep$}}$ be a Dirac (type) operator (see below) that acts on
${\goth Sec}(M,\ep)$. The fermionic action is defined in terms of the smooth function:
$\langle\psi,\,\DDD_{\!\!\mbox{\tiny$\ep$}}\psi\rangle_{\!\mbox{\tiny$\ep$}}$,
with $\langle\cdot,\cdot\rangle_{\!\mbox{\tiny$\ep$}}$ being the Hermitian form on $\ep$.
Clearly, nothing changes when a (tricky) null is added, i.e.
\bb
\label{motiv. pauli-type dop}
\langle\psi,\,\DDD_{\!\!\mbox{\tiny$\ep$}}\psi\rangle_{\!\mbox{\tiny$\ep$}}
&\equiv&
\langle\psi,\,\DDD_{\!\!\mbox{\tiny$\ep$}}\psi\rangle_{\!\mbox{\tiny$\ep$}} +
\langle\psi,\Phi_{\!\mbox{\tiny$\ep$}}\psi\rangle_{\!\mbox{\tiny$\ep$}} -
\langle\psi,\,\Phi_{\!\mbox{\tiny$\ep$}}\psi\rangle_{\!\mbox{\tiny$\ep$}}\nonumber\\[.1cm]
&=:&
\mbox{\small$\left\langle\left(\!\!
              \begin{array}{c}
                \psi \\
                \psi \\
              \end{array}
            \!\!\right),\,
\left(\!\!
  \begin{array}{cc}
    \DDD_{\!\!\mbox{\tiny$\ep$}} & -\Phi_{\!\mbox{\tiny$\ep$}} \\
    \Phi_{\!\mbox{\tiny$\ep$}} & \DDD_{\!\!\mbox{\tiny$\ep$}} \\
  \end{array}
\!\!\right)\!
\left(\!\!
\begin{array}{c}
\psi \\
\psi \\
\end{array}
\!\!\right)
\right\rangle$}_{\!\!\!\!\mbox{\tiny$\!\phantom{\ep}^2\ep$}}\,.
\ee
Here, $\Phi_{\!\mbox{\tiny$\ep$}}\in{\goth Sec}(M,\enep)$ denotes an arbitrary zero-order operator and $\hspace{-.2cm}\phantom{\ep}^2\ep := \ep\op\ep$ the ``doubling'' of $\ep$ with
an appropriately induced Hermitian form and Clifford structure.

This apparently trivial observation may become meaningful, actually, if the bosonic
action is also defined in terms of Dirac operators. In fact, it has been shown that both the fermionic part and the bosonic part of the Standard Model action -- the latter also including the Einstein-Hilbert functional -- can be geometrically described in terms of a single Dirac operator (c.f., for instance, in \cite{AT:96} and \cite{Tol:98} with respect to the combined Einstein-Hilbert-Yang-Mills and the Einstein-Hilbert-Yang-Mills-Higgs action in terms of the non-commutative residue):
\bb
\label{pauli type dop}
{/\!\!\!\!P}_{\!\!\mbox{\tiny D}} &=& \left(
         \begin{array}{cc}
           i\dda + \tau_{\!\mbox{\tiny$\ep$}}\circ\phi_{\!\mbox{\tiny$\ep$}} &
           -{/\!\!\!\! F}_{\!\!\mbox{\tiny D}} \\
           {/\!\!\!\! F}_{\!\!\mbox{\tiny D}} &
           i\dda + \tau_{\!\mbox{\tiny$\ep$}}\circ\phi_{\!\mbox{\tiny$\ep$}} \\
         \end{array}
       \right)\nonumber\\[.1cm]
       &\equiv&
       i\dda + \tau_{\!\mbox{\tiny$\ep$}}\circ\phi_{\!\mbox{\tiny$\ep$}} +
       {\cal I}_{\mbox{\tiny$\ep$}}\circ{/\!\!\!\! F}_{\!\!\mbox{\tiny D}}\,.
\ee
Here, respectively, the Dirac operator
\bb
\label{simple type dop}
\DDD_{\!\!\mbox{\tiny$\ep$}} \equiv
i\dda + \tau_{\!\mbox{\tiny$\ep$}}\circ\phi_{\!\mbox{\tiny$\ep$}}
\ee
belongs to the distinguished class of Dirac operators of simple type on the Clifford module
(\ref{cliff mod}) and $\,{/\!\!\!\! F}_{\!\!\mbox{\tiny D}}$ is the ``quantized'' relative curvature of (\ref{simple type dop}). The details of these and the following notions will be summarized in the next section.

The specific class of Dirac operators (\ref{simple type dop}) will play a crucial role in what follows (see also, for example, \cite{Qui:85}, \cite{Bis:86} for the role of simple type Dirac operators in the case of the family index theorem and \cite{Con:94} of non-commutative geometry).
When evaluated with respect to (\ref{pauli type dop}), the ``total Dirac action''
(see below)
\bb
\label{total dirac act}
{\cal I}_{\mbox{\tiny D,tot}} :=
\int_M\left(\langle\Psi,\,{/\!\!\!\!P}_{\!\!\mbox{\tiny D}}
\Psi\rangle_{\!\mbox{\tiny$\!\!\!\phantom{x}^{2}\!\ep$}}
+ {\rm tr}_\gamma(curv(\,{/\!\!\!\!P}_{\!\!\mbox{\tiny D}}) -
\varepsilon{\rm ev}_{\!g}(\omega^2_{\!\mbox{\tiny D}}))\right)dvol_{\mbox{\tiny M}}
\ee
decomposes into the various parts of the Standard Model action, including gravity described in terms of the Einstein-Hilbert functional. In particular, the fermionic part reduces to the
usual Dirac-Yukawa action:
\bb
\label{ferm part dirac act}
\int_M\langle\Psi,\,{/\!\!\!\!P}_{\!\!\mbox{\tiny D}}
\Psi\rangle_{\!\mbox{\tiny$\!\!\!\phantom{x}^{2}\!\ep$}}\,
dvol_{\mbox{\tiny M}}
 =
\int_M\langle\psi,(i\dda + \phi_{\!\mbox{\tiny$\ep$}})\psi\rangle_{\!\mbox{\tiny$\ep$}}\,
dvol_{\mbox{\tiny M}}\,,
\ee
provided the sections $\Psi\in{\goth Sec}(M,\!\!\!\!\phantom{\ep}^2\ep)$
on the doubled Clifford module
\bb
\label{doubled cliff mod}
(\!\!\!\phantom{\ep}^2\ep\equiv\ep\op\ep,\tau_{\mbox{\tiny$\!\!\!\phantom{x}^{2}\!\ep$}}
\equiv\tau_{\!\mbox{\tiny$\ep$}}\ominus\tau_{\!\mbox{\tiny$\ep$}},
\gamma_{\mbox{\tiny$\!\!\!\phantom{x}^{2}\!\ep$}}
\equiv\gamma_{\!\mbox{\tiny$\ep$}}\op\gamma_{\!\mbox{\tiny$\ep$}})\twoheadrightarrow
(M,g_{\mbox{\tiny M}})
\ee
are restricted to ``diagonal sections'' $\Psi = (\psi,\psi)$ and the sections
$\psi\in{\goth Sec}(M,\ep)$ are restricted, furthermore, to the ``physical sub-bundle''
$\ep_{\!\mbox{\tiny phys}}\hookrightarrow\ep\twoheadrightarrow M$ of the underlying
Clifford module (c.f. \cite{ToTh:05} and the corresponding references therein).

The specific form of the Dirac operator (\ref{pauli type dop}), acting on the sections of the doubled Clifford module, parallels the first order differential operator
\bb
\label{dop plus pauli}
i\dda - m - i{/\!\!\!\! F}_{\!\!\!\mbox{\tiny A}}\,,
\ee
with $F_{\!\!\mbox{\tiny A}}$ being the electromagnetic field strength that was introduced
to account for the anomalous magnetic moment of the proton at a time when it was not yet clear that the proton is a composite of quarks but considered as ``elementary'' (see, for example, Chapter 2-2-3 in \cite{IZ:87}). However, when the quarks entered the stage of particle physics the {\it Pauli term} $i{/\!\!\!\! F}_{\!\!\!\mbox{\tiny A}}$ became superfluous. Moreover, the additional fermionic interaction caused by the Pauli term rendered the quantum field theory based upon (\ref{dop plus pauli}) non-renormalizable.

It is a remarkable feature of ``Dirac type gauge theories'' that the complete Standard Model action (including gravity) can be geometrically described in terms of the ``Pauli type Dirac operator'' (\ref{pauli type dop}). It has been shown that this description of the Standard Model allows to make a prediction for the value of the mass of the Higgs boson which is
consistent with all the otherwise known data from the Standard Model. In other words,
the geometrical description of the Standard Model based upon the geometry of
$\,{/\!\!\!\!P}_{\!\!\mbox{\tiny D}}$
renders the Standard Model even more predictive than it is the case with respect to its usual description (c.f. \cite{ToTh:06}, where one can also find a brief comparison to similar results presented in \cite{CCM:06}, see also \cite{CM:07}). For this matter it seems worth investigating more closely the specific form of Pauli type Dirac operators and the restrictions made with respect to the fermionic sector that guarantee the Pauli like term
${\cal I}_{\mbox{\tiny$\ep$}}\circ{/\!\!\!\! F}_{\!\!\mbox{\tiny D}}$ to only contribute to the bosonic part of the total Dirac action (\ref{total dirac act}).

In this paper, we carefully discuss the fact that in the bosonic part of
(\ref{total dirac act}) only curvature terms enter, whereas the fermionic part is determined by connections, only. This subtle interplay between the fermionic and the bosonic part of the total Dirac action permits to geometrically regard the Yang-Mills action as a ``covariant generalization'' of the Einstein-Hilbert action and the Standard Model action as a natural generalization of the Einstein-Hilbert action with cosmological constant. Moreover, the geometrical analysis of the operator (\ref{pauli type dop}) permits to also naturally include the notion of Majorana masses within the scheme of Dirac type gauge theories. It will be shown that the thus described Majorana masses dynamically contribute to the bosonic part
of (\ref{total dirac act}) in the form of Einstein's ``biggest blunder''.

Some of the features presented seem close to the geometrical description of the Standard Model in terms of A. Connes' non-commutative geometry (c.f., for example, \cite{Con:95}, \cite{Con:96}, \cite{CCM:06} and \cite{CM:07}). However, the geometrical setup presented is different in various respects. For example, the relation between Dirac operators and connections is based upon the canonical first order decomposition of any Dirac (type) operator (c.f. Section 2). As a consequence, the Higgs boson is intimately tied to gravity in the setup presented. Indeed, the Higgs boson is shown to generalize the Yang-Mills connection via the metric. Furthermore, the bosonic part of the total Dirac action (\ref{total dirac act}) is based upon the canonical second order decomposition of any Dirac (type) operator. This, indeed, provides a canonical generalization of the Einstein-Hilbert action with cosmological constant (c.f. Section 3). These basic features of Dirac type gauge theories will be the starting point of everything that follows.

The paper is organized as follows: The following section provides a summary of some of the basic notions already used in the introduction. Also, some motivation for the ensuing constructions are presented. In the third section, we present the geometrical
picture that underlies Dirac type gauge theories. In particular we discuss the Einstein-Hilbert action from the point of view of Dirac operators.
In the fourth section, we discuss Pauli type Dirac operators in view of ``real, $\zz_2-$bi-graded Clifford modules'' (``real Clifford modules'', for short, see the work \cite{ABS:64}, which may serve as a kind of standard reference). We present some examples of particular interest. In the fifth section, we discuss the geometrical description of Majorana masses within Dirac type gauge theories. In particular, we discuss a generalization of the STM action when Majorana masses are taken into account. The sixth section is devoted to
a discussion of the Standard Model (STM) action in terms of real Dirac operators of simple type. This will provide a new geometrical picture of the STM action and how the latter is related to the Einstein-Hilbert functional of General Relativity. Finally, the last section summarizes the main conclusions. Before we get started, however, it might be worth presenting a brief summary of the main results obtained.

The presented geometrical discussion of the operators (\ref{pauli type dop}), defining the bosonic part of the total Dirac action (\ref{total dirac act}), is based upon a careful analysis of the geometrical background of the Dirac equation and the Majorana equation:
\bb
\label{dirac equation}
i\ddd\chi &=& m_{\mbox{\tiny D}}\chi\;\;\,\quad\Leftrightarrow\quad
\left\{\begin{array}{ccc}
         i\ddd\chi_{\mbox{\tiny R}} & = & m_{\mbox{\tiny D}}\chi_{\mbox{\tiny L}}\,, \\
         i\ddd\chi_{\mbox{\tiny L}} & = & m_{\mbox{\tiny D}}\chi_{\mbox{\tiny R}}\,,
       \end{array}
\right.
\\[.25cm]
\label{majorana equation}
i\ddd\chi &=& m_{\mbox{\tiny M}}\chi^{\mbox{\tiny cc}}\quad\Leftrightarrow\quad
\left\{\begin{array}{ccc}
         i\ddd\chi_{\mbox{\tiny R}} & = &
         m_{\mbox{\tiny M}}\chi_{\mbox{\tiny R}}^{\mbox{\tiny cc}}\,, \\
         i\ddd\chi_{\mbox{\tiny L}} & = &
         m_{\mbox{\tiny M}}\chi_{\mbox{\tiny L}}^{\mbox{\tiny cc}}\,.
       \end{array}
\right.
\ee
Here, respectively, $\chi_{\mbox{\tiny R}},\,\chi_{\mbox{\tiny L}}$ are the ``chiral''
eigen sections, $m_{\mbox{\tiny D}}$ is the ``Dirac mass'', $m_{\mbox{\tiny M}}$ the ``Majorana mass'' and ``$cc$'' has the physical meaning of ``charge conjugation''. It will be shown how a specific interplay between the two basic $\zz_2-$gradings, realized in nature by chirality and charge conjugation, allows one to overcome the issue of fermion doubling. The latter is usually encountered in the description of the fermionic action in terms of simple type Dirac operators. Furthermore, the interplay between parity and charge conjugation will also give the Pauli-Dirac operators their geometrical meaning. The geometrical background of Pauli-Dirac operators in terms of real Clifford modules has been partially discussed before
in \cite{Tol:09}. There, however, only the reduced Dirac action:
\bb
{\cal I}_{\mbox{\tiny D,red}} := \int_M{\rm tr}_\gamma curv(\,\DDD)\,dvol_{\mbox{\tiny M}}
\ee
was used. Also, the requirements imposed on ``particle-anti-particle modules'' (c.f. Definition 3 in loc. site) turn out to be too restrictive and do not allow to geometrically describe, for example, Dirac's first order differential operator
$i\ddd - m_{\mbox{\tiny D}}$ in terms of simple type Dirac operators. It thus does not account for the issue of the fermion doubling already mentioned. This drawback is remedied in this work in terms of Dirac modules associated with Majorana modules (c.f. Section 4). Also, we take the
opportunity to generalize formulae 110 and 113 of Lemma 1 in loc. site, which hold true only in the special case of $\Phi\in{\goth Sec}(M,{\rm End}_\gamma(\ep))$  (c.f. the
corresponding formulae 110 and 111 of Lemma 4.1).

In this work emphasis is put on {\it real Dirac operators of simple type}, which turn out to yield an appropriate geometrical description of both the fermionic and the bosonic action of the Standard Model. Indeed, on the basis of real Dirac operators of simple type, the Standard Model action will be shown to be described by the Einstein-Hilbert action with a ``cosmological constant'' term (c.f. Section 6):
\bb
{\cal I}_{\mbox{\tiny EHYMH}} =
\int_M{\rm tr}_\gamma\!\left[curv(\,\ddd_{\!\!\!\mbox{\tiny${\cal A}$}}) +
\Phi_{\mbox{\tiny YMH}}^2\right]dvol_{\mbox{\tiny M}}\,.
\ee

This may be regarded as a generalization of Lovelock's Theorem (c.f. \cite{Lov:72}
and Section 3). In contrast to this theorem, however,
\bb
\Lambda_{\mbox{\tiny YMH}} := {\rm tr}_\gamma\!\left(\Phi_{\mbox{\tiny YMH}}^2\right)
\ee
also depends on the metric and, in fact, is shown to coincide with the (Hodge dual of the) STM Lagrangian density ${\cal L}_{\mbox{\tiny YMH}}\in\Omega^n(M)$ plus a ``true'' cosmological constant term that is determined by Dirac and Majorana masses of an otherwise non-interacting
species of particles, collectively called ``cosmological neutrinos'' (c.f. Section 4):
\bb
\Lambda_{\mbox{\tiny YMH}} &=& \ast{\cal L}_{\mbox{\tiny YMH}} -
\Lambda_{\mbox{\tiny DM,$\nu$}}\,,
\nonumber\\[.1cm]
\Lambda_{\mbox{\tiny DM,$\nu$}} &\equiv&
a'{\rm tr}_{\!\mbox{\tiny$\www_{\nu}$}}m_{\mbox{\tiny D,$\nu$}}^4
- b'{\rm tr}_{\!\mbox{\tiny$\www_{\nu}$}}m_{\mbox{\tiny D,$\nu$}}^2
+ a'{\rm tr}_{\!\mbox{\tiny$\www_{\nu}$}}m_{\mbox{\tiny M,$\nu$}}^4
- b'{\rm tr}_{\!\mbox{\tiny$\www_{\nu}$}}m_{\mbox{\tiny M,$\nu$}}^2
\nonumber\\
&&
-\;
2a'{\rm tr}_{\!\mbox{\tiny$\www_{\nu}$}}
(m_{\mbox{\tiny D,$\nu$}}\circ m_{\mbox{\tiny M,$\nu$}})^2\,.
\ee
Here, $a',\,b' > 0$ are numerical constants that are determined by the dimension of the underlying (space-time) manifold (c.f. Sections 4 and 5). Though not discussed in detail
in this work, the point to be emphasized here is that, due to its specific form, the cosmological constant $\Lambda_{\mbox{\tiny DM,$\nu$}}$ can be arbitrarily small, though,
for example, the contribution of the Majorana masses $m_{\mbox{\tiny M,$\nu$}}$ to the
 ``dark matter/energy'' of the universe can be very high.

\section{Preliminaries}
For the sake of self-consistency and for the convenience of the reader, we summarize some facts about general Clifford modules although
later on we shall be mainly concerned with the case of twisted Grassmann bundles. Nonetheless, it seems worth starting with the general case to clarify the general scheme.
Afterwards, we shall introduce some facts concerning the case of twisted Grassmann bundles (resp. sub-bundles thereof).

The bundle of Grassmann and Clifford algebras with respect to $(M,g_{\mbox{\tiny M}})$ are supposed to be generated by the cotangent bundle of $M$. In what follows, however, we shall
be mainly concerned with their complexifications. In particular, all Clifford modules are considered as complex vector bundles.

\subsection{General Clifford modules}
To get started let
$(\ep,\gamma_{\!\mbox{\tiny$\ep$}})\twoheadrightarrow(M,g_{\mbox{\tiny M}})$ be a general bundle of
Clifford modules over a smooth, orientable (semi-)Riemannian manifold of even dimension
$n = p + q$ and signature $s = p - q.$ Let $Cl_{\mbox{\tiny M}}\twoheadrightarrow M$ be the
algebra bundle of (complexified) Clifford algebras with respect to the (semi-)metric $g_{\mbox{\tiny M}}$
that is generated by the cotangent bundle $T^*\!M\twoheadrightarrow M$. The mapping
$T^*\!M\mapright{\gamma_{\!\mbox{\tiny$\ep$}}}{\rm End}(\ep)$ denotes a Clifford mapping:
\bb
\gamma_{\!\mbox{\tiny$\ep$}}(\alpha)^2 = \varepsilon\,g_{\mbox{\tiny M}}(\alpha,\alpha)
\,{\rm id}_{\mbox{\tiny$\ep$}}\,,
\ee
for all $\alpha\in T^*\!M$. Here, the use of $\varepsilon\in\{\pm 1\}$ allows to treat both signatures $\pm s$ simultaneously. Especially, it takes into account that both signatures are
physically indistinguishable.

By abuse of notation, Clifford mappings also denote the induced representations of the Clifford bundle on the corresponding algebra bundles of endomorphisms ${\rm End}(\ep)\twoheadrightarrow M.$ Also, we do not distinguish between
(semi-)metrics of signature $s = p - q$ on $M$ and sections of the ``Einstein-Hilbert bundle''
\bb
\ep_{\mbox{\tiny EH}}:=
{\cal F}_{\!\!\mbox{\tiny M}}\times_{\rm GL(n)}{\rm GL(n)}/{\rm SO(p,q)}\twoheadrightarrow M
\ee
that is associated with the frame bundle ${\cal F}_{\!\!\mbox{\tiny M}}\twoheadrightarrow M$
of $M.$ Finally, the scalar products on the tangent and the cotangent bundle of $M$ are also denoted by $g_{\mbox{\tiny M}}.$

On every Clifford module there exists a canonical one-form
$\Theta\in\Omega^1(M,{\rm End}(\ep))$, which locally reads:
\bb
\label{canonical one-form}
\Theta \stackrel{\rm loc.}{=} {\mbox{\small$\frac{\varepsilon}{n}$}}\,e^k\ot\gamma_{\!\mbox{\tiny$\ep$}}(e_k^\flat)\,.
\ee
Here, $(e_1,\ldots,e_n)$ is a local (orthonormal) basis of $TM\twoheadrightarrow M$ and $(e^1,\ldots,e^n)$ its dual. The mappings: $\phantom{x}^{\flat/\sharp}:\,TM\rightleftarrows T^*\!M$, are the ``musical'' isomorphisms induced by $g_{\mbox{\tiny M}}.$

The canonical one-form (\ref{canonical one-form}) is thus the (normalized)
soldering form of the frame bundle of $M$ lifted to $\ep\twoheadrightarrow M.$ It also plays
a basic role in the definition of the twistor operator in conformal geometry\footnote{The
author would like to thank M. Schneider for pointing him out this relation.} (c.f. page 164, Lecture 6 in \cite{Bra:04}).
Indeed, the canonical one-form provides a right inverse of the restriction of the canonical mapping:
\bb
\delta_\gamma:\,\Omega^*(M,{\rm End}(\ep))&\longrightarrow&{\goth Sec}(M,{\rm End}(\ep))\cr
\omega\ot\chi &\mapsto&\gamma_{\!\mbox{\tiny$\ep$}}(\sigma^{-1}_{\mbox{\tiny Ch}}(\omega))\circ\chi
\ee
to one-forms via
\bb
ext_\Theta:\,\Omega(M,{\rm End}(\ep))&\longrightarrow&\Omega^1(M,{\rm End}(\ep))\cr
\Phi &\mapsto& \Theta\wedge\Phi\,.
\ee
Here,
\bb
\label{symbold map}
\sigma_{\mbox{\tiny Ch}}:\,Cl_{\mbox{\tiny M}}&\longrightarrow&\Lambda_{\mbox{\tiny M}}\cr
a &\mapsto& \gamma_{\mbox{\tiny Ch}}(a)1_{\mbox{\tiny$\Lambda$}}
\ee
denotes Chevalley's canonical linear isomorphism between the Clifford bundle and the
Grassmann bundle $\Lambda_{\mbox{\tiny M}}\twoheadrightarrow M$ of $M.$ It is based upon the Clifford mapping:
\bb
\gamma_{\mbox{\tiny Ch}}:\,
T^*\!M &\longrightarrow& {\rm End}(\Lambda_{\mbox{\tiny M}})\nonumber\\[.1cm]
\alpha &\mapsto& \left\{
\begin{array}{ccc}
  \Lambda_{\mbox{\tiny M}} & \longrightarrow & \hspace{-3cm}\Lambda_{\mbox{\tiny M}} \\
  \omega & \mapsto & \varepsilon\,int_g(\alpha)\omega + ext(\alpha)\omega\,.
\end{array}
\right.
\ee
Here, respectively, $int_g(\alpha)\omega$ is the contraction (inner derivative) of $\omega$ with respect to $\alpha^\sharp\in
TM$ and $ext(\alpha)\omega$ denotes the exterior multiplication of $\omega$ with respect to $\alpha\in T^*\!M$.

The isomorphism (\ref{symbold map}) is referred to as ``symbol map'' and its inverse
as ``quantization map''. Although misleading from a physical point of view, we shall
still use this common term and call the section
$\,{/\!\!\!\!\alpha}\equiv\delta_\gamma(\alpha)\in{\goth Sec}(M,{\rm End}(\ep))$
the ``quantization'' of the ``non-commutative super-field''
$\alpha\in\Omega^*(M,{\rm End}(\ep)) =
\bigoplus_{k=0}^n{\goth Sec}(M,\Lambda^{\!k}\,T^*\!M\ot{\rm End}(\ep)).$

On the affine set ${\cal A}(\ep)$ of all (linear) connections on $\ep\twoheadrightarrow M$
there exists a distinguished affine subset, consisting of what is called {\it Clifford connections}. This subset may be
characterized as follows:
\bb
\label{clifford connections}
{\cal A}_{\mbox{\tiny Cl}}(\ep) := \{\da\in{\cal A}(\ep)\,|\,
\da^{\mbox{\tiny$T^*\!M\!\ot\!\enep$}}\Theta\equiv 0\}\,.
\ee

We call a first order differential operator $\,\DDD,$ acting on sections of $\ep\twoheadrightarrow M,$ of {\it Dirac type},
provided it fulfills:
\bb
[\,\DDD,f] = \gamma_{\!\mbox{\tiny$\ep$}}(df)\,,
\ee
for all $f\in{\cal C}^\infty(M).$ The set of all such operators is denoted by $\dep.$

An odd Dirac type operator $\,\DDD\in\dep$ on a $\zz_2-$graded Clifford module bundle
$(\ep,\tau_{\!\mbox{\tiny$\ep$}},\gamma_{\!\mbox{\tiny$\ep$}})$ is called a {\it Dirac operator}. Here,
$\tau_{\!\mbox{\tiny$\ep$}}\in\enep$ denotes the underlying grading
involution, such that $\,\DDD\in\dep$ is a Dirac operator provided it satisfies: $\,\DDD\circ\tau_{\!\mbox{\tiny$\ep$}} = -\tau_{\!\mbox{\tiny$\ep$}}\circ\,\DDD$.

At this point, we have to warn the reader. Usually, every Dirac type operator is assumed to be odd. For reasons that will become clear in the next section, however, we want to distinguish between Dirac type operators and Dirac operators on a Clifford module. Clearly,
every Dirac operator is of Dirac type. Moreover, every Dirac type operator may be written as
\bb
\DDD = \dda + \Phi_{\!\mbox{\tiny A}}\,,
\ee
where $\dda\equiv\delta_\gamma\circ\da$ and $\Phi_{\!\mbox{\tiny A}}\in
{\goth Sec}(M,{\rm End}(\ep)),$ in general, will depend on the choice of the Clifford connection $\da\in{\cal A}_{\mbox{\tiny Cl}}(\ep).$

Every Dirac type operator $\DDD\in\dep$ has a unique first order and second order decomposition:
\bb
\label{1st/2nd decomp. of dop}
\DDD &=& \ddb + \Phi_{\mbox{\tiny D}}\,,\\
\DDD^2 &=& \triangle_{\mbox{\tiny B}} + V_{\!\mbox{\tiny D}}\,.
\ee
Here, $\ddb \equiv \delta_\gamma\circ\db$ is the quantization of the {\it Bochner connection}
$\db\in{\cal A}(\ep)$ that is uniquely defined by $\DDD$ via
\bb
2\,ev_{\!g}(df,\db\psi) := \varepsilon\left([\,\DDD^2,f] -
\delta_{\!\mbox{\tiny$g$}}df\right)\!\psi\,,
\ee
for all $f\in{\cal C}^\infty(M)$ and $\psi\in{\goth Sec}(M,\ep)$. The second order operator:
\bb
\triangle_{\mbox{\tiny B}}:=\varepsilon\, ev_{\!g}\left(\db^{\mbox{\tiny$T^*\!M\!\ot\!\ep$}}\circ\db\right)\,,
\ee
is the induced {\it Bochner-Laplacian} (or ``trace/connection Laplacian'', see, for example, in \cite{BG:90} and \cite{Gil:95}, as well as in Chapter 2.1 in  \cite{BGV:96}).

With every Dirac type operator $\,\DDD\in\dep$ there is naturally associated with a connection $\dD\in{\cal A}(\ep),$ such that $\ddD\equiv\delta_\gamma\circ\dD = \,\DDD.$ This {\it Dirac connection} is given by (c.f. \cite{ToTh:05})
\bb
\label{dirac connection}
\dD := \db + ext_\Theta(\Phi_{\mbox{\tiny D}})\,.
\ee
We call the one-form $\omega_{\mbox{\tiny D}}:=ext_\Theta(\,\DDD - \ddb),$ uniquely associated with $\,\DDD\in\dep$, the {\it Dirac form} and the tangent vector field:
$\xi_{\mbox{\tiny D}}:=
-\varepsilon({\rm tr}_{\!\mbox{\tiny$\ep$}}\omega_{\mbox{\tiny D}})^\sharp$, the
{\it Dirac vector field}.

It follows that
\bb
\label{dirac function}
{\rm tr}_{\!\mbox{\tiny$\ep$}}V_{\!\mbox{\tiny D}} =
{\rm tr}_\gamma\!\left(curv(\,\DDD) - \varepsilon ev_g(\omega^2_{\mbox{\tiny D}})\right) +
div\xi_{\mbox{\tiny D}}\,,
\ee
where
\bb
curv(\,\DDD) := \dD\wedge\dD\in\Omega^2(M,{\rm End}(\ep))
\ee
is the curvature of the Dirac type operator $\,\DDD\in\dep$ and
${\rm tr}_\gamma\equiv{\rm tr}_{\!\mbox{\tiny$\ep$}}\circ\delta_\gamma$ is the ``quantized trace''.

We call the Hodge dual of the smooth function (\ref{dirac function}) the
{\it universal Dirac-Lagrangian}:
\bb
{\cal L}_{\!\mbox{\tiny D}} := \ast{\rm tr}_{\!\mbox{\tiny$\ep$}}V_{\!\mbox{\tiny D}}\,.
\ee
Its cohomology class is generated by the top-form:
${\rm tr}_\gamma\!\left(curv(\,\DDD) - \varepsilon ev_g(\omega^2_{\mbox{\tiny D}})\right)
dvol_{\mbox{\tiny M}}$.

It follows that two Dirac type operators $\,\DDD,\,\DDD'\in\dep$ define the same
Bochner connection, provided $\,\DDD' - \,\DDD$ anti-commutes with the Clifford connection
(c.f. Corollary \ref{korrollar simple type dop}). Therefore, on every $\zz_2-$graded Clifford module there is a distinguished class of Dirac type operators, depending on whether $\Phi_{\mbox{\tiny D}} = \,\DDD - \ddb$ even or odd. In particular, we call a Dirac operator of {\it simple type} if it reads:
\bb
\label{dops of simple type}
\DDD = \dda + \tau_{\!\mbox{\tiny$\ep$}}\circ\phi_{\!\mbox{\tiny$\ep$}}\,,
\ee
where $\phi_{\!\mbox{\tiny$\ep$}}\in{\rm Sec}(M,{\rm End}^-_\gamma(\ep))$. Here,
${\rm End}_\gamma(\ep)\hookrightarrow{\rm End}(\ep)\twoheadrightarrow M$ is the algebra sub-bundle of all $\gamma_{\!\mbox{\tiny$\ep$}}-$invariant endomorphisms and $\tau_{\!\mbox{\tiny$\ep$}}$ is the underlying grading involution on
$\ep\twoheadrightarrow M.$

The specific form (\ref{dops of simple type}) of simple type Dirac operators is determined by the condition
that the corresponding Bochner connections are given by Clifford connections. Of course, every $\dda\in\dep$
is of simple type.

Likewise, one may consider Dirac type operators of the form
\bb
\label{ymh-dop}
\DDD &:=& \dda + \Phi_{\!\mbox{\tiny H}}\,,\nonumber\\[.1cm]
\Phi_{\!\mbox{\tiny H}}\!\!\!
&\in&
\!\!\!{\rm Sec}(M,{\rm End}_\gamma(\ep))\,.
\ee
These Dirac type operators have the property that their Dirac connections read:
\bb
\label{ymh-connection}
\dD &=& \db + ext_\Theta(\Phi_{\mbox{\tiny D}})\cr
&=&
\da + ext_\Theta(n\varepsilon\,\Phi_{\!\mbox{\tiny H}}) +
ext_\Theta[(1 - n\varepsilon)\Phi_{\!\mbox{\tiny H}}]\cr
&=&
\da + \Phi_{\!\mbox{\tiny H}}\Theta\cr
&\equiv&
\da + H\,.
\ee
We call the one-form
\bb
\label{higgs gauge pot.}
H &:=& \Phi_{\!\mbox{\tiny H}}\Theta\nonumber\\[.1cm]
&\stackrel{loc.}{=}&
\mbox{\small$\frac{\varepsilon}{n}$}\,
e^k\ot\gamma_{\mbox{\tiny$\ep$}}(e_k^\flat)\circ\Phi_{\!\mbox{\tiny H}}
\ee
the {\it Higgs gauge potential} and the connections:
$\partial_{\mbox{\tiny YMH}} := \da + H \stackrel{loc.}{=}\partial + A + H$,
{\it Yang-Mills-Higgs connections} on the Clifford module $\ep\twoheadrightarrow M.$

We remark that for $\Phi_{\!\mbox{\tiny H}}\in{\rm Sec}(M,{\rm End}^+_\gamma(\ep))$, the Yang-Mills-Higgs connections are odd connections. They have the property that the (locally defined) Yang-Mills gauge potentials $A\in\Omega^1(M,{\rm End}^+_\gamma(\ep))$ provide connections which respect the sub-bundles $\ep^\pm\twoheadrightarrow M$. The Yang-Mills part of a Yang-Mills-Higgs connection is thus ``chirality preserving''. In contrast, the Higgs gauge potential $H\in\Omega^1(M,{\rm End}^-(\ep))$ provides a connection between these sub-bundles of $\ep\twoheadrightarrow M$ and thus constitutes the ``chirality violating'' part of the Yang-Mills-Higgs gauge potential. This is similar to the geometrical interpretation of the connections constructed within the original Connes-Lott description of the Yang-Mills-Higgs sector of the Standard Model (c.f. \cite{CL:90}, \cite{GV:93}, \cite{Con:94}, \cite{SZ:95}, \cite{KS:96}; see also \cite{MO:94} and \cite{MO:96} in the case of alternative approaches). However, in contrast to non-commutative geometry, where mainly the algebraic structure of connections is taken into account, Dirac connections are always related to the underlying geometry that is encoded within the canonical one-form (\ref{canonical one-form}). In particular, within the scheme presented, the Higgs gauge potential is intimately related to gravity.

\begin{definition}
A Clifford module bundle $(\ep,\gamma_{\!\mbox{\tiny$\ep$}})\twoheadrightarrow(M,g_{\mbox{\tiny M}})$ is said to be
``flat'', provided there is a Clifford connection $\da\in{\cal A}_{\mbox{\tiny Cl}}(\ep)$
fulfilling
\bb
\label{flat clifford con.}
curv(\dda)\, = \,{/\!\!\!\!R}iem(g_{\mbox{\tiny M}})\,,
\ee
where $\,{/\!\!\!\!R}iem(g_{\mbox{\tiny M}})\in\Omega^2(M,{\rm End}(\ep))$ is the Riemann curvature with respect to $g_{\mbox{\tiny M}}$ lifted to the Clifford module.
\end{definition}

We call a Clifford connection satisfying (\ref{flat clifford con.}) a
{\it flat Clifford connection} and denote it by $\partial\in{\cal A}_{\mbox{\tiny Cl}}(\ep).$

Let $(\ep,\gamma_{\!\mbox{\tiny$\ep$}},\gamma_{\!\mbox{\tiny$\ep$,op}})\twoheadrightarrow M$ be a {\it Clifford bi-module}. Besides the Clifford left action, provided by the Clifford mapping $\gamma_{\!\mbox{\tiny$\ep$}}:\,T^*\!M\rightarrow\enep,$ there is also a right action of $Cl_{\mbox{\tiny M}}\twoheadrightarrow M$ on $\ep\twoheadrightarrow M$. Accordingly, there is a Clifford left action of the bundle of {\it opposite Clifford algebras}: $Cl_{\mbox{\tiny M}}^{\mbox{\tiny op}}\twoheadrightarrow M$ that is induced by a Clifford mapping $\gamma_{\!\mbox{\tiny$\ep$,op}}:\,T^*\!M\rightarrow\enep$. This left action
of $Cl_{\mbox{\tiny M}}^{\mbox{\tiny op}}$ on $\ep$ is again denoted by $\gamma_{\!\mbox{\tiny$\ep$,op}}.$ Note that
\bb
\gamma_{\!\mbox{\tiny$\ep$,op}}({\goth a})\in{\rm End}_\gamma(\ep)
\ee
for all ${\goth a}\in Cl_{\mbox{\tiny M}}^{\mbox{\tiny op}}$.

On a Clifford bi-module there exists a distinguished class of connections.
\begin{definition}
Let $(\ep,\gamma_{\!\mbox{\tiny$\ep$}},\gamma_{\!\mbox{\tiny$\ep$,op}})\twoheadrightarrow M$ be a Clifford bi-module. A connection $\naep\in{\cal A}(\ep)$ is called ``$S-$reducible'',
provided it is a right Clifford connection:
\bb
\nabla^{\mbox{\tiny${\rm End}(\ep)$}}_{\!\!\xi}
\gamma_{\!\mbox{\tiny$\ep$,op}}({\goth a}) &:=&
[\naep_{\!\!\!\!\!\xi},\gamma_{\!\mbox{\tiny$\ep$,op}}({\goth a})]
\nonumber\\[.1cm]
&=&
\gamma_{\!\mbox{\tiny$\ep$,op}}(
\nabla_{\!\!\xi}^{\mbox{\tiny${\rm Cl}^{\rm op}$}}{\goth a})\,,
\ee
for all ${\goth a}\in{\goth Sec}(M,Cl_{\mbox{\tiny M}}^{\mbox{\tiny op}})$ and
$\xi\in{\goth Sec}(M,TM)$. The (affine) sub-space of $S-$reducible connections is denoted by
${\cal A}_{\mbox{\tiny S}}(\ep)\subset{\cal A}(\ep).$
\end{definition}

We make use of the following (common) terminology: A ``Clifford module'' generically means
a Clifford left module. Accordingly, ``Clifford connections'' always refer to the appropriate
left action. Likewise, the notion of ``Dirac (type) operators'' also refers to this left action. However, on a Clifford bi-module every Dirac (type) operator $\,\DDD\in\dep$ may
be considered to act on ${\goth Sec}(M,\ep)$ either from the left, or from the right. That is, one has to distinguish between ``left-Dirac (type) operators'' and ``right-Dirac (type) operators''. In the sequel, ``Dirac (type) operators'' always mean left-Dirac (type) operators. Note that for every right-Dirac (type) operator there is a unique Dirac (type) operator $\,\DDD_{\!\!\mbox{\tiny op}}\in\dep$, which acts from the left on
${\goth Sec}(M,\ep)$ via $\gamma_{\!\mbox{\tiny$\ep$,op}}$. Clearly, every $\,\DDD\in\dep$ uniquely defines an appropriate
$\,\DDD_{\!\!\mbox{\tiny op}}\in\dep$ and vice versa. We thus call
$\,\DDD_{\!\!\mbox{\tiny op}}$ the {\it opposite Dirac (type) operator} associated with
$\,\DDD\in\dep$.

Note that $S-$reducible connections on a Clifford bi-module may also be characterized by the
requirement
\bb
\nabla^{\mbox{\tiny$T^*\!M\!\ot\!{\rm End}(\ep)$}}\Theta_{\!\mbox{\tiny op}} = 0\,.
\ee
Here, $\Theta_{\!\mbox{\tiny op}}\stackrel{loc.}{=}\mbox{\small$\frac{\varepsilon}{n}$}\,
e^k\ot\gamma_{\!\mbox{\tiny$\ep$,op}}(e_k^\flat)$ is the canonical one-form represented
on the Clifford bi-module via $\gamma_{\!\mbox{\tiny$\ep$,op}}.$ Hence,
$\delta_{\gamma_{\mbox{\tiny op}}}\circ ext_{\Theta_{\!\mbox{\tiny op}}} = \mbox{\small$\frac{\varepsilon}{n}$}\,\gamma_{\!\mbox{\tiny$\ep$,op}}(e_k^\flat\, e^k) =
{\rm id}_{\mbox{\tiny$\ep$}}$.

Clearly, the Grassmann bundle provides the archetypical example of a Clifford bi-module.

\begin{definition}
A Dirac (type) operator $\,\DDD\in\dep$ on a Clifford bi-module is called ``$S-$re\-ducible'', provided its Dirac connection is
$S-$reducible: $\dD\in{\cal A}_{\mbox{\tiny S}}(\ep).$ The set of all S-reducible Dirac (type) operators is denoted by
${\cal D}_{\mbox{\tiny S}}(\ep).$
\end{definition}

In this section, we summarized some basic notions with respect to general Clifford (bi-) modules. In the sequel, we shall
restrict our further discussion mainly to the more specific case of twisted Grassmann bundles. On the one hand, this case
is broad enough to geometrically describe most of the cases encountered in physics. On the other hand, it is topologically
less restrictive than the case of a twisted spinor bundle.

\subsection{Twisted Grassmann bundles}
Basically, the advantage in restricting to twisted Grassmann bundles is provided by the fact that each section of the Einstein-Hilbert bundle then yields a natural Clifford action and thus turns the twisted Grassmann bundle into a Clifford module bundle. This Clifford
action, of course, is given by Chevalley's canonical isomorphism (\ref{symbold map}), which only takes the metric structure into account.

Therefore, let $E\twoheadrightarrow M$ be any Hermitian and (maybe trivially)
$\zz_2-$graded complex vector bundle of finite rank. We then consider
\bb
\label{twisted grassmann bdl.}
{\cal S} := \Lambda_{\mbox{\tiny M}}\ot_\cc E \twoheadrightarrow M\,.
\ee

Any section $g_{\mbox{\tiny M}}\in{\goth Sec}(M,\ep_{\mbox{\tiny EH}})$ turns (\ref{twisted grassmann bdl.}) into a bundle of Clifford left modules according to the action:
\bb
Cl_{\mbox{\tiny M}}\times_M{\cal S} &\twoheadrightarrow& {\cal S}\cr
(a, \omega\ot\chi) &\mapsto&
\sigma_{\!\mbox{\tiny Ch}}(a\sigma^{-1}_{\!\mbox{\tiny Ch}}(\omega))\ot\chi\,,
\ee
where the Clifford multiplication is denoted by juxtaposition. The underlying Clifford mapping is denoted, again, by $\gamma_{\mbox{\tiny Ch}}$ and no distinction is made between the Clifford mapping and its induced homomorphism.

Note that
\bb
{\rm End}_\gamma(\sss) = Cl^{\mbox{\tiny op}}_{\mbox{\tiny M}}\times_M{\rm End}(E)\,,
\ee
with $Cl^{\mbox{\tiny op}}_{\mbox{\tiny M}}\twoheadrightarrow M$ acting from the right on $\sss\twoheadrightarrow M.$ Also, note that any
$g_{\mbox{\tiny M}}\in{\goth Sec}(M,\ep_{\mbox{\tiny EH}})$ turns $\sss\twoheadrightarrow M$ into a Hermitian vector bundle:
\bb
\langle\omega_1\ot\chi_1,\omega_2\ot\chi_2\rangle_{\!\mbox{\tiny$\sss$}} \,=\,
g_{\!\mbox{\tiny$\Lambda{\rm M}$}}(\omega_1,\omega_2)
\langle\chi_1,\chi_2\rangle_{\mbox{\tiny E}}\,.
\ee
Here, $\langle\cdot,\cdot\rangle_{\mbox{\tiny E}}$ is the Hermitian product on
$E\twoheadrightarrow M$ and $g_{\!\mbox{\tiny$\Lambda{\rm M}$}}$ is the induced (semi-)metric
on the Grassmann bundle $\Lambda_{\mbox{\tiny M}}\twoheadrightarrow M.$

A first order differential operator $\DDD$, acting on sections of $\sss\twoheadrightarrow M$,
is called of Dirac type, if there is a section
$g_{\mbox{\tiny M}}\in{\goth Sec}(M,\ep_{\mbox{\tiny EH}})$ such that
\bb
[\,\DDD,f] = \gamma_{\mbox{\tiny Ch}}(df)\,,
\ee
for all $f\in{\cal C}^\infty(M).$ The set of all of these operators is denoted by
${\cal D}(\sss).$ Similar to the general case, a Dirac operator on a twisted Grassmann bundle is defined to be an odd Dirac type operator with respect to the particular grading involution:
\bb
\tau_{\mbox{\tiny$\sss$}} &=& \tau_{\mbox{\tiny M}}\circ\iota_{\!\mbox{\tiny$\sss$}}\,,\cr
\iota_{\!\mbox{\tiny$\sss$}} &:=& {\rm id}_{\mbox{\tiny$\Lambda$}}\ot\tau_{\mbox{\tiny E}}\,.
\ee
Here, $\tau_{\mbox{\tiny E}}$ is a grading involution on $E\twoheadrightarrow M$ that can also be trivial. The {\it chirality} involution $\tau_{\mbox{\tiny M}}\in{\rm End}(\ep)$ is defined by
\bb
\label{chirality invol.}
\tau_{\mbox{\tiny M}} := {\mbox{\small$\sqrt{(-1)^{n(n-1)/2 + q}}$}}\;
\delta_\gamma(dvol_{\mbox{\tiny M}})\,,
\ee
with $dvol_{\mbox{\tiny M}}\in\Omega^n(M)$ being the metric induced volume form. We call
$\iota_{\!\mbox{\tiny$\sss$}}\in{\rm End}_\gamma(\sss)$ the ``inner involution''.

The {\it universal Dirac action} is (formally) defined by the
functional\footnote{In general, the domain of definition of the universal Dirac
action is an appropriate subset of $\dep$, only, for $M$ is not supposed to be compact.}:
\bb
\label{universal dirac action}
{\cal I}_{\mbox{\tiny D}}:\,{\cal D}(\sss) &\longrightarrow&\cc\nonumber\\[.1cm]
\DDD\,&\mapsto& \langle\!\langle[M],\,[{\cal L}_{\mbox{\tiny D}}]\rangle\!\rangle
\,\equiv\,
\int_M\!{\cal L}_{\mbox{\tiny D}}\,.
\ee
The {\it total Dirac action} is given by the functional:
\bb
\label{total dirac action}
{\cal I}_{\mbox{\tiny D,tot}}:\,{\goth Sec}(M,\sss)\times{\cal D}(\sss)
&\longrightarrow&\cc\nonumber\\[.1cm]
(\psi,\,\DDD) &\mapsto& \langle\!\langle\psi,\,\DDD\psi\rangle\!\rangle\; +\;
{\cal I}_{\mbox{\tiny D}}(\,\DDD)\,,
\ee
with
\bb
\label{fermionic action}
\langle\!\langle\psi,\,\DDD\psi\rangle\!\rangle :=
\int_M\langle\psi,\,\DDD\psi\rangle_{\mbox{\tiny$\sss$}}\,dvol_{\mbox{\tiny M}}\,.
\ee

For every (symmetric) $\DDD\in{\cal D}(\sss)$ the functional (\ref{fermionic action}) is considered as a (real-valued) quadratic form on ${\goth Sec}(M,\sss)$. It is called the
{\it fermionic} part of the total Dirac action. Accordingly, the universal Dirac action (\ref{universal dirac action}) is referred to as the {\it bosonic} part of the total Dirac action.

It follows that the {\it gauge group of the total Dirac action} is given by the semi-direct product:
\bb
\label{gauge group tda}
{\cal G}_{\mbox{\tiny D}} = {\cal D}{\rm iff}(M)\ltimes{\cal G}_{\mbox{\tiny$\sss$}}\,,
\ee
with ${\cal G}_{\mbox{\tiny$\sss$}}$ being the gauge group of $\sss\twoheadrightarrow M.$
For every section $g_{\mbox{\tiny M}}$ this gauge group explicitly reads:
\bb
{\cal G}_{\mbox{\tiny$\sss$}} = {\cal G}_{\mbox{\tiny EH}}\times{\cal G}_{\mbox{\tiny YM}}
\ee
Here, ${\cal G}_{\mbox{\tiny YM}}\simeq{\rm Aut}_\gamma(\sss)$ is the subgroup of all
bundle automorphisms on $\sss\twoheadrightarrow M$ which are $\gamma-$invariant and
${\cal G}_{\mbox{\tiny EH}}$ is the gauge group of the SO(p,q)-reduced frame bundle.

In contrast, the {\it gauge group of the universal Dirac action} is provided by the affine group:
\bb
\label{gauge group uda}
{\cal P}_{\!\mbox{\tiny D}} = {\cal G}_{\mbox{\tiny D}}\ltimes{\cal T}_{\mbox{\tiny D}}\,,
\ee
with the {\it translational group} being given by
\bb
{\cal T}_{\mbox{\tiny D}}\simeq\Omega^1(M,{\rm End}_\gamma(\sss))\,.
\ee
Its action on ${\cal D}(\sss)$ reads: $\DDD\,\mapsto\,\DDD\, + \,{/\!\!\!\!\alpha}.$ We stress that the universal Dirac-Lagrangian is invariant with respect to this action.

We close this section with the following remarks concerning the case of {\it twisted spinor bundles}\footnote{The author would like to thank V. Soucek for appropriate remarks.}.
For this let $M$ be an even-dimensional, orientable spin-manifold. In this case, every Clifford module bundle $\ep\twoheadrightarrow M$ is equivalent to a twisted spinor bundle $\sss\ot\www\twoheadrightarrow M.$ Here, the (total space of the) vector bundle
$\www := {\rm Hom}_\gamma(\sss,\ep)\twoheadrightarrow M$ is defined by the $\gamma-$equivariant
homomorphisms (c.f. \cite{ABS:64} and Sec. 3.3 in \cite{BGV:96}). Basically, this follows from Wedderburn's structure theorems about invariant linear mappings (see, for example, Chap. 11
in \cite{Gre:78}). Accordingly, the above mentioned equivalence is provided by the evaluation map. Although the spinor module carries a canonical Clifford action according to the identification $Cl_{\mbox{\tiny M}}\ot\cc\simeq{\rm End}(\sss)$, there are usually different
(i.e. inequivalent) spin-structures for given
$g_{\mbox{\tiny M}}\in{\goth Sec}(M,\ep_{\mbox{\tiny EH}})$
(see, for example, Chap. 3 in \cite{BGV:96} and Sec. 1.8 in \cite{Jos:98}). This
makes the actual domain of definition of the Dirac action geometrically more interesting,
for one may ask how the Dirac action changes with a change of the spin-structure (c.f., for example, \cite{Bau:81} and \cite{Fri:84}). One may also take into account what is called ``generalized spin structures'', or ``canonically generalized spin structures'' (c.f., for example, \cite{AI:80}, \cite{Hes:94} and \cite{Hes:96}). In fact, in the case of real representations the appropriate discussions presented in \cite{Hes:96} seem to fit with the discussion presented in this work.  Clearly, if the spin-structure is basically unique, then the case of twisted spinor bundles can be similarly treated to the case of twisted Grassmann bundles.

\section[The geometrical picture]{The geometrical picture of Dirac type operators and the Einstein-Hilbert action}
In this section, we briefly discuss the geometrical picture that underpins the Einstein-Hilbert action, ${\cal I}_{\mbox{\tiny EH}}$, when the latter is expressed in
terms of Dirac type operators. From the usual Lichnerowicz/Schr\"odinger decomposition of $\,\dda^2$ (c.f. \cite{Schr:32} and \cite{Lich:63}):
\bb
\label{lichnerowic-formula}
\dda^2 - \varepsilon ev_{\!g}(\da^{\mbox{\tiny$T^*\!M\!\ot\!\sss$}}\circ\da)
&=&
\delta_\gamma(curv(\,\dda))\nonumber\\[.1cm]
&=& -\mbox{\small$\frac{\varepsilon}{4}$}\,scal(g_{\mbox{\tiny M}}) +
\delta_\gamma(F_{\!\!\mbox{\tiny A}})\,,
\ee
it follows that
\bb
\label{EH-action}
{\cal I}_{\mbox{\tiny EH}}(g_{\mbox{\tiny M}}) \thicksim
\int_M\!\ast{\rm tr}_\gamma(curv(\dda))\,.
\ee
Note that the ``relative curvature'': $F_{\!\!\mbox{\tiny A}}:= curv(\,\dda) - \,{/\!\!\!\!R}iem(g_{\mbox{\tiny M}})$, of Clifford connections\footnote{In the case of Clifford connections, the relative curvature of $\,\dda$ is also called ``twisting curvature''.} has the peculiar property that $F_{\!\!\mbox{\tiny A}}\in\Omega^2(M,{\rm End}^+_\gamma(\sss))$. Therefore, ${\rm tr}_\gamma(F_{\!\!\mbox{\tiny A}})\equiv 0$.

This description of the Einstein-Hilbert action allows us to point out a subtle difference between the fermionic and the bosonic actions, usually not taken into account. This difference provides the geometrical origin of the difference between the respective gauge groups of the fermionic and the bosonic part of the total Dirac action. The discussion presented in this section will eventually yield some motivation for the ``Pauli map'' that is introduced in the next section, which permits to interpret the Yang-Mills and the Standard Model action as natural generalizations of the Einstein-Hilbert action with a ``cosmological constant term''.

Let
$\sss_{\mbox{\tiny D}}:=\ep_{\mbox{\tiny EH}}\times_M{\rm End}(\sss)\twoheadrightarrow M.$
We consider the quotient
\bb
\Gamma_{\!\mbox{\tiny D}}:=
{\goth Sec}(M,\sss_{\mbox{\tiny D}})/{\cal T}_{\mbox{\tiny D}}\,,
\ee
with the equivalence relation given by
\bb
(g_{\mbox{\tiny M}},\Phi) \thicksim (g'_{\mbox{\tiny M}},\Phi')\quad:\Leftrightarrow\quad
\left\{\begin{array}{ccc}
         g'_{\mbox{\tiny M}} & = & \hspace{-.7cm} g_{\mbox{\tiny M}}\,, \\
         \Phi' & = & \Phi + \,{/\!\!\!\!\alpha}\,.
       \end{array}
\right.
\ee

It follows that $\Gamma_{\!\mbox{\tiny D}}\simeq{\cal D}(\sss)/{\cal T}_{\mbox{\tiny D}}$. Therefore, the restriction to $S-$reducible Dirac connections yields a principal fibering
\bb
\label{principal fibering}
{\cal D}_{\mbox{\tiny S}}(\sss)&\twoheadrightarrow&\Gamma_{\!\mbox{\tiny D}}\cr
\DDD = \dda + \Phi_{\!\mbox{\tiny A}} &\mapsto&
[(g_{\mbox{\tiny M}},\Phi_{\!\mbox{\tiny A}})]\,,
\ee
with typical fiber given by the abelian group $\Omega^1(M,{\rm End}(E)).$

The principal fibering (\ref{principal fibering}) is clearly trivial but only in a non-canonical way unless the twisting part of $\sss\twoheadrightarrow M$ is given by the trivial bundle
$E = M\times\cc^{\mbox{\tiny N}}\twoheadrightarrow M$. This holds true also in the case where
$\sss\twoheadrightarrow M$ is supposed to be flat for every $g_{\mbox{\tiny M}}.$ Indeed, every choice of a connection on $E\twoheadrightarrow M$ yields a trivializing section:
\bb
\label{trivializing section}
\sigma_{\!\mbox{\tiny A}}:\,\Gamma_{\!\mbox{\tiny D}}&\longrightarrow&
{\cal D}_{\mbox{\tiny S}}(\sss)\cr
[(g_{\mbox{\tiny M}},\Phi)] &\mapsto& \dda + \Phi\,.
\ee

It follows that $\sigma_{\!\mbox{\tiny A}}^*{\cal I}_{\mbox{\tiny D}}$ is independent of the choice of the trivializing section, because of the translational invariance of the universal Dirac action. In particular, when restricted to the distinguished subset:
\bb
\Gamma_{\!\mbox{\tiny EH}} &:=& \{\,[(g_{\mbox{\tiny M}},\Phi)]\in
\Gamma_{\!\mbox{\tiny D}}\,|\,\Phi\thicksim\,{/\!\!\!\!\alpha}\,\}\nonumber\\[.1cm]
&\simeq&
{\goth Sec}(M,\ep_{\mbox{\tiny EH}})\,,
\ee
every trivializing section (\ref{trivializing section}) yields the Einstein-Hilbert functional:
\bb
\sigma_{\!\mbox{\tiny A}}^*{\cal I}_{\mbox{\tiny D}}:\,{\goth Sec}(M,\ep_{\mbox{\tiny EH}})
&\longrightarrow&\cc\cr
g_{\mbox{\tiny M}} &\mapsto& {\cal I}_{\mbox{\tiny EH}}(g_{\mbox{\tiny M}})\,.
\ee

The sections $g_{\mbox{\tiny M}}\in{\goth Sec}(M,\ep_{\mbox{\tiny EH}})$ are thus geometrically represented on ${\cal D}_{\mbox{\tiny S}}(\sss)$ by the trivializing sections
(\ref{trivializing section}):
\bb
\sigma_{\!\mbox{\tiny A}}(g_{\mbox{\tiny M}}) = \dda =
d_{\!\mbox{\tiny A}} + \varepsilon\delta_{\!\mbox{\tiny$g,A$}}\,.
\ee
Here, $\delta_{\!\mbox{\tiny$g,A$}}$ denotes the formal adjoint of the exterior covariant derivative $d_{\!\mbox{\tiny A}}$ that is defined with respect to some Clifford connection.

Accordingly, the geometrical meaning of these gauge sections is to make the metric on $M$ ``covariant'' on $\sss$. This geometrical view of the gauge sections becomes most apparent
for flat modules (i.e. for flat $E\twoheadrightarrow M$). In this case, one gets:
\bb
\label{triv. sec., flat modules}
\sigma_{\!\mbox{\tiny A}}(g_{\mbox{\tiny M}}) &=& \ddd\, + \,{/\!\!\!\!A}\cr
&=&
d + \varepsilon\delta_{\!\mbox{\tiny$g$}} + \,{/\!\!\!\!A}\,,
\ee
with the Gauss-Bonnet-Hodge-de Rham operator,
$d + \varepsilon\delta_{\!\mbox{\tiny$g$}}$, being determined by the (semi-)metric $g_{\mbox{\tiny M}}$.

We emphasize that this geometrical picture of the (semi-)metric is provided by the translational invariance of the universal Dirac-Lagrangian. Finally, any Dirac (type) operator $\DDD\in{\cal D}(\sss)$ may be locally regarded as a ``generalized covariance'' of its underlying (semi-)Riemannian metric $g_{\mbox{\tiny M}}$:
\bb
\sigma_{\mbox{\tiny$\Phi_{\!\!A}$}}(g_{\mbox{\tiny M}})\equiv
\sigma_{\!\mbox{\tiny A}}([(g_{\mbox{\tiny M}},\Phi)]) \;\stackrel{\rm loc.}{=}\;
d + \varepsilon\delta_{\!\mbox{\tiny$g$}} + \Phi_{\!\mbox{\tiny A}}\,,
\ee
where locally: $\Phi_{\!\mbox{\tiny A}} := \Phi + \,{/\!\!\!\!A}.$

So far, we discussed the geometrical picture of the Einstein-Hilbert action, when the latter is described in terms of the universal Dirac action. It is natural to ask for the appropriate substitute of the Einstein-Hilbert action with a cosmological constant $\Lambda\in\rr$:
\bb
\label{einstein-bb}
{\cal I}_{\mbox{\tiny${\rm EH},\Lambda$}}(g_{\mbox{\tiny M}}) =
\int_M \ast(scal(g_{\mbox{\tiny M}}) + \Lambda)\,.
\ee

To answer this question, we take into account (\ref{EH-action}) that expresses the Einstein-Hilbert action in terms of the curvature of the quantized Yang-Mills connection
$\partial_{\mbox{\tiny YM}} \equiv \da \stackrel{\rm loc.}{=} \partial + A$. It may
thus not come as a big surprise that the functional (\ref{einstein-bb}) turns
out to be expressible in terms of the curvature of the quantized Yang-Mills-Higgs connection
$\partial_{\mbox{\tiny YMH}} \stackrel{\rm loc.}{=}\partial + A + H$:
\bb
\label{EH-action plus cosm. const.}
{\cal I}_{\mbox{\tiny${\rm EH},\Lambda$}}(g_{\mbox{\tiny M}}) &\sim&
\int_M\ast{\rm tr}_\gamma\!\left(curv(\,\ddd_{\!\mbox{\tiny YMH}}) -
\varepsilon ev_{\!g}(\omega_{\!\mbox{\tiny D}}^2)\right)\nonumber\\[.1cm]
&=&
\int_M\ast\!\left({\rm tr}_\gamma curv(\,\ddd_{\!\mbox{\tiny YM}}) -
\Lambda_{\mbox{\tiny H}}\right)\,,
\ee
whereby the ``cosmological constant'' reads:
\bb
\label{cosmological const. vers. Higgs}
\Lambda \equiv \Lambda_{\mbox{\tiny H}} &:=& \lambda\,{\rm tr}_g H^2\cr
&=&
\lambda'\,{\rm tr}_{\!\mbox{\tiny$\sss$}}\Phi_{\!\mbox{\tiny H}}^2\,.
\ee
Here, $\lambda,\,\lambda'\in\rr$ are numerical constants determined by the dimension of $M.$ Notice that the right-hand side of (\ref{cosmological const. vers. Higgs}) is indeed independent of the metric although the Higgs gauge potential itself is metric dependent.

The point to be emphasized is that the Einstein equations do not demand the Higgs gauge potential $H = ext_\Theta(\Phi_{\!\mbox{\tiny H}})$ itself to be constant but only to take values on the sphere bundle of radius $\Lambda/\lambda'$. Consequently, if the Yang-Mills gauge group ${\cal G}_{\mbox{\tiny YM}}\subset{\cal P}_{\mbox{\tiny D}}$ is supposed to act
transitively on the sphere bundle (like in the case of the ordinary Higgs potential), then
the Higgs gauge potential becomes gauge equivalent to the one-form
$im_{\mbox{\tiny D}}\Theta,$ with
$m_{\mbox{\tiny D}}\in{\goth Sec}(M,{\rm End}_\gamma(\sss))$ being a constant section of length $\Lambda/\lambda'.$ Clearly, such a section exists if and only if the Yang-Mills gauge group is reducible to the isotropy group of $m_{\mbox{\tiny D}}.$ The rank of the
reduced gauge group is determined by the co-dimension of the sphere bundle depending on the representation of the Yang-Mills gauge group
on the Clifford module. This completely parallels the usual Higgs mechanism used in the Standard Model description of particle physics and indicates how spontaneous symmetry breaking can be described when gravity is taken into account. In fact, we claim that the Higgs potential is only needed to provide the Higgs boson itself with mass but
the symmetry reduction is triggered by gravity in the way indicated. As mentioned earlier,
this geometrical interpretation of spontaneous symmetry breaking is based upon the intimate relation between gravity and the Higgs that is formally provided by the geometrical construction of Dirac connections in terms of the canonical one-form. We also point to the fact that the Higgs mass-term (\ref{cosmological const. vers. Higgs}) is of the same physical dimension as the scalar curvature, as opposed to the fourth order term in the usual Higgs potential, which is dimensionless like the quadratic Yang-Mills-Lagrangian (in four dimensions). The same holds true for the ``kinetic term'' of the Higgs. Hence, from a geometrical point of view one may regard the Higgs sector of the Standard Model as the sum of various terms having different geometrical origin. This is most clearly exhibited when the Einstein-Hilbert action with cosmological constant is expressed in terms of the Yang-Mills-Higgs connection $\partial_{\mbox{\tiny YMH}}$ and when also the {\it Yang-Mills-Higgs curvature}:
\bb
\label{ymh curvature}
F_{\!\mbox{\tiny YMH}} &:=& curv(\,\ddd_{\mbox{\tiny YMH}}) -
\,{/\!\!\!\!R}iem(g_{\mbox{\tiny M}})\nonumber\\[.1cm]
&=&
F_{\!\mbox{\tiny YM}} + d_{\mbox{\tiny A}}H + H\wedge H
\ee
is taken into account, where $F_{\!\mbox{\tiny YM}}\equiv F_{\!\mbox{\tiny A}}
\in\Omega^2(M,{\rm End}^+_\gamma(\ep))$ is the usual {\it Yang-Mills curvature} (``twisting curvature'').

Note that
\bb
d_{\mbox{\tiny A}}H + H\wedge H =
\left(d_{\mbox{\tiny A}}\Phi_{\mbox{\tiny H}} + \Phi^2_{\mbox{\tiny H}}\Theta\right)
\wedge\Theta\,.
\ee
Therefore, the Yang-Mills-Higgs connection is not flat, in general, even if the Yang-Mills connection is supposed to be flat and $\Phi_{\mbox{\tiny H}} = im_{\mbox{\tiny D}}$ is a constant section, for in this case
\bb
F_{\!\mbox{\tiny YMH}} = -m_{\mbox{\tiny D}}^2\Theta\wedge\Theta\,.
\ee
The (square of the) Dirac mass may be geometrically interpreted as curvature.

To summarize: We briefly discussed how the Einstein-Hilbert action and the
Einstein-Hilbert action with a cosmological constant can be geometrically described in terms of, respectively, Yang-Mills and
Yang-Mills-Higgs connections. Because of the translational invariance of the universal Dirac-Lagrangian, the latter does not
depend on the chosen Yang-Mills connection. The Higgs part of the Yang-Mills-Higgs connection may serve to provide a symmetry
reduction of the underlying Yang-Mills gauge group and thus contributes only by
a constant section. Therefore, the bosonic part of the total Dirac action does not explicitly depend on the choice of the
Yang-Mills part of the Yang-Mills-Higgs connection. It is (up to gauge) locally determined by
\bb
\label{extended connection}
\ddd_{\mbox{\tiny YMH}} \stackrel{\rm loc.}{=}
d + \varepsilon\delta_{\!\mbox{\tiny$g$}} + im_{\mbox{\tiny D}}\,,
\ee
which is but the general relativistic analogue of Dirac's original first order differential operator $i\ddd - m$. We point out that the Dirac connection of (\ref{extended connection}) is basically identical to the notion of the ``extended connection'' in terms of a ``frame field'' as discussed, for example, in \cite{CF:08} and the corresponding references cited therein.
In fact, the local term $-i\gamma_\mu\,M/4$ (c.f. the beginning of page 547 in loc. site) is
but a special case of a (locally defined) Dirac form
$\omega_{\!\mbox{\tiny D}}=ext_\Theta(\,\DDD - \,\ddb)$ (c.f. Sec. 2.1).

When the fermionic part of the total Dirac action is taken into account, the translational symmetry of the bosonic part is broken. As a quadratic form that is determined by $\,\DDD\in{\cal D}(\sss),$ this gauge reduction occurs since the fermionic part of the total action only depends on the choice of the Dirac operator. In contrast, the universal Dirac action is defined in terms of the corresponding curvature of the chosen Dirac operator. This subtle interplay between the fermionic and the bosonic part of the total Dirac action will be geometrically analyzed more carefully in the following section in terms of ``real Clifford modules'' and the ``Pauli map''.

\section{Real Clifford modules and the Pauli map}
Let $(\ep,\gamma_{\mbox{\tiny$\ep$}})\twoheadrightarrow(M,g_{\mbox{\tiny M}})$ be a Hermitian Clifford module. The Hermitian product is denoted by $\langle\cdot,\cdot\rangle_{\!\mbox{\tiny$\ep$}}$.
\begin{definition}
A Hermitian Clifford module is called a ``real $\zz_2-$bi-graded Hermitian Clifford module'' (``real Clifford module'' for short), if it is endowed, in addition, with a $\cc-$linear involution $\tau_{\!\mbox{\tiny$\ep$}}$, making $\ep = \ep^+\op\ep^-\twoheadrightarrow M$ $\zz_2-$graded, and
a $\cc-$anti-linear involution $J_{\!\mbox{\tiny$\ep$}}$, making
$\ep = {\cal M}_{\!\mbox{\tiny$\ep$}}\ot\cc\twoheadrightarrow M$ real, such that
\bb
\tau_{\!\mbox{\tiny$\ep$}}\circ\gamma_{\!\mbox{\tiny$\ep$}}(\alpha) &=&
-\gamma_{\!\mbox{\tiny$\ep$}}(\alpha)\circ\tau_{\!\mbox{\tiny$\ep$}}\,,\cr
J_{\!\mbox{\tiny$\ep$}}\circ\gamma_{\!\mbox{\tiny$\ep$}}(\alpha) &=&
\pm\gamma_{\!\mbox{\tiny$\ep$}}(\alpha)\circ J_{\!\mbox{\tiny$\ep$}}\,,\cr
J_{\!\mbox{\tiny$\ep$}}\circ\tau_{\!\mbox{\tiny$\ep$}} &=&
\pm\tau_{\!\mbox{\tiny$\ep$}}\circ J_{\!\mbox{\tiny$\ep$}}\,,\cr
\langle J_{\!\mbox{\tiny$\ep$}}(z),J_{\!\mbox{\tiny$\ep$}}(w)\rangle_{\!\mbox{\tiny$\ep$}}
&=&
\pm\langle w,z\rangle_{\!\mbox{\tiny$\ep$}}\,,
\ee
for all $\alpha\in T^*\!M$ and $z,w\in\ep$.

A real Clifford module is called a ``Majorana module'', provided
that
\bb
J_{\!\mbox{\tiny$\ep$}}\circ\tau_{\!\mbox{\tiny M}} =
-\tau_{\!\mbox{\tiny M}}\circ J_{\!\mbox{\tiny$\ep$}}\,.
\ee
\end{definition}

We make use of the following abbreviation: $B^{\mbox{\tiny cc}}\equiv
J_{\!\mbox{\tiny$\ep$}}\circ B\circ J_{\!\mbox{\tiny$\ep$}}$, for all $B\in{\rm End}(\ep).$
Similarly, $\,\DDD_{\!\mbox{\tiny$\ep$}}^{\mbox{\tiny cc}}\equiv
J_{\!\mbox{\tiny$\ep$}}\circ\,\DDD_{\!\mbox{\tiny$\ep$}}\circ J_{\!\mbox{\tiny$\ep$}}$,
for all $\,\DDD_{\!\mbox{\tiny$\ep$}}\in\dep.$ An operator $B\in{\rm End}(\ep)$ is called
``real'' (resp. ``imaginary''), if $B^{\mbox{\tiny cc}} =  B$ (resp.
$B^{\mbox{\tiny cc}} =  -B$). We denote by
${\cal D}_{\!\mbox{\tiny real}}(\ep)\subset\dep$ the subset of all real Dirac (type) operators: $\,\DDD_{\!\mbox{\tiny$\ep$}}^{\mbox{\tiny cc}} = \,\DDD_{\!\mbox{\tiny$\ep$}}.$

Let
\bb
\label{cliff-mod}
(\ep,\langle\cdot,\cdot\rangle_{\!\mbox{\tiny$\ep$}},\tau_{\!\mbox{\tiny$\ep$}},
\gamma_{\!\mbox{\tiny$\ep$}},J_{\!\mbox{\tiny$\ep$}})
\ee
be a real Clifford module bundle over $(M,g_{\mbox{\tiny M}}),$ such that
\bb
\tau_{\!\mbox{\tiny$\ep$}}^{\mbox{\tiny cc}} &=& \pm\tau_{\!\mbox{\tiny$\ep$}}\,,\\
\gamma_{\!\mbox{\tiny$\ep$}}^{\mbox{\tiny cc}} &=& +\gamma_{\!\mbox{\tiny$\ep$}}\,.
\ee

We denote by
\bb
({\cal P},\langle\cdot,\cdot\rangle_{\!\mbox{\tiny${\cal P}$}},
\tau_{\!\mbox{\tiny${\cal P}$}},\gamma_{\!\mbox{\tiny${\cal P}$}})
\ee
the doubling of the Clifford module $(\ep,\langle\cdot,\cdot\rangle_{\!\mbox{\tiny$\ep$}},\tau_{\!\mbox{\tiny$\ep$}},
\gamma_{\!\mbox{\tiny$\ep$}}).$ That is,
\bb
{\cal P} &:=& \hspace{-.3cm}\phantom{\ep}^2\ep\; \equiv \begin{array}{c}
                                        \ep \\
                                        \op \\
                                        \ep
                                      \end{array} =\; \ep\ot\cc^2\,,\\
\langle\cdot,\cdot\rangle_{\!\mbox{\tiny${\cal P}$}} &:=&
\mbox{\small$\frac{1}{2}$}(\langle\cdot,\cdot\rangle_{\!\mbox{\tiny$\ep$}} +
\langle\cdot,\cdot\rangle_{\!\mbox{\tiny$\ep$}})\,,\\
\tau_{\!\mbox{\tiny${\cal P}$}} &:=&
\tau_{\!\mbox{\tiny$\ep$}}\ot\tau_{\!\mbox{\tiny 2}}\,,\\
\gamma_{\!\mbox{\tiny${\cal P}$}} &:=&
\gamma_{\!\mbox{\tiny$\ep$}}\ot{\bf 1}_{\!\mbox{\tiny 2}}\,.
\ee
Here and in the sequel: ${\bf 1}_{\!\mbox{\tiny 2}}\in\cc(2)$ and
$\tau_{\!\mbox{\tiny 2}},\,\varepsilon_{\!\mbox{\tiny 2}},\,{\rm I}_{\mbox{\tiny 2}}\in\cc(2)$ denote, respectively, the two-by-two unit matrix and
\bb
\tau_{\mbox{\tiny 2}} \equiv\left(\!\!
                                  \begin{array}{cc}
                                    1 & \phantom{-}0 \\
                                    0 & -1 \\
                                  \end{array}
                                  \!\!\right)\,,\quad
\varepsilon_{\mbox{\tiny 2}} \equiv\left(\!\!
                                  \begin{array}{cc}
                                    0 & 1 \\
                                    1 & 0 \\
                                  \end{array}
                                \!\!\right)\,,\quad
{\rm I}_{\mbox{\tiny 2}} \equiv\left(\!\!
                                  \begin{array}{cc}
                                    0 & -1 \\
                                    1 & \phantom{-}0 \\
                                  \end{array}
                                \!\!\right)\,.
\ee

The real structure on $\ep$ then allows to also introduce a real structure on
the doubled Clifford module ${\cal P} = \!\!\!\!\phantom{\ep}^2\ep$:
\bb
J_{\!\mbox{\tiny${\cal P}$}} := J_{\!\mbox{\tiny$\ep$}}\ot\varepsilon_{\!\mbox{\tiny 2}}\,,
\ee
such that
\bb
\label{cliff-mod2}
({\cal P},\langle\cdot,\cdot\rangle_{\!\mbox{\tiny${\cal P}$}},
\tau_{\!\mbox{\tiny${\cal P}$}},\gamma_{\!\mbox{\tiny${\cal P}$}},
J_{\!\mbox{\tiny${\cal P}$}})
\ee
becomes, again, a real Clifford module over $(M,g_{\mbox{\tiny M}}).$ It follows that
\bb
\tau_{\!\mbox{\tiny${\cal P}$}}^{\mbox{\tiny cc}} &=&
\pm\tau_{\!\mbox{\tiny${\cal P}$}}\quad\Leftrightarrow\quad
\tau_{\!\mbox{\tiny$\ep$}}^{\mbox{\tiny cc}} = \mp\tau_{\!\mbox{\tiny$\ep$}}\,,\\
\gamma_{\!\mbox{\tiny${\cal P}$}}^{\mbox{\tiny cc}} &=& +\gamma_{\!\mbox{\tiny${\cal P}$}}\,.
\ee

With respect to the real structure $J_{\!\mbox{\tiny$\ppp$}}$ the doubled Clifford module  may be regarded as the complexification of the real vector bundle:
\bb
{\cal M}_{\mbox{\tiny${\cal P}$}} :=
\left\{\mbox{\small$\left(\!\!
         \begin{array}{c}
           {\goth z} \\
           {\goth z}^{\mbox{\tiny cc}} \\
         \end{array}
       \!\!\right)$}\in{\cal P}\,|\,{\goth z}\in\ep
\right\}\twoheadrightarrow M\,.
\ee

This real vector bundle contains a distinguished real sub-vector bundle:
\bb
{\cal V}_{\!\mbox{\tiny${\cal P}$}} :=
\left\{\mbox{\small$\left(\!\!
         \begin{array}{c}
           {\goth z} \\
           {\goth z} \\
         \end{array}
       \!\!\right)$}\in{\cal P}\,|\,{\goth z}\in{\cal M}_{\mbox{\tiny$\ep$}}
\right\}\twoheadrightarrow M\,,
\ee
whoses complexification ${\cal V}_{\!\mbox{\tiny${\cal P}$}}^\cc$ of the total space
may be identified with the diagonal embedding
\bb
\ep&\hookrightarrow&\!\!\!\!\phantom{\ep}^2\ep\nonumber\\[.1cm]
{\goth z}&\mapsto&\mbox{\small$\left(\!\!
         \begin{array}{c}
           {\goth z} \\
           {\goth z} \\
         \end{array}
       \!\!\right)$}\,.
\ee
Here, ${\cal M}_{\mbox{\tiny$\ep$}}:=\{{\goth z}\in\ep\,|\,
J_{\!\mbox{\tiny$\ep$}}({\goth z}) = {\goth z}\}\subset\ep$ is the (total space) of the induced real sub-vector bundle, such that $\ep = {\cal M}_{\mbox{\tiny$\ep$}}^\cc$.

A general real Dirac operator on the real Clifford module (\ref{cliff-mod2}) reads (for a proof see, for example, in \cite{Tol:09}, Theorem 1):
\bb
\DDD_{\!\mbox{\tiny${\cal P}$}} =
\left(
  \begin{array}{cc}
    \DDD_{\!\mbox{\tiny$\ep$}} & \phi_{\mbox{\tiny$\ep$}} -
    {\cal F}_{\!\!\mbox{\tiny$\ep$}} \\
    \phi_{\mbox{\tiny$\ep$}} +
    {\cal F}_{\!\!\mbox{\tiny$\ep$}} & \DDD_{\!\mbox{\tiny$\ep$}}^{\mbox{\tiny cc}} \\
  \end{array}
\right)\,.
\ee
Here, respectively, $\DDD_{\!\mbox{\tiny$\ep$}}\in\dep$ is any Dirac operator on
(\ref{cliff-mod}) and
\bb
\phi_{\mbox{\tiny$\ep$}}^{\mbox{\tiny cc}} &=& +\phi_{\mbox{\tiny$\ep$}}\,,\cr
{\cal F}_{\!\!\mbox{\tiny$\ep$}}^{\mbox{\tiny cc}} &=& -{\cal F}_{\!\!\mbox{\tiny$\ep$}}
\ee
are general sections of ${\rm End}^+(\ep)\twoheadrightarrow M$.

The (affine) set of all real Dirac operators on the doubled Clifford module (\ref{cliff-mod2}) contains a distinguished (affine) sub-set, consisting of those Dirac operators where in addition $\,\DDD_{\!\mbox{\tiny$\ep$}}^{\mbox{\tiny cc}} = \,\DDD_{\!\mbox{\tiny$\ep$}}$ is a real Dirac operator on (\ref{cliff-mod}). In particular, the Dirac operators
\bb
\DDD_{\!\mbox{\tiny${\cal P}$}} &=&
\left(
  \begin{array}{cc}
    \DDD_{\!\mbox{\tiny$\ep$}} & \phi_{\mbox{\tiny$\ep$}}\\
    \phi_{\mbox{\tiny$\ep$}} & \DDD_{\!\mbox{\tiny$\ep$}} \\
  \end{array}
\right)\nonumber\\[.1cm]
&=&
\DDD_{\!\mbox{\tiny$\ep$}}\ot{\bf 1}_{\!\mbox{\tiny 2}} +
\phi_{\mbox{\tiny$\ep$}}\ot\varepsilon_{\!\mbox{\tiny 2}}
\ee
also preserve ${\goth Sec}(M,{\cal V}_{\!\mbox{\tiny${\cal P}$}}).$

In contrast, one may consider the distinguished class of Dirac operators on
the doubled real Clifford module (\ref{cliff-mod2}) which are
determined already by the real Dirac operators on (\ref{cliff-mod}):
\bb
\label{paulidop}
{/\!\!\!\!P}_{\!\!\mbox{\tiny D}} &:=&
\left(
  \begin{array}{cc}
    \DDD_{\!\mbox{\tiny$\ep$}} & -{\cal F}_{\!\!\mbox{\tiny$\ep$}} \\
    {\cal F}_{\!\!\mbox{\tiny$\ep$}} & \DDD_{\!\mbox{\tiny$\ep$}} \\
  \end{array}
\right)\nonumber\\[.1cm]
&=&
\DDD_{\!\mbox{\tiny$\ep$}}\ot{\bf 1}_{\!\mbox{\tiny 2}} +
{\cal F}_{\!\!\mbox{\tiny$\ep$}}\ot{\rm I}_{\mbox{\tiny 2}}\,,
\ee
with ${\cal F}_{\!\!\mbox{\tiny$\ep$}}$ being defined by the (relative) curvature of
$\,\DDD_{\!\mbox{\tiny$\ep$}}:$
\bb
{\cal F}_{\!\!\mbox{\tiny$\ep$}}\, :=
\,{/\!\!\!\!{\cal F}}_{\!\!\!\mbox{\tiny D}} &\equiv&
i\delta_\gamma(curv(\,\DDD_{\!\mbox{\tiny$\ep$}}) - \,{/\!\!\!\!R}iem(g_{\mbox{\tiny M}}))\cr
&=&
i\,{/\!\!\!\!F}_{\!\!\mbox{\tiny D}}\,.
\ee

Note that ${/\!\!\!\!F}_{\!\!\mbox{\tiny D}}\in{\goth Sec}(M,{\rm End}^+(\ep))$ is even and real for real (or imaginary) Dirac operators $\,\DDD_{\!\mbox{\tiny$\ep$}}\in\dep.$ Whence,
$\,{/\!\!\!\!{\cal F}}_{\!\!\!\mbox{\tiny D}}^{\mbox{\tiny cc}} =
-\,{/\!\!\!\!{\cal F}}_{\!\!\!\mbox{\tiny D}}.$

By a slight abuse of notation, we rewrite the {\it Pauli type} Dirac operators (\ref{paulidop}) as
\bb
{/\!\!\!\!P}_{\!\!\mbox{\tiny D}} := \DDD_{\!\mbox{\tiny$\ep$}} +
\iota\,{/\!\!\!\!{\cal F}}_{\!\!\!\mbox{\tiny D}}
\ee
to bring them most closely to Dirac's first order operator including the Pauli term
$i\,{/\!\!\!\!F}$. Here,
\bb
\iota\,\,{/\!\!\!\!{\cal F}}_{\!\!\!\mbox{\tiny D}}\equiv
\left(\!\!
  \begin{array}{cc}
    0 & -{\rm id}_{\mbox{\tiny$\ep$}} \\
    {\rm id}_{\mbox{\tiny$\ep$}} & 0 \\
  \end{array}
\!\!\right)\circ
\left(\!\!
  \begin{array}{cc}
    \,{/\!\!\!\!{\cal F}}_{\!\!\!\mbox{\tiny D}} & 0 \\
    0 & \,{/\!\!\!\!{\cal F}}_{\!\!\!\mbox{\tiny D}} \\
  \end{array}
\!\!\right)\,.
\ee

In the sequel, we shall consider Pauli type Dirac operators on the doubled Clifford module
(\ref{cliff-mod2}) as mappings:
\bb
{/\!\!\!\!P}_{\!\!\mbox{\tiny D}}:\,{\goth Sec}(M,{\cal V}_{\!\mbox{\tiny${\cal P}$}}^\cc)
&\longrightarrow&{\goth Sec}(M,{\cal P})\nonumber\\[.1cm]
\phantom{\psi}^{\mbox{\tiny 2}}\psi = \mbox{\small$\left(\!\!
               \begin{array}{c}
                 \psi \\
                 \psi \\
               \end{array}
             \!\!\right)$} &\mapsto&
\mbox{\small$\left(\!\!
               \begin{array}{c}
                 \DDD_{\!\mbox{\tiny$\ep$}}\psi -
                 \,{/\!\!\!\!{\cal F}}_{\!\!\!\mbox{\tiny D}}\psi  \\
                 \DDD_{\!\mbox{\tiny$\ep$}}\psi +
                 \,{/\!\!\!\!{\cal F}}_{\!\!\!\mbox{\tiny D}}\psi \\
               \end{array}
             \!\!\right)$}\,.
\ee

Therefore, the restriction of our original Pauli type Dirac operators to the diagonal embedding $\ep\hookrightarrow\!\!\!\!\phantom{\ep}^2\ep$ may formally be interpreted
as the restriction of the real Dirac operators (\ref{paulidop}) to the sections of the distinguished sub-bundle
\bb
\label{paulibdl}
{\cal V}_{\!\mbox{\tiny${\cal P}$}}^\cc\hookrightarrow{\cal P}\twoheadrightarrow M\,.
\ee
For this matter, we also call this bundle the {\it Pauli bundle} associated with the real Clifford bundle (\ref{cliff-mod}).

We put emphasize on the following fact: The Lagrangian density that is defined by the smooth function
$\langle\Psi,\,{/\!\!\!\!P}_{\!\!\mbox{\tiny D}}\Psi\rangle_{\!\mbox{\tiny${\cal P}$}}$
reduces to $\langle\psi,\,\DDD_{\!\mbox{\tiny$\ep$}}\psi\rangle_{\!\mbox{\tiny$\ep$}}$,
when the Pauli type Dirac operators are restricted to the sections of the Pauli bundle.
Therefore, the two fermionic functionals:
\bb
{\cal I}_{\mbox{\tiny D,ferm}}:\,{\goth Sec}(M,\ep)\times\dep&\longrightarrow&\cc
\nonumber\\[.1cm]
(\psi,\,\DDD_{\!\mbox{\tiny$\ep$}}) &\mapsto&
\int_M\langle\psi,\,\DDD_{\!\mbox{\tiny$\ep$}}\psi\rangle_{\!\mbox{\tiny$\ep$}}
\,dvol_{\mbox{\tiny M}}\,,
\ee
\bb
{\cal I}'_{\mbox{\tiny D,ferm}}:\,{\goth Sec}(M,\ep)\times\dep&\longrightarrow&\cc
\nonumber\\[.1cm]
(\psi,\,\DDD_{\!\mbox{\tiny$\ep$}}) &\mapsto&
\int_M\langle\hspace{-.2cm}\phantom{\psi}^{\mbox{\tiny 2}}\!\psi,
\,{/\!\!\!\!P}_{\!\!\mbox{\tiny D}}
\hspace{-.25cm}\phantom{\psi}^{\mbox{\tiny 2}}\!\psi
\rangle_{\!\mbox{\tiny${\cal P}$}}\,dvol_{\mbox{\tiny M}}
\ee
contain the same information, actually.

The (generalized) Pauli term thus does not alter the fermionic action. In particular,
the fermionic action is fully determined by the (Dirac) connections on the vector bundle
$\ep\twoheadrightarrow M$ and not, in addition, by the curvature of these (Dirac) connections. As mentioned earlier, this fact is known to play a fundamental role in quantizing the fermionic action. Of course, when the functional ${\cal I}'_{\mbox{\tiny D,ferm}}$ is actually regarded as being a functional on ${\goth Sec}(M,{\cal V}_{\!\mbox{\tiny${\cal P}$}}^\cc),$ then a stationary point of this functional has to satisfy the more restrictive condition:
\bb
\psi\in{\rm ker}(\,\DDD_{\!\mbox{\tiny$\ep$}})\cap
{\rm ker}(\,{/\!\!\!\!{\cal F}}_{\!\!\!\mbox{\tiny D}})\,.
\ee

The equivalence of the two fermionic actions ${\cal I}_{\mbox{\tiny D,ferm}}$ and
${\cal I}'_{\mbox{\tiny D,ferm}}$ (when both are regarded as being functionals on
the same domain) is very basic for the structure of Dirac type gauge theories. Indeed,
these equivalent geometrical descriptions of the fermionic action seem to provide a deep relation between the fermionic part of the total Dirac action and its corresponding
bosonic part.

To formalize the above discussed equivalence of the functionals
${\cal I}_{\mbox{\tiny D,ferm}}$ and ${\cal I}'_{\mbox{\tiny D,ferm}},$ we introduce the
following
\begin{definition}
Let
$(\ep,\langle\cdot,\cdot\rangle_{\!\mbox{\tiny$\ep$}},\tau_{\!\mbox{\tiny$\ep$}},
\gamma_{\!\mbox{\tiny$\ep$}},J_{\!\mbox{\tiny$\ep$}})$ be a real Clifford module bundle over $(M,g_{\mbox{\tiny M}})$ satisfying the requirements imposed on (\ref{cliff-mod}). Also, let
${\cal D}_{\!\mbox{\tiny real}}(\ep)\subset\dep$ be the (affine) set of real Dirac
operators acting on ${\goth Sec}(M,\ep)$.

We call the mapping
\bb
\label{pauli-map}
{\cal P}_{\!\mbox{\tiny D}}:\,
{\cal D}_{\!\mbox{\tiny real}}(\ep) &\longrightarrow&
{\cal D}_{\!\mbox{\tiny real}}({\cal P})\cr
\DDD_{\!\mbox{\tiny$\ep$}}&\mapsto&\,{/\!\!\!\!P}_{\!\!\mbox{\tiny D}}\,,
\ee
which associates with every real Dirac operator on $\ep\twoheadrightarrow M$ the appropriate
Pauli type Dirac operator on the doubled Clifford module
${\cal P}=\!\!\!\!\phantom{\ep}^2\ep\twoheadrightarrow M$, the ``Pauli map''.
\end{definition}

Geometrically, one may regard the fermionic action as a mapping from $\dep$ into the
quadratic forms on ${\goth Sec}(M,\ep):$
\bb
\label{quadratic form}
{\cal I}_{\mbox{\tiny D,ferm}}:\,\dep &\longrightarrow&
{\goth Map}({\goth Sec}(M,\ep),\cc)\nonumber\\[.1cm]
\DDD_{\!\mbox{\tiny$\ep$}} &\mapsto&
\left\{\begin{array}{ccc}
         {\goth Sec}(M,\ep) & \longrightarrow & \hspace{-1.8cm}\cc \\
         \psi & \mapsto & {\cal I}_{\mbox{\tiny D,ferm}}(\psi,\DDD_{\!\mbox{\tiny$\ep$}})
       \end{array}
\right.
\ee

When restricted to ${\cal D}_{\!\mbox{\tiny real}}(\ep),$ the Pauli map thus allows to
lift the quadratic form ${\cal I}_{\mbox{\tiny D,ferm}}$ on ${\goth Sec}(M,\ep)$ to
the quadratic form ${\cal I}'_{\mbox{\tiny D,ferm}}$ on ${\goth Sec}(M,{\cal P})$:
\bb
{\cal I}'_{\mbox{\tiny D,ferm}} =
{\cal I}_{\mbox{\tiny D,ferm}}\circ{\cal P}_{\!\mbox{\tiny D}}\,,
\ee
such that ${\cal P}_{\!\mbox{\tiny D}}$ acts like the identity when
${\cal I}'_{\mbox{\tiny D,ferm}}$ is restricted to
${\goth Sec}(M,{\cal V}_{\!\mbox{\tiny${\cal P}$}}^\cc)\subset{\goth Sec}(M,{\cal P}).$

As mentioned earlier, on every Clifford module there exists a distinguished class of Dirac operators called of simple type. Explicitly, they read:
\bb
\DDD_{\!\mbox{\tiny$\ep$}} = \dda + \tau_{\!\mbox{\tiny$\ep$}}\circ\phi_{\mbox{\tiny D}}
\ee
where $\phi_{\mbox{\tiny D}}\in{\goth Sec}(M,{\rm End}^-_\gamma(\ep)).$ In general, however,
these Dirac operators are not real. Therefore, our original Pauli type Dirac operators
(\ref{pauli type dop}) fail to be real and our geometrical understanding of this class of Dirac operators in terms of the Pauli map (\ref{pauli-map}) is not yet complete.

Of course, this flaw may most straightforwardly be remedied by giving up the restriction of the Pauli map to real Dirac operators. However, this will then not yield any new insight concerning the structure of the original Pauli type operators (\ref{pauli type dop}). Even worse, one loses significant information as will be shown in the next section. Indeed, there it will be shown
that the Pauli map (\ref{pauli-map}) allows to naturally include the geometrical description
of ``Majorana masses'' in terms of real Dirac operators of simple type.

\subsection{Majorana masses and real Dirac operators of simple type}
Let $(\sss,\langle\cdot,\cdot\rangle_{\!\mbox{\tiny$\sss$}},
\tau_{\!\mbox{\tiny$\sss$}}, \gamma_{\!\mbox{\tiny$\sss$}},
J_{\!\mbox{\tiny$\sss$}})$ be a real Clifford module bundle over $(M,g_{\mbox{\tiny M}})$. We put
\bb
\ep &:=& \hspace{-.3cm}\phantom{\sss}^2\sss = \sss\ot\cc^2\,,\\[.1cm]
\langle\cdot,\cdot\rangle_{\!\mbox{\tiny$\ep$}} &:=&
\mbox{\small$\frac{1}{2}$}(\langle\cdot,\cdot\rangle_{\!\mbox{\tiny$\sss$}} +
\langle\cdot,\cdot\rangle_{\!\mbox{\tiny$\sss$}})\,,\\[.1cm]
\tau_{\!\mbox{\tiny$\ep$}} &:=& \left(\!\!
                                   \begin{array}{cc}
                                     \tau_{\mbox{\tiny$\sss$}} & \phantom{-}0 \\
                                     0 & -\tau_{\mbox{\tiny$\sss$}} \\
                                   \end{array}
                                 \!\!\right) =
                                 \tau_{\mbox{\tiny$\sss$}}\ot\tau_{\!\mbox{\tiny 2}}
                                 \,,\\[.1cm]
\gamma_{\!\mbox{\tiny$\ep$}} &:=& \left(\!\!
                                   \begin{array}{cc}
                                     \gamma_{\mbox{\tiny$\sss$}} & 0 \\
                                     0 & \gamma_{\mbox{\tiny$\sss$}}^{\mbox{\tiny cc}} \\
                                   \end{array}
                                 \!\!\right)
                                 \,,\\[.1cm]
J_{\!\mbox{\tiny$\ep$}} &:=& \left(\!\!
                                   \begin{array}{cc}
                                     0 & J_{\mbox{\tiny$\sss$}} \\
                                     J_{\mbox{\tiny$\sss$}} & 0 \\
                                   \end{array}
                                 \!\!\right) =
                                 J_{\mbox{\tiny$\sss$}}\ot\varepsilon_{\!\mbox{\tiny 2}}\,.
\ee

It follows that
\bb
\tau_{\!\mbox{\tiny$\ep$}}^{\mbox{\tiny cc}} &=& \pm\tau_{\!\mbox{\tiny$\ep$}}\quad
\Leftrightarrow\quad\tau_{\!\mbox{\tiny$\sss$}}^{\mbox{\tiny cc}} = \mp\tau_{\!\mbox{\tiny$\sss$}}\,,\cr
\gamma_{\!\mbox{\tiny$\ep$}}^{\mbox{\tiny cc}} &=& +\gamma_{\!\mbox{\tiny$\ep$}}\,.
\ee

\begin{definition}
Let $\,\DDD_{\!\mbox{\tiny$\sss$}}\in{\cal D}(\sss)$ be a Dirac type operator on the real Clifford module $\sss\twoheadrightarrow M$. The real Dirac type operator on the induced real Clifford module $\ep\twoheadrightarrow M:$
\bb
\DDD_{\!\mbox{\tiny$\ep$}} &:=& \left(
                                \begin{array}{cc}
                                  \DDD_{\!\mbox{\tiny$\sss$}} & 0 \\
                                  0 & \DDD_{\!\mbox{\tiny$\sss$}}^{\mbox{\tiny cc}} \\
                                \end{array}
                              \right)\,,\nonumber\\[.1cm]
                              &\equiv&
                             \DDD_{\!\mbox{\tiny$\sss$}} \op \,\DDD_{\!\mbox{\tiny$\sss$}}^{\mbox{\tiny cc}}\,,
\ee
is called the ``real form'' of $\,\DDD_{\!\mbox{\tiny$\sss$}}\,$.
\end{definition}

\begin{proposition}\label{prop of real dop of simple type}
The most general real Dirac operator of simple type, acting on ${\goth Sec}(M,\ep)$,
explicitly reads:
\bb
\DDD_{\!\mbox{\tiny$\ep$}} &=& \ddd_{\!\!\!\mbox{\tiny${\cal A}$}} + \tau_{\mbox{\tiny$\ep$}}\circ\phi_{\mbox{\tiny$\ep$}}\,,
\ee
whereby $\,\ddd_{\!\!\!\mbox{\tiny${\cal A}$}} := \dda\, \op \,\dda^{\mbox{\tiny cc}}$
is the real form of $\,\dda$ and
\bb
\phi_{\!\mbox{\tiny$\ep$}} := \left(\!\!
                                    \begin{array}{cc}
                                     \chi_{\mbox{\tiny$\sss$}}  &
                                     \pm\phi^{\mbox{\tiny cc}}_{\!\mbox{\tiny$\sss$}} \\
                                      -\phi_{\!\mbox{\tiny$\sss$}} &
                                      \mp\chi_{\mbox{\tiny$\sss$}}^{\mbox{\tiny cc}} \\
                                    \end{array}
                                  \!\!\right)\,,
\ee
depending on whether $\tau^{\mbox{\tiny cc}}_{\mbox{\tiny$\sss$}} =
\pm\tau_{\mbox{\tiny$\sss$}}$. Moreover,
$\phi_{\!\mbox{\tiny$\sss$}}\in{\goth Sec}(M,{\rm End}_\gamma^+(\sss))$ is explicitly
given by
\bb
\phi_{\!\mbox{\tiny$\sss$}} \equiv
\left\{\begin{array}{cc}
\chi'_{\mbox{\tiny$\sss$}} + \tau_{\mbox{\tiny$\sss$}}\circ
\delta_\gamma(\sigma_{\!\mbox{\tiny$\sss$}})\,, &  \;\mbox{for}\quad
\gamma_{\mbox{\tiny$\sss$}}^{\mbox{\tiny cc}} = +\gamma_{\!\mbox{\tiny$\sss$}}\,,\\[.1cm]
\tau_{\mbox{\tiny$\sss$}}\circ\mu_{\mbox{\tiny M}} +
\delta_\gamma(\sigma_{\!\mbox{\tiny$\sss$}})\,, & \;\mbox{for}\quad
\gamma_{\mbox{\tiny$\sss$}}^{\mbox{\tiny cc}} = -\gamma_{\!\mbox{\tiny$\sss$}}\,.
\end{array}
\right.
\ee
Here, $\mu_{\mbox{\tiny M}},\,\chi'_{\mbox{\tiny$\sss$}}
\in\Omega^0(M,{\rm End}^+_\gamma(\sss))$, $\,\chi_{\mbox{\tiny$\sss$}}
\in\Omega^0(M,{\rm End}^-_\gamma(\sss))$ and $\sigma_{\!\mbox{\tiny$\sss$}}\in
\Omega^1(M,{\rm End}^-_\gamma(\sss))$.
\end{proposition}

The proof of the above statement is based upon the following statement and a corollary
thereof. Both of which are interesting in its own and will be useful also later on.

\begin{lemma}\label{lemma lichnerowicz formula}
Let $(\ep,\gamma_{\mbox{\tiny$\ep$}})\twoheadrightarrow(M,g_{\mbox{\tiny M}})$ be a
general Clifford module over a smooth (semi-)Rie\-mannian manifold. Also, let
$\,\DDD_{\!k} + \Phi_k\in\dep\;(k= 1,2)$ be two Dirac type operators, acting on ${\goth Sec}(M,\ep).$ The Laplace type operator
\bb
H := (\,\DDD_{\!1} + \Phi_1)\circ(\,\DDD_{\!2} + \Phi_2)
\ee
has the explicit Lichnerowicz decomposition:
$H = \triangle_{\mbox{\tiny H}} + V_{\!\mbox{\tiny H}},$ where the second
order part is defined in terms of the connection:
\bb
\label{H-connection}
\partial_{\mbox{\tiny H}} &:=& \db + \alpha_{\mbox{\tiny H}}\,,\cr
\alpha_{\mbox{\tiny H}}(v) &:=& \mbox{\small$\frac{\varepsilon}{2}$}\left(\gamma_{\mbox{\tiny$\ep$}}(v^\flat)\circ\Phi_2 +
\Phi_1\circ\gamma_{\mbox{\tiny$\ep$}}(v^\flat) + (\,\DDD_{\!1} - \,\DDD_{\!2})
\circ\gamma_{\mbox{\tiny$\ep$}}(v^\flat)\right)\,,
\ee
for all $v\in TM.$ The zero order part explicitly reads:
\bb
\label{H-potential}
V_{\!\mbox{\tiny H}} \;:= \hspace{7cm}\nonumber\\[.1cm]
V_{\!\mbox{\tiny D}} +
\delta_\gamma(\db\Phi_2) -
\varepsilon ev_{\!g}(\partial_{\mbox{\tiny H}}\alpha_{\mbox{\tiny H}}) -
\varepsilon ev_{\!g}(\alpha^2_{\mbox{\tiny H}}) + \Phi_{\mbox{\tiny D}}\circ\Phi_2 +
(\Phi_1 + (\,\DDD_{\!1} - \,\DDD_{\!2}))\circ(\Phi_2 + \Phi_{\mbox{\tiny D}})\,.
\ee
Here, $\db\in{\cal A}(\ep)$ denotes the Bochner connection that is defined by $\,\DDD_{\!2}\,\equiv\,\DDD\,$ and
\bb
V_{\!\mbox{\tiny D}} &:=& \,\DDD^2\, - \triangle_{\mbox{\tiny B}}\,,\cr
\Phi_{\mbox{\tiny D}} &:=& \DDD - \ddb\,.
\ee
\end{lemma}

\noindent
{\bf Proof:} First, we again put $\,\DDD\,\equiv\,\DDD_{\!2}$ and abbreviate
$\Phi_{12} \equiv\,\DDD_{\!1}\,-\,\DDD_{\!2}$ to re-write $H$ as
\bb
H = \DDD^2 + [\,\DDD,\Phi_2]\, +  (\Phi_1 + \Phi_2 + \Phi_{12})\circ\,\DDD +
(\Phi_1 + \Phi_{12})\circ\Phi_2\,.
\ee
It follows that for all $f\in{\cal C}^\infty(M):$
\bb
[[\,\DDD,\Phi_2],f] = [\delta_\gamma(df),\Phi_2]
\ee
and thus
\bb
[H,f] = [\,\DDD^2,f] + \delta_\gamma(df)\circ\Phi_2 + \Phi_1\circ\delta_\gamma(df) +
\Phi_{12}\circ\delta_\gamma(df)\,.
\ee
This yields the explicit formula (\ref{H-connection}) for the connection $\partial_{\mbox{\tiny H}}.$

The explicit formula (\ref{H-potential}) for the zero order term is then obtained
from the identity $V_{\mbox{\tiny H}} = H - \triangle_{\mbox{\tiny H}}$ and
\bb
\triangle_{\mbox{\tiny H}} &\equiv&
\varepsilon ev_{\!g}(\partial_{\mbox{\tiny H}}\circ\partial_{\mbox{\tiny H}})\cr
&=&
\triangle_{\mbox{\tiny D}} +
\varepsilon ev_{\!g}(\partial_{\mbox{\tiny H}}\alpha_{\mbox{\tiny H}}) +
\varepsilon ev_{\!g}(\alpha^2_{\mbox{\tiny H}}) +
2\,\varepsilon ev_{\!g}(\alpha_{\mbox{\tiny H}},\db)\,.
\ee

\noindent
This proves the statement.\hfill$\Box$

\vspace{.5cm}

For later convenience we consider $V_{\mbox{\tiny H}}\equiv V_{\mbox{\tiny D}}$
in the case where $H = \,\DDD^2$ and $\,\DDD = \dda + \Phi$. From Lemma
\ref{lemma lichnerowicz formula} it follows for
$\,\DDD_{\!1}\,=\,\DDD_{\!2}\,\equiv\,\dda$ and $\Phi_1 = \Phi_2 \equiv \Phi$ that
\bb
\label{dirac potential}
V_{\mbox{\tiny D}} = \delta_\gamma(curv(\dda)) +
\delta_\gamma(\da\Phi) +
\Phi^2 -
\varepsilon ev_{\!g}(\alpha^2_{\mbox{\tiny D}}) -
\varepsilon ev_{\!g}(\db\alpha_{\mbox{\tiny D}})\,,
\ee
whereby $\db = \da + \alpha_{\mbox{\tiny D}}$ and
\bb
\label{anti-com. relation}
\alpha_{\mbox{\tiny D}}(v) &:=& \mbox{\small$\frac{\varepsilon}{2}$}\left\{\gamma_{\mbox{\tiny$\ep$}}(v^\flat),\Phi\right\}\,.
\ee

Clearly, Lemma (\ref{lemma lichnerowicz formula}) generalizes the well-known
formula by Lichnerowicz/Schr\"odinger (\ref{lichnerowic-formula}) with respect to $\,\dda^2$ to general Laplacians which can be factorized by arbitrary Dirac type operators. The next statement yields an easy characterization of simply type Dirac operators.

\begin{korollar}\label{korrollar simple type dop}
A Dirac operator $\DDD$ on a $\zz_2-$graded Clifford module $(\ep,\gamma_{\mbox{\tiny$\ep$}})\twoheadrightarrow(M,g_{\mbox{\tiny M}})$ is of simple type
if and only if
\bb
\{\,\DDD\,-\,\ddb\,,\gamma_{\mbox{\tiny$\ep$}}(\alpha)\} \equiv 0\,,
\ee
for all $\alpha\in T^*\!M.$ Here, $\ddb\,\equiv\delta_\gamma\circ\db$ is the quantized Bochner connection that is defined by $\,\DDD\in\dep.$
\end{korollar}

\noindent
{\bf Proof:}
It follows from Lemma \ref{lemma lichnerowicz formula} that two Dirac type operators
$\,\DDD'\,,\DDD\in\dep$ share the same Bochner connection if and only if the zero-order
operator $\,\DDD'\,-\,\DDD$ anti-commutes with the Clifford action (c.f. formula (\ref{anti-com. relation})). Whence, $\,\DDD$ and $\,\ddb$ have the same Bochner connection
$\db$ if and only if $\,\DDD\,-\,\ddb$ anti-commutes with the Clifford action. However, Clifford connections $\da\in{\cal A}_{\mbox{\tiny Cl}}(\ep)$ are the only connections with
the property that the three notions of Dirac connection, Clifford connection and Bochner connection coincide, i.e.:
\bb
\dD = \da = \db\,.
\ee
Whence, the Dirac type operator $\,\ddb\in\dep$ yields the Bochner connection $\db$ if and only if $\db\in{\cal A}_{\mbox{\tiny Cl}}(\ep).$ This proves the statement.\hfill$\Box$

\vspace{.5cm}

We now turn back to the proof of Proposition \ref{prop of real dop of simple type}.

\vspace{.5cm}

\noindent
{\bf Proof of Proposition \ref{prop of real dop of simple type}:}
The most general real Dirac operator, acting on ${\goth Sec}(M,\ep)$, reads:
\bb
\DDD'_{\!\!\mbox{\tiny$\ep$}} = \left(\!\!
                                    \begin{array}{cc}
                                     \DDD_{\!\mbox{\tiny$\sss$}}  & \Phi_{\!\mbox{\tiny$\sss$}}^{\mbox{\tiny cc}} \\
                                      \Phi_{\!\mbox{\tiny$\sss$}} &
                                      \DDD_{\!\mbox{\tiny$\sss$}}^{\mbox{\tiny cc}} \\
                                    \end{array}
                                  \!\!\right)\,,
\ee
whereby $\Phi_{\!\mbox{\tiny$\sss$}}\in{\goth Sec}(M,{\rm End}^+(\sss)).$ We may re-write this real Dirac operator as
\bb
\DDD'_{\!\!\mbox{\tiny$\ep$}} = \,\DDD_{\!\!\mbox{\tiny$\ep$}} + \Phi'_{\!\mbox{\tiny$\ep$}}
\ee
with $\,\DDD_{\!\!\mbox{\tiny$\ep$}}$ being the real form of $\DDD_{\!\mbox{\tiny$\sss$}}$
and
\bb
\Phi'_{\!\mbox{\tiny$\ep$}} \equiv \left(\!\!
                                    \begin{array}{cc}
                                     0  & \Phi_{\!\mbox{\tiny$\sss$}}^{\mbox{\tiny cc}} \\
                                      \Phi_{\!\mbox{\tiny$\sss$}} &
                                      0 \\
                                    \end{array}
                                  \!\!\right)\,.
\ee

Let, respectively, $\partial_{\!\mbox{\tiny${\rm B}'$}}$ and $\db$ be the Bochner connections of $\,\DDD'_{\!\!\mbox{\tiny$\ep$}}$ and $\,\DDD_{\!\!\mbox{\tiny$\ep$}}.$ Then, Lemma \ref{lemma lichnerowicz formula} implies that
\bb
\partial_{\!\mbox{\tiny${\rm B}'$}} &=& \db + \alpha_{\mbox{\tiny${\rm D}'$}}\,,
\nonumber\\[.1cm]
\alpha_{\mbox{\tiny${\rm D}'$}}(v) &=& \mbox{\small$\frac{\varepsilon}{2}$}\left\{\gamma_{\mbox{\tiny$\ep$}}(v^\flat),
\Phi'_{\!\mbox{\tiny$\ep$}}\right\}\,.
\ee

By assumption $\partial_{\!\mbox{\tiny${\rm B}'$}}\in{\cal A}_{\mbox{\tiny Cl}}(\ep).$
We show that also $\db$ is a Clifford connection and thus $\alpha_{\mbox{\tiny${\rm D}'$}}$
has to commute with the Clifford action. This condition will eventually give us the explicit form of the zero order operator $\Phi'_{\!\mbox{\tiny$\ep$}}.$

Indeed, it follows that
\bb
\DDD'_{\!\!\mbox{\tiny$\ep$}} &=& \ddd_{\!\mbox{\tiny${\rm B}'$}} +
\Phi_{\mbox{\tiny${\rm D}'$}}\cr
&=&
\ddb + \,{/\!\!\!\!\alpha}_{\mbox{\tiny${\rm D}'$}} + \Phi_{\mbox{\tiny${\rm D}'$}}\cr
&=&
\ddb + \Phi_{\mbox{\tiny D}} + \Phi'_{\!\mbox{\tiny$\ep$}}\,,
\ee
where $\Phi_{\mbox{\tiny D}} = \,\DDD_{\!\!\mbox{\tiny$\ep$}} - \ddb.$ Therefore,
\bb
\Phi_{\mbox{\tiny${\rm D}'$}}&=&\tau_{\mbox{\tiny$\ep$}}\circ\phi_{\mbox{\tiny${\rm D}'$}}\cr
&=&
\Phi_{\mbox{\tiny D}} + \Phi'_{\!\mbox{\tiny$\ep$}} -
\,{/\!\!\!\!\alpha}_{\mbox{\tiny${\rm D}'$}}\,.
\ee
Whence,
\bb
\phi_{\mbox{\tiny${\rm D}'$}} = \tau_{\mbox{\tiny$\ep$}}\circ\Phi_{\mbox{\tiny D}} +
\tau_{\mbox{\tiny$\ep$}}\circ(\Phi'_{\!\mbox{\tiny$\ep$}} -
\,{/\!\!\!\!\alpha}_{\mbox{\tiny${\rm D}'$}})
\ee
and the condition
$\phi_{\mbox{\tiny${\rm D}'$}}\in{\goth Sec}(M,{\rm End}^-_\gamma(\ep))$ yields the equivalence:
\bb
\label{simple type cond.}
[\phi_{\mbox{\tiny${\rm D}'$}},\gamma_{\mbox{\tiny$\ep$}}(\alpha)] = 0
\quad\Leftrightarrow\quad\left\{
\begin{array}{ccc}
  \{\Phi_{\mbox{\tiny D}},\gamma_{\mbox{\tiny$\ep$}}(\alpha)\} & = & 0\,, \\[.1cm]
  \{(\Phi'_{\!\mbox{\tiny$\ep$}} -
\,{/\!\!\!\!\alpha}_{\mbox{\tiny${\rm D}'$}}),\gamma_{\mbox{\tiny$\ep$}}(\alpha)\} & = & 0\,,
\end{array}
\right.
\ee
for all $\alpha\in T^*\!M.$

According to Corollary \ref{korrollar simple type dop}, the first relation of
(\ref{simple type cond.}) implies that also $\db\in{\cal A}_{\mbox{\tiny Cl}}(\ep).$ Whence, $\,\DDD_{\!\!\mbox{\tiny$\ep$}}$ is of simple type:
\bb
\DDD_{\!\!\mbox{\tiny$\ep$}} = \left(
                                 \begin{array}{cc}
                                   \dda + \tau_{\mbox{\tiny$\sss$}}\circ\chi_{\mbox{\tiny$\sss$}} & 0 \\
                                   0 & (\dda + \tau_{\mbox{\tiny$\sss$}}\circ
                                   \chi_{\mbox{\tiny$\sss$}})^{\mbox{\tiny cc}} \\
                                 \end{array}
                               \right)\,,
\ee
with $\chi_{\mbox{\tiny$\sss$}}\in{\goth Sec}(M,{\rm End}^-_\gamma(\sss)).$

Moreover, being the difference of two Clifford connections it follows that
\bb
\label{real simple type condition}
[\alpha_{\mbox{\tiny${\rm D}'$}}(v),\gamma_{\mbox{\tiny$\ep$}}(\alpha)] \equiv 0\,,
\ee
for all $v\in TM$ and $\alpha\in T^*\!M.$ Taking into account the explict form of
$\alpha_{\mbox{\tiny${\rm D}'$}}$, the condition (\ref{real simple type condition}) is
seen to be equivalent to
\bb
\left[[\Phi_{\mbox{\tiny$\sss$}},\gamma_{\mbox{\tiny$\sss$}}(\alpha_1)]_\pm,
\gamma_{\mbox{\tiny$\sss$}}(\alpha_2)\right]_\mp \equiv 0\,,
\ee
for all $\alpha_1, \alpha_2\in T^*\!M.$ Here, $[x,y]_\pm \equiv xy \pm yx$, with the relative sign referring to $\gamma_{\mbox{\tiny$\sss$}}^{\mbox{\tiny cc}} = \pm\gamma_{\mbox{\tiny$\sss$}}.$

It follows that
\bb
\label{solution of simple type cond.}
\Phi_{\mbox{\tiny$\sss$}} = \left\{
\begin{array}{ccc}
  \delta_\gamma(\sigma_{\!\mbox{\tiny$\sss$}}) + \tau_{\mbox{\tiny$\sss$}}\circ\chi'_{\mbox{\tiny$\sss$}}\,, & \mbox{for} &  \gamma_{\mbox{\tiny$\sss$}}^{\mbox{\tiny cc}} = +\gamma_{\mbox{\tiny$\sss$}}\,,\\[.1cm]
  \mu_{\mbox{\tiny M}} + \tau_{\mbox{\tiny$\sss$}}\circ\delta_\gamma(\sigma_{\!\mbox{\tiny$\sss$}})\,,& \mbox{for} &
   \gamma_{\mbox{\tiny$\sss$}}^{\mbox{\tiny cc}} = -\gamma_{\mbox{\tiny$\sss$}}\,,
\end{array}
\right.
\ee
with $\chi'_{\mbox{\tiny$\sss$}},
\mu_{\mbox{\tiny M}}\in{\goth Sec}(M,{\rm End}^+_\gamma(\sss))$ and $\sigma_{\!\mbox{\tiny$\sss$}}\in\Omega^1(M,{\rm End^-_\gamma(\sss)})$.

For reasons of consistency we still have to verify the second relation of
(\ref{simple type cond.}) in order to complete the proof of Proposition
\ref{prop of real dop of simple type}. However, this is done straightforwardly taking
the explicit solution (\ref{solution of simple type cond.}) of
(\ref{real simple type condition}) into account. \hfill$\Box$

The significance of Proposition (\ref{prop of real dop of simple type}) is given by
generalizing the notion of simple type Dirac operators to those which are also real.
These are certainly distinguished Dirac operators on the real Clifford module
$\ep = \hspace{-.3cm}\phantom{\sss}^2\sss\twoheadrightarrow M$ on which one
may then apply the Pauli map (\ref{pauli-map}). Even more, these real simple
type Dirac operators also allow to incorporate Majorana masses within the scheme of
Dirac type gauge theories. For this, let $(\sss,\partial)$ be a flat
Majorana module with an imaginary Clifford action and grading involution. The stationary points of the fermionic action
${\cal I}_{\mbox{\tiny D,ferm}}$, which is defined by the real Dirac
operator of simple type
\bb
\DDD_{\!\mbox{\tiny M}} := \left(\!\!
                             \begin{array}{cc}
                               \ddd & i\mu_{\mbox{\tiny M}} \\
                               -i\mu_{\mbox{\tiny M}} & -\ddd \\
                             \end{array}
                           \!\!\right)\in{\cal D}_{\!\mbox{\tiny real}}(\ep)
\ee
with $\mu_{\mbox{\tiny M}}\in{\goth Sec}(M,{\rm End}^+_\gamma(\sss))$ being real, fulfil
the Majorana equations:
\bb
\label{majoeq}
i\ddd\chi &=& \mu_{\mbox{\tiny M}}\chi^{\mbox{\tiny cc}}\,,\cr
i\ddd\chi^{\mbox{\tiny cc}} &=& \mu_{\mbox{\tiny M}}\chi\,.
\ee

We note that the (total space of the) real sub-vector bundle
${\cal M}_{\mbox{\tiny$\ep$}}\twoheadrightarrow M,$ whose complexification equals $\ep\twoheadrightarrow M$, reads:
\bb
{\cal M}_{\mbox{\tiny$\ep$}} = \left\{\mbox{\small$\left(\!\!
                                                       \begin{array}{c}
                                                         z \\
                                                         z^{\mbox{\tiny cc}} \\
                                                       \end{array}
                                                     \!\!\right)
$}\in\ep\,|\,z\in\sss\right\}\,.
\ee
Hence, $\DDD_{\!\mbox{\tiny M}}$ leaves the real module
${\goth Sec}(M,{\cal M}_{\mbox{\tiny$\ep$}})$ invariant.

The equations (\ref{majoeq}) are diagonal with respect to the grading involution $\tau_{\!\mbox{\tiny$\sss$}}$.
In particular, they are diagonal with respect to the chirality involution $\tau_{\!\mbox{\tiny M}}$:
\bb
\DDD_{\!\mbox{\tiny M}}\psi = 0\qquad\Leftrightarrow\qquad
\left\{\begin{array}{ccc}
         i\ddd\chi_{\mbox{\tiny R}} & = &
         \mu_{\mbox{\tiny M}}\chi_{\mbox{\tiny R}}^{\mbox{\tiny cc}}\,, \\
         i\ddd\chi_{\mbox{\tiny L}} & = &
         \mu_{\mbox{\tiny M}}\chi_{\mbox{\tiny L}}^{\mbox{\tiny cc}}\,, \\
       \end{array}
\right.
\ee
plus the corresponding conjugate equations. Here, we have put
$\psi = \mbox{\small$\left(\!\!\begin{array}{c}
\chi \\
\chi^{\mbox{\tiny cc}} \\
\end{array}
\!\!\right)$}\in{\goth Sec}(M,{\cal M}_{\mbox{\tiny$\ep$}})$ and the chiral eigen sections
of $\tau_{\!\mbox{\tiny M}}$ are again denoted by
$\chi_{\mbox{\tiny R}},\chi_{\mbox{\tiny L}}\in{\goth Sec}(M,\sss),$ such that
$\chi = \chi_{\mbox{\tiny R}} + \chi_{\mbox{\tiny L}}$.

In this section, we discussed how the Majorana equations can be described in terms of real Dirac operators of simple type on real Clifford modules. We turn now to the corresponding discussion of the {\it Dirac-Yukawa equation}:
\bb
\label{diracyukawaeq}
i\dda\chi = \varphi_{\mbox{\tiny D}}\chi\quad\Leftrightarrow\quad
\left\{\begin{array}{ccc}
         i\dda\chi_{\mbox{\tiny R}} & = & \varphi_{\mbox{\tiny D}}\chi_{\mbox{\tiny L}}\,, \\
         i\dda\chi_{\mbox{\tiny L}} & = & \varphi_{\mbox{\tiny D}}\chi_{\mbox{\tiny R}}\,.
       \end{array}
\right.
\ee
The {\it Yukawa (coupling) term} $\varphi_{\mbox{\tiny D}}$ generalizes in a gauge covariant manner the usual mass term $m_{\mbox{\tiny D}}$ of the Dirac equation
(\ref{dirac equation}) with help of the Higgs field.

\subsection{Dirac masses and real Dirac operators of simple type}

In the last section we have shown how Majorana masses can be geometrically described in terms
of a real Clifford module if the latter is considered as being the doubling of a Majorana module. In order to also geometrically describe Dirac masses within Dirac type gauge theories we have to consider special Majorana modules $\sss\twoheadrightarrow M$, called {\it Dirac modules}. More precisely, we make the following

\begin{definition}
A real Clifford module
\bb
(\sss,\langle\cdot,\cdot\rangle_{\!\mbox{\tiny$\sss$}},
\tau_{\!\mbox{\tiny$\sss$}}, \gamma_{\!\mbox{\tiny$\sss$}},
J_{\!\mbox{\tiny$\sss$}})
\ee
is called a ``Dirac module'', provided there is a Majorana module
$(\www,\langle\cdot,\cdot\rangle_{\!\mbox{\tiny$\www$}},
\tau_{\!\mbox{\tiny$\www$}}, \gamma_{\!\mbox{\tiny$\www$}},
J_{\!\mbox{\tiny$\www$}})$ over $(M,g_{\mbox{\tiny M}})$, such that
\bb
\sss = \hspace{-.4cm}\phantom{\www}^2\www = \www\ot\cc^2
\ee
and
\bb
\tau_{\!\mbox{\tiny$\sss$}} &=& \left(\!\!
                                   \begin{array}{cc}
                                     {\rm id}_{\mbox{\tiny$\www$}} & \phantom{-}0 \\
                                     0 & -{\rm id}_{\mbox{\tiny$\www$}} \\
                                   \end{array}
                                 \!\!\right) =
                                 {\rm id}_{\mbox{\tiny$\www$}}\ot\tau_{\!\mbox{\tiny 2}}
                                 \,,\\[.1cm]
\gamma_{\mbox{\tiny$\sss$}} &=& \left(\!\!
                                   \begin{array}{cc}
                                     0 & \gamma_{\mbox{\tiny$\www$}} \\
                                     \gamma_{\mbox{\tiny$\www$}} & 0 \\
                                   \end{array}
                                 \!\!\right) =
                                 \gamma_{\mbox{\tiny$\www$}}\ot\varepsilon_{\!\mbox{\tiny 2}}
                                 \,,\\[.1cm]
J_{\mbox{\tiny$\sss$}} &=& \left(\!\!
                                   \begin{array}{cc}
                                     0 & J_{\mbox{\tiny$\www$}} \\
                                     J_{\mbox{\tiny$\www$}} & 0 \\
                                   \end{array}
                                 \!\!\right) =
                                 J_{\mbox{\tiny$\www$}}\ot\varepsilon_{\!\mbox{\tiny 2}}\,.
\ee
Finally,
\bb
\left\langle\mbox{\small$\left(\!\!
         \begin{array}{c}
           u_1 \\
           v_1 \\
         \end{array}
       \!\!\right)$},
       \mbox{\small$\left(\!\!
         \begin{array}{c}
           u_2 \\
           v_2 \\
         \end{array}
       \!\!\right)$}
\right\rangle_{\!\!\!\mbox{\tiny$\sss$}} =
\langle u_1,v_2\rangle_{\!\mbox{\tiny$\www$}} \pm
\langle v_1,u_2\rangle_{\!\mbox{\tiny$\www$}}\,,
\ee
depending on whether
$\langle J_{\mbox{\tiny$\www$}}(u),J_{\mbox{\tiny$\www$}}(v)\rangle_{\!\mbox{\tiny$\www$}}
=\pm\langle v,u\rangle_{\!\mbox{\tiny$\www$}}$, for all $u,v\in\www$.
\end{definition}

It follows that
\bb
\tau_{\!\mbox{\tiny$\sss$}}^{\mbox{\tiny cc}} &=& -\tau_{\!\mbox{\tiny$\sss$}}\,,\cr
\gamma_{\mbox{\tiny$\sss$}}^{\mbox{\tiny cc}} &=& \pm\gamma_{\mbox{\tiny$\sss$}}\quad
\Leftrightarrow\quad\gamma_{\mbox{\tiny$\www$}}^{\mbox{\tiny cc}} = \pm\gamma_{\mbox{\tiny$\www$}}\,.
\ee

Let $\dda\in{\cal D}(\www)$ and $\varphi_{\mbox{\tiny D}}\in
{\goth Sec}(M,{\rm End}_\gamma(\www))$. Furthermore, we assume that
$\dda \pm i\varphi_{\mbox{\tiny D}}$ are $S-$reducible if $\www\hookrightarrow\Lambda_{\mbox{\tiny M}}\ot E\twoheadrightarrow M$. We put
\bb
\label{Dirac-Yukawa op}
\DDD_{\!\mbox{\tiny D}} := \left(\!\!
                             \begin{array}{cc}
                               0 & \dda - i\varphi_{\mbox{\tiny D}} \\
                               \dda + i\varphi_{\mbox{\tiny D}} & 0 \\
                             \end{array}
                           \!\!\right)\in{\cal D}(\sss)\,,
\ee
which is easily seen to be of simple type. In fact, one may rewrite $\DDD_{\!\mbox{\tiny D}}$ as
\bb
\label{simple type dop on dirac mod.}
\DDD_{\!\mbox{\tiny D}} = \,\dda + i\mu_{\mbox{\tiny D}}\,,
\ee
with
\bb
\mu_{\mbox{\tiny D}} &\equiv& -\tau_{\!\mbox{\tiny$\sss$}}\circ\phi_{\mbox{\tiny D}}\,,\cr
\phi_{\mbox{\tiny D}} &:=& \varphi_{\mbox{\tiny D}}\ot\varepsilon_{\!\mbox{\tiny 2}}\in
{\goth Sec}(M,{\rm End}^-_\gamma(\sss))\,.
\ee

Here, by a slight abuse of notation we identify $\da\in{\cal A}_{\mbox{\tiny Cl}}(\www)$
with
\bb
\da = \left(\!\!\begin{array}{cc}
\da & 0 \\
0 & \da \\
\end{array}
\!\!\right)\in{\cal A}_{\mbox{\tiny Cl}}(\sss)\,,
\ee
such that always $\,\dda = \delta_\gamma\circ\da$ and, respectively,
$\,\dda\in{\cal D}(\www)$, or $\,\dda\in{\cal D}(\sss)$, depending on whether ``$\gamma$''
denotes either $\gamma_{\mbox{\tiny$\www$}}$, or $\gamma_{\mbox{\tiny$\sss$}}$.

Note that the simple type Dirac operator $\,\DDD_{\!\mbox{\tiny D}}\in{\cal D}(\sss)$ is not real. Also, the first order operators $\,\dda \pm i\varphi_{\mbox{\tiny D}}\in{\cal D}(\www)$ are not Dirac operators, in general. Clearly, for constant sections $\varphi_{\mbox{\tiny D}} = m_{\mbox{\tiny D}}$, these two operators are but the complex factors of the
{\it Klein-Gordon operator}:
$\,\DDD_{\!\mbox{\tiny D}}^2 = \,\dda^2 + m^2_{\mbox{\tiny D}}$.

Also note that the most general Dirac operators on a Dirac module read:
\bb
\DDD_{\!\mbox{\tiny$\sss$}} := \left(\!\!
                             \begin{array}{cc}
                               0 & \DDD_{\!\mbox{\tiny$\www$,1}}\\
                               \DDD_{\!\mbox{\tiny$\www$,2}} & 0 \\
                             \end{array}
                           \!\!\right)\in{\cal D}(\sss)\,,
\ee
where, respectively, $\,\DDD_{\!\mbox{\tiny$\www$,1}},\,\DDD_{\!\mbox{\tiny$\www$,2}}
\in{\cal D}(\www)$ are of Dirac type. In particular, the most general real Dirac operators
on a Dirac module are given by
\bb
\DDD_{\!\mbox{\tiny$\sss$}} := \left(\!\!
                             \begin{array}{cc}
                               0 & \,\DDD_{\!\mbox{\tiny$\www$}}^{\mbox{\tiny cc}} \\
                               \DDD_{\!\mbox{\tiny$\www$}} & 0 \\
                             \end{array}
                           \!\!\right)\in{\cal D}_{\!\mbox{\tiny real}}(\sss)\,,
\ee
In either case, the Dirac operators on a Dirac module are thus parameterized by general first order differential operators, acting on ${\goth Sec}(M,\www)$, such that their principal symbols are determined by the Clifford action of the underlying Majorana module. Then, our
Lemma (\ref{lemma lichnerowicz formula}) provides an explicit (global) formula for the
corresponding Lichnerowicz/Schr\"odinger decomposition of any such Dirac operator $\,\DDD_{\!\mbox{\tiny$\sss$}}\in{\cal D}(\sss)$ in terms of the underlying Dirac operators
$\,\DDD_{\!\mbox{\tiny$\www$,1}},\,\DDD_{\!\mbox{\tiny$\www$,2}}\in{\cal D}(\www)$.

Finally, the most general Dirac operator of simple type, acting on sections of a Dirac module, takes the form
\bb
\DDD_{\!\mbox{\tiny$\sss$}} := \left(\!\!
                             \begin{array}{cc}
                               0 & \dda - \Phi_{\mbox{\tiny$\www$}} \\
                               \dda + \Phi_{\mbox{\tiny$\www$}} & 0 \\
                             \end{array}
                           \!\!\right)\in{\cal D}(\sss)\,,
\ee
with $\Phi_{\mbox{\tiny$\www$}}\in{\goth Sec}(M,{\rm End}_\gamma(\www))$ being a general section. Indeed, for
\bb
\mu_{\mbox{\tiny$\sss$}} := \Phi_{\mbox{\tiny$\www$}}\ot{\rm I}_{\mbox{\tiny 2}}\in
{\goth Sec}(M,{\rm End}(\sss))
\ee
one obtains that for all $\alpha\in T^*\!M$:
\bb
\{\gamma_{\mbox{\tiny$\sss$}}(\alpha),\mu_{\mbox{\tiny$\sss$}}\} &=&
\gamma_{\mbox{\tiny$\www$}}(\alpha)\circ\Phi_{\mbox{\tiny$\www$}}\ot
\{\varepsilon_{\mbox{\tiny 2}},{\rm I}_{\mbox{\tiny 2}}\}\cr
&=&
0\,.
\ee
Therefore, $\,\DDD_{\!\mbox{\tiny$\sss$}}\equiv\,\dda + \mu_{\mbox{\tiny$\sss$}}$ is of simple type, whereby
\bb
\mu_{\mbox{\tiny$\sss$}} &:=& -\tau_{\mbox{\tiny$\sss$}}\circ\phi_{\mbox{\tiny$\sss$}}\,,\cr
\phi_{\mbox{\tiny$\sss$}} &:=& \Phi_{\mbox{\tiny$\www$}}\ot\varepsilon_{\mbox{\tiny 2}}\in
{\goth Sec}(M,{\rm End}^-_\gamma(\sss))\,.
\ee
Note that real Dirac operators on a Dirac module cannot be of simple type and vice versa.

To clarify how the Dirac-Yukawa equation (\ref{diracyukawaeq}) may arise from the Dirac functional ${\cal I}'_{\mbox{\tiny D,ferm}}$, one simply considers the real form of the (symmetric) simple type Dirac operator $\,\DDD_{\!\mbox{\tiny D}} = \dda +
i\mu_{\mbox{\tiny D}}$ on the Dirac module $\sss\twoheadrightarrow M,$ thereby defining a real Dirac operator of simple type on the associated real Clifford module
$\ep\twoheadrightarrow M$. Clearly,
\bb
{\cal I}_{\mbox{\tiny D,ferm}}({\cal P}_{\!\mbox{\tiny D}}(\,\DDD_{\!\mbox{\tiny D}} \op \,\DDD_{\!\mbox{\tiny D}}^{\mbox{\tiny cc}}))(\!\!\!\phantom{\psi}^{\mbox{\tiny 2}}\Psi) \;=\;
{\cal I}_{\mbox{\tiny D,ferm}}(\psi,\,\DDD_{\!\mbox{\tiny D}})\,.
\ee
Here,
\bb
\psi = \left(\!\!
         \begin{array}{c}
           \chi_1 \\
           \chi_2 \\
         \end{array}
       \!\!\right)\in{\goth Sec}(M,\sss)\,,\quad
\Psi = \left(\!\!
         \begin{array}{c}
           \psi \\
           \psi^{\mbox{\tiny cc}} \\
         \end{array}
       \!\!\right)\in{\goth Sec}(M,\ep)\,,
\ee
where $\chi_1,\,\chi_2\in{\goth Sec}(M,\www)$ are arbitrary sections, which are in one-to-one
correspondence with arbitrary eigen sections of the involution $\tau_{\mbox{\tiny$\sss$}}$ (not of $\tau_{\!\mbox{\tiny$\www$}}\,$!). Hence, to recover the Dirac-Yukawa equation (\ref{diracyukawaeq}) one may restrict to the eigen sections of $\tau_{\mbox{\tiny$\sss$}}$, corresponding to the eigen value equal to $+1$. That is,
\bb
i\dda\chi = \varphi_{\mbox{\tiny D}}\chi\quad\Leftrightarrow\quad\left\{
\begin{array}{ccc}
  \DDD_{\!\mbox{\tiny D}}\psi & = & 0\,, \\
  \tau_{\mbox{\tiny$\sss$}}\psi & = & \psi\,.
\end{array}
\right.
\ee
This ``solves'' the issue of fermion doubling already mentioned in the
introduction (and carefully discussed, for example, in \cite{LMMS:96}, \cite{LMMS:97}, \cite{GIS:97} and \cite{ToTh:05}; see also \cite{CM:07}).

For $\varphi_{\mbox{\tiny D}}\in{\goth Sec}(M,{\rm End}_\gamma(\www))$ the Dirac-Yukawa equation is odd with respect to the chirality involution (\ref{chirality invol.}):
\bb
i\dda\chi = \varphi_{\mbox{\tiny D}}\chi\quad\Leftrightarrow\quad\left\{
\begin{array}{ccc}
  i\dda\chi_{\mbox{\tiny R}} & = & \varphi_{\mbox{\tiny D}}\chi_{\mbox{\tiny L}}\,, \\
  i\dda\chi_{\mbox{\tiny L}} & = & \varphi_{\mbox{\tiny D}}\chi_{\mbox{\tiny R}}\,.
\end{array}
\right.
\ee
Again, $\chi_{\mbox{\tiny R}},\,\chi_{\mbox{\tiny L}}\in{\goth Sec}(M,\www)$
denote the chiral eigen sections: $\tau_{\mbox{\tiny M}}\chi_{\mbox{\tiny R/L}} =
\pm\chi_{\mbox{\tiny R/L}}$, such that $\chi = \chi_{\mbox{\tiny R}} +
\chi_{\mbox{\tiny L}}$.

For $\varphi_{\mbox{\tiny D}}\in{\goth Sec}(M,{\rm End}^+_\gamma(\www))$ the Dirac-Yukawa equation is gauge covariant only if the
chiral eigen sections carry the same representation of the underlying gauge group (i.e. the fermions are considered
``left-right gauge symmetric'').
Otherwise, the Yukawa coupling term has to be odd:
$\varphi_{\mbox{\tiny D}}\in{\goth Sec}(M,{\rm End}^-_\gamma(\www))$.

The first order differential operators $\dda \pm i\varphi_{\mbox{\tiny D}}$ on the Majorana module $\www\twoheadrightarrow M$
are of Dirac type, in general.  In contrast, the induced first order operator $\,\DDD_{\!\mbox{\tiny D}}\in{\cal D}(\sss)$ is
always a Dirac operator
(of simple type) on the corresponding Dirac module. We stress once more that in any case both Dirac type operators
$\dda \pm i\varphi_{\mbox{\tiny D}}$ are needed, actually, to define a simple type Dirac operator on the Dirac module,
thereby excluding the reality of
$\,\DDD_{\!\mbox{\tiny D}}$ . Finally, the Dirac-Yukawa equation is clearly diagonal with respect to the action of
``charge conjugation'' $J_{\mbox{\tiny$\www$}}$.

In the next section, we discuss the combined Dirac-Yukawa-Majorana equation and its implication for the Dirac action. We also briefly discuss the bundle structure that allows to regard the Majorana masses as constant sections of the Dirac module bundle associated with a Majorana module.

\section[The Pauli map and DYM operators]
{The Pauli map of the combined Dirac-Yukawa and Dirac-Majorana operator}
In the following, let $(\www,\partial)$ be a (partially) flat Majorana module over
$(M,g_{\mbox{\tiny M}})$, such that $\gamma_{\!\mbox{\tiny$\www$}}^{\mbox{\tiny cc}} = -\gamma_{\!\mbox{\tiny$\www$}}$.

We may complement the Majorana operator $\DDD_{\!\mbox{\tiny M}}\in{\cal D}(\ep)$
by the replacement of the real Dirac operator $\,\ddd\ominus\ddd$ on the Dirac module $\ep\twoheadrightarrow M$ by the real form of
$\,\DDD_{\!\mbox{\tiny D}} = \,\dda + i\mu_{\mbox{\tiny D}}\in{\cal D}(\sss)$ to obtain
the following real Dirac operator of simple type:
\bb
\label{diracyukawamajoop}
\DDD_{\!\mbox{\tiny YM}} &:=& \left(\!\!
                              \begin{array}{cc}
                                \dda + i\mu_{\mbox{\tiny D}} &  i\mu_{\mbox{\tiny M}}\\
                                -i\mu_{\mbox{\tiny M}} &
                                (\dda + i\mu_{\mbox{\tiny D}})^{\mbox{\tiny cc}} \\
                              \end{array}
                            \!\!\right)\nonumber\\[.1cm]
&\equiv&
\ddd_{\!\!\!\mbox{\tiny${\cal A}$}} + i\mu_{\mbox{\tiny YM}}\in
{\cal D}_{\!\mbox{\tiny real}}(\ep)\,.
\ee
Here, respectively,
\bb
\ddd_{\!\!\!\mbox{\tiny${\cal A}$}} &:=& \dda\op\,\dda^{\mbox{\tiny cc}}\in
{\cal D}_{\!\mbox{\tiny real}}(\ep)
\ee
is the real form of $\dda\in{\cal D}(\sss)$ and
\bb
\mu_{\mbox{\tiny YM}} &:=& \tau_{\!\mbox{\tiny$\ep$}}\circ\phi_{\mbox{\tiny YM}}\,,\\[.1cm]
\phi_{\mbox{\tiny YM}} &:=& \left(\!\!
                              \begin{array}{cc}
                                \tau_{\mbox{\tiny$\sss$}}\circ\mu_{\mbox{\tiny D}} & \tau_{\mbox{\tiny$\sss$}}\circ\mu_{\mbox{\tiny M}} \\
                                \tau_{\mbox{\tiny$\sss$}}\circ\mu_{\mbox{\tiny M}} & -(\tau_{\mbox{\tiny$\sss$}}\circ
                                \mu_{\mbox{\tiny D}})^{\mbox{\tiny cc}} \\
                              \end{array}
                            \!\!\right)\nonumber\\[.1cm]
                            &\equiv&
                            \left(\!\!
                              \begin{array}{cc}
                                -\phi_{\mbox{\tiny D}} &
                                \phi_{\mbox{\tiny M}} \\
                                \phantom{-}\phi_{\mbox{\tiny M}} &
                                \phi_{\mbox{\tiny D}}^{\mbox{\tiny cc}} \\
                              \end{array}
                            \!\!\right)\in{\goth Sec}(M,{\rm End}^-_\gamma(\ep))\,.
\ee
We call in mind that the {\it Majorana mass operator}
 $\mu_{\mbox{\tiny M}}\in{\goth Sec}(M,{\rm End}_\gamma^+(\sss))$ is supposed to be real.
In contrast, no such reality assumption is imposed on the {\it Dirac mass operator}
$\mu_{\mbox{\tiny D}}\in{\goth Sec}(M,{\rm End}^-(\sss))$, which has to fulfil the
requirement:
\bb
\{\mu_{\mbox{\tiny D}},\gamma_{\!\mbox{\tiny$\sss$}}(\alpha)\} = 0\,,
\ee
for all $\alpha\in T^*\!M$.

We call $\,\DDD_{\!\mbox{\tiny YM}} =
\,\ddd_{\!\!\!\mbox{\tiny${\cal A}$}} + i\mu_{\mbox{\tiny YM}}\in
{\cal D}_{\!\mbox{\tiny real}}(\ep)$ the {\it Dirac-Yukawa-Majorana operator (DYM)}.

Let $\chi\in{\goth Sec}(M,\www)$ and $\psi = \mbox{\small$\left(\!\!
                                               \begin{array}{c}
                                                 \chi \\
                                                 0 \\
                                               \end{array}
                                             \!\!\right)$}\in{\goth Sec}(M,\sss)$
be the associated eigen section of $\tau_{\!\mbox{\tiny$\sss$}}$ that corresponds to the
eigen value equal to $+1$. Also let
\bb
{/\!\!\!\!P}_{\!\!\mbox{\tiny DYM}} :=
{\cal P}_{\!\mbox{\tiny D}}(\,\DDD_{\!\mbox{\tiny YM}})\in
{\cal D}_{\!\mbox{\tiny real}}({\cal P})
\ee
and $\Psi = \mbox{\small$\left(\!\!\begin{array}{c}
\psi \\
\psi^{\mbox{\tiny cc}} \\
\end{array}
\!\!\right)$}\in{\goth Sec}(M,\ep).$ Note that $\psi^{\mbox{\tiny cc}} = \mbox{\small$\left(\!\!
\begin{array}{c}
0 \\
\chi^{\mbox{\tiny cc}} \\
\end{array}
\!\!\right)$}\in{\goth Sec}(M,\sss)$.

Clearly,
\bb
\langle\!\!\!\!\phantom{\Psi}^{\mbox{\tiny 2}}\Psi,\,
{/\!\!\!\!P}_{\!\!\mbox{\tiny DYM}}\!\!\!\!\!\phantom{\Psi}^{\mbox{\tiny 2}}
\Psi\rangle_{\mbox{\tiny$\ppp$}} \;=\;
\langle\Psi,\,\DDD_{\!\mbox{\tiny YM}}\Psi\rangle_{\!\mbox{\tiny$\ep$}}
\ee
and
\bb
\DDD_{\!\mbox{\tiny YM}}\Psi = \left(\!\!
\begin{array}{c}
(\dda + i\mu_{\mbox{\tiny D}})\psi + i\mu_{\mbox{\tiny M}}\psi^{\mbox{\tiny cc}}\\[.1cm]
(\dda + i\mu_{\mbox{\tiny D}})^{\mbox{\tiny cc}}\,\psi^{\mbox{\tiny cc}} -
i\mu_{\mbox{\tiny M}}\psi\\
\end{array}
\!\!\right)\,.
\ee
Whence,
\bb
\langle\!\!\!\!\phantom{\Psi}^{\mbox{\tiny 2}}\Psi,\,
{/\!\!\!\!P}_{\!\!\mbox{\tiny DYM}}\!\!\!\!\!\phantom{\Psi}^{\mbox{\tiny 2}}
\Psi\rangle_{\mbox{\tiny$\ppp$}} &=&
\mbox{\small$\frac{1}{2}$}(
\langle\psi,(\dda + i\mu_{\mbox{\tiny D}})\psi\rangle_{\mbox{\tiny$\sss$}} +
\langle\psi,i\mu_{\mbox{\tiny M}}\psi^{\mbox{\tiny cc}}\rangle_{\mbox{\tiny$\sss$}} +
\nonumber\\[.1cm]
&&
+\;
\langle\psi^{\mbox{\tiny cc}},
(\dda + i\mu_{\mbox{\tiny D}})^{\mbox{\tiny cc}}\psi^{\mbox{\tiny cc}}
\rangle_{\mbox{\tiny$\sss$}} -
\langle\psi^{\mbox{\tiny cc}},i\mu_{\mbox{\tiny M}}\psi\rangle_{\mbox{\tiny$\sss$}})
\nonumber\\[.2cm]
&=&
\mbox{\small$\frac{1}{2}$}(
\langle\chi,(\dda + i\varphi_{\mbox{\tiny D}})\chi\rangle_{\mbox{\tiny$\www$}} +
\langle\chi,im_{\mbox{\tiny M}}\chi^{\mbox{\tiny cc}}\rangle_{\mbox{\tiny$\www$}} +
\nonumber\\[.1cm]
&&
+\;
\langle\chi^{\mbox{\tiny cc}},
(\dda + i\varphi_{\mbox{\tiny D}})^{\mbox{\tiny cc}}\chi^{\mbox{\tiny cc}}
\rangle_{\mbox{\tiny$\www$}} -
\langle\chi^{\mbox{\tiny cc}},im_{\mbox{\tiny M}}\chi\rangle_{\mbox{\tiny$\www$}})\,,
\ee
where we have put
\bb
\mu_{\mbox{\tiny M}} \equiv \left(\!\!
                              \begin{array}{cc}
                                m_{\mbox{\tiny M}} & 0 \\
                                0 & m_{\mbox{\tiny M}} \\
                              \end{array}
                            \!\!\right)\,,\qquad
                            m_{\mbox{\tiny M}}\in{\goth Sec}(M,{\rm End}_\gamma(\www))\;
                            {\mbox{real and constant}}
\ee
according to the definition (and the physical interpretation) of the Majorana mass operator.

Thus, the quadratic form
${\cal I}_{\mbox{\tiny D,ferm}}(\,{/\!\!\!\!P}_{\!\!\mbox{\tiny DYM}})$ on
${\goth Sec}(M,\ep)$ yields the Euler-Lagrange equations:
\bb
\label{comb.dymeq.on S}
i\dda\psi &=& \mu_{\mbox{\tiny D}}\psi + \mu_{\mbox{\tiny M}}\psi^{\mbox{\tiny cc}}\,,\\
\label{comb.dymeq.cc.on S}
(i\dda\psi)^{\mbox{\tiny cc}} &=&
\mu_{\mbox{\tiny D}}^{\mbox{\tiny cc}}\psi^{\mbox{\tiny cc}} + \mu_{\mbox{\tiny M}}\psi\,.
\ee
When restricted to $\tau_{\!\mbox{\tiny$\sss$}}\psi = \psi,$ these equations
become equivalent to:
\bb
\label{comb.dymeq.on W}
i\dda\chi &=& \varphi_{\mbox{\tiny D}}\chi + m_{\mbox{\tiny M}}\chi^{\mbox{\tiny cc}}\,,\\
\label{comb.dymeq.cc.on W}
(i\dda\chi)^{\mbox{\tiny cc}} &=&
\varphi_{\mbox{\tiny D}}^{\mbox{\tiny cc}}\chi^{\mbox{\tiny cc}} + m_{\mbox{\tiny M}}\chi\,.
\ee

In order to geometrically describe the Yukawa coupling term (and thus the
``Dirac mass'' after spontaneous symmetry breaking) in terms of (real) Dirac operators of simple type one simply has to go from the underlying
Majorana module to the corresponding Dirac module quite similar to how the ``Pauli matrices'' are lifted to
the ``Dirac matrices'' (the latter being considered in the Majorana representation). The doubling of the Dirac module then allows to also geometrically describe the characteristic
``particle-anti-particle'' coupling that arises by the Majorana mass term, in terms of real, simple type Dirac operators. In fact, this is where the real structure necessarily enters the scheme, thereby turning the Clifford module into a real Clifford module. Finally, the Pauli map of the DYM, which describes both the ``left-right'' coupling and the
``particle-anti-particle'' coupling on the same geometrical footing, then allows to geometrically describe the Pauli term in a way that does not alter the fermionic
action. For reasons of renormalization, this is actually necessary.

Before we discuss the bosonic part of the full Dirac action with respect to the
real Dirac operator ${/\!\!\!\!P}_{\!\!\mbox{\tiny DYM}}$, we still comment on the
gauge invariance of the equations (\ref{comb.dymeq.on W}--\ref{comb.dymeq.cc.on W}).
Of course, this is related to the dynamical discrepancy between the fermionic
left-right coupling, provided by the Dirac mass, and the particle-anti-particle coupling that is invoked on the fermions by the Majorana mass.

For the sake of gauge invariance, the underlying Majorana module
$\www\twoheadrightarrow M$ has to be {\it partially flat} when Majorana masses are taken into account. In this case: $\da \not= \partial$ only for $\chi\in ker(m_{\mbox{\tiny M}}).$  In geometrical terms this and the constant Majorana mass operator may be described by the assumption that the Majorana module splits:
\bb
\www = \begin{array}{c}
         \www_{\!\mbox{\tiny$\nu$}} \\
         \op \\
         \www_{\!\mbox{\tiny$e$}}
       \end{array} \twoheadrightarrow M\,,
\ee
where the sub-bundle $\www_{\!\mbox{\tiny$\nu$}}\twoheadrightarrow M$ carries the trivial
representation of the Yang-Mills gauge group
${\cal G}_{\mbox{\tiny YM}}\subset{\cal G}_{\mbox{\tiny D}}$ and
\bb
m_{\mbox{\tiny M}} \equiv \left(\!\!
                       \begin{array}{cc}
                         m_{\mbox{\tiny M,$\nu$}} & 0 \\
                         0 & 0 \\
                       \end{array}
                     \!\!\right)\,.
\ee

Accordingly,  $\da\in{\cal A}_{\mbox{\tiny Cl}}(\www)$ and
$\varphi_{\mbox{\tiny D}}\in{\goth Sec}(M,{\rm End}_\gamma(\www))$ may be decomposed as
\bb
\da &\equiv&
\left(\!\!
\begin{array}{cc}
\partial & 0 \\
0 & \da \\
\end{array}
\!\!\right)\,,
\nonumber\\[.1cm]
\varphi_{\mbox{\tiny D}} &\equiv&
\left(\!\!
\begin{array}{cc}
m_{\mbox{\tiny D,$\nu$}} & 0 \\
0 & \varphi_{\!\mbox{\tiny$e$}} \\
\end{array}
\!\!\right)\,,
\ee
with $m_{\mbox{\tiny M,$\nu$}},\;m_{\mbox{\tiny D,$\nu$}}\in{\goth Sec}(M,{\rm End_\gamma(\www_{\!\mbox{\tiny$\nu$}})})$
being real with respect to $J_{\mbox{\tiny$\www$}}$ and constant. Furthermore,
$\varphi_{\!\mbox{\tiny$e$}}\in{\goth Sec}(M,{\rm End_\gamma(\www_{\!\mbox{\tiny$e$}})})$.

The combined Dirac-Majorana equations (\ref{comb.dymeq.on W}--\ref{comb.dymeq.cc.on W})
become equivalent to
\bb
\label{diracmajomasseq}
i\ddd\nu &=& m_{\mbox{\tiny D,$\nu$}}\nu + m_{\mbox{\tiny M,$\nu$}}\nu^{\mbox{\tiny cc}}\,,\\
\label{diracmasseq}
i\dda e & = & \varphi_{\!\mbox{\tiny$e$}}e\,,
\ee
together with the corresponding complex (or charge) conjugate equations.
Here, we have put $\chi \equiv (\nu, e)\in {\goth Sec}(M,\www_{\!\mbox{\tiny$\nu$}}\!
\op\www_{\!\mbox{\tiny$e$}})$ for the ``uncharged sections'' and the ``charged sections'', respectively, of the Majorana module $\www\twoheadrightarrow M$. Generically, the uncharged sections $\nu\in{\goth Sec}(M,\www_{\!\mbox{\tiny$\nu$}})$ are referred to as
{\it ``cosmological neutrinos''}. They are carriers of Dirac and/or Majorana masses or are massless, depending on $ker(m_{\mbox{\tiny D,$\nu$}})$ and $ker(m_{\mbox{\tiny M,$\nu$}})$. Clearly, in the case of {\it Majorana neutrinos}: $\nu^{\mbox{\tiny cc}} = \nu\in
{\goth Sec}(M,{\cal M}_{\mbox{\tiny$\www,\nu$}})\subset
{\goth Sec}(M,{\cal M}_{\mbox{\tiny$\www$}})$ (whereby
$\www = {\cal M}^\cc_{\mbox{\tiny$\www$}}$), the notions of Dirac and Majorana masses coincide and (\ref{diracmajomasseq}) reduces to
\bb
i\ddd\nu = m_{\mbox{\tiny$\nu$}}\nu\,,\quad(\nu^{\mbox{\tiny cc}} = \nu)\,.
\ee

Only the sub-module
\bb
ker(m_{\mbox{\tiny M}}) = \www_{\!\mbox{\tiny$e$}}\hookrightarrow\www\twoheadrightarrow M
\ee
of the Majorana module carries a non-trivial representation of the Yang-Mills gauge sub-group of ${\cal G}_{\mbox{\tiny D}}$.

In the case of the Standard Model, the cosmological neutrinos should not be confounded with the electrically neutral (left-handed) component of
$e\in{\goth Sec}(M,\www_{\!\mbox{\tiny$e$}})$ after the mechanism of spontaneous symmetry
break has been established. Indeed, the sections
$\nu\in{\goth Sec}(M,\www_{\!\mbox{\tiny$\nu$}})$ represent a kind of new species
of particles which do not contribute to any yet known kind of interaction besides gravity. This ``ghost like species'' of particles may thus serve as candidates for ``dark matter'' (resp. ``dark energy''). Of course, the masses of the cosmological neutrinos cannot be dynamically generated by the mechanism of spontaneous symmetry breaking since the
cosmological neutrinos only carry the trivial representation of the Yang-Mills gauge group.
This is certainly unsatisfying but may change with the upcoming experiments made at the Large Hadron Collider (LHC) at CERN/Swiss.

The Dirac mass matrix is known to only couple particles of different chirality but
respects the particle-anti-particle grading. This is opposed to the Majorana mass matrix. Since the latter is ``non-dynamical'' one may wonder to what extent the Majorana masses may nonetheless dynamically contribute, for example, to the Standard Model? A partial answer to this question within Dirac type gauge theories will be discussed next.

\subsection{The Dirac action concerning DYM}
So far, we have carefully discussed the fermionic action of the total Dirac action. In this section we discuss the bosonic part of the latter with respect to the corresponding DYM. Since the Dirac-Yukawa-Majorana operator $\,\DDD_{\!\mbox{\tiny YM}} =
\,\ddd_{\!\!\!\mbox{\tiny${\cal A}$}} + i\mu_{\mbox{\tiny YM}}\in
{\cal D}_{\!\mbox{\tiny real}}(\ep)$ is of simple type it lifts to
${\cal D}_{\!\mbox{\tiny real}}({\cal P}=\!\!\!\!\phantom{\ep}^{\mbox{\tiny 2}}\!\ep)$ via the Pauli map. It thereby generalizes the operator (\ref{pauli type dop}). The latter operator has been shown earlier to yield the Standard Model (STM) action including gravity (c.f. \cite{Tol:98}, \cite{ToTh:05} and \cite{ToTh:06}). This time, however, also Majorana masses are taken into account. We therefore summarize the basic steps allowing to express the Lagrangian density
\bb
{\cal L}_{\mbox{\tiny DYM}} := \ast{\rm tr}_\gamma\!
\left(curv(\,{/\!\!\!\!P}_{\!\!\mbox{\tiny DYM}}) -
\varepsilon ev_g(\omega^2_{\mbox{\tiny D}})\right)
\ee
in terms of the sections given by the metric $g_{\mbox{\tiny M}},$
the Yang-Mills gauge field $A,$ the Higgs field $\varphi_{\mbox{\tiny D}}$ (resp. $\varphi_{\mbox{\tiny$e$}}$)  and the Majorana (Dirac) masses $m_{\mbox{\tiny M}}$ ($m_{\mbox{\tiny D}}$) which altogether parameterize the Dirac-Yukawa-Majorana operator $\,\DDD_{\!\mbox{\tiny YM}}\in{\cal D}_{\mbox{\tiny S,real}}(\ep).$

Following the calculation, it will be shown that this density is automatically real and thus
takes values in $\Omega^n(M).$ Furthermore, the calculation will also allow to reveal a subtle relation between simple type Dirac operators and the ``kinetic term'' of the Higgs field within the STM action.

To get started, we put $\,{/\!\!\!\!P}_{\!\!\mbox{\tiny DYM}} =
\,\ddd_{\!\!\!\mbox{\tiny${\cal A}$}} + i\Phi_{\!\mbox{\tiny DYM}}$, where
\bb
\Phi_{\!\mbox{\tiny DYM}} \equiv \left(\!\!
                                     \begin{array}{cc}
                                       \mu_{\mbox{\tiny YM}} &
                                       -\,{/\!\!\!\!F}_{\!\!\mbox{\tiny DYM}} \\
                                       {/\!\!\!\!F}_{\!\!\mbox{\tiny DYM}} &
                                       \mu_{\mbox{\tiny YM}} \\
                                     \end{array}
                                   \!\!\right)\,,\qquad
\mu_{\mbox{\tiny YM}} := \left(\!\!
                              \begin{array}{cc}
                                \mu_{\mbox{\tiny D}} &
                                \mu_{\mbox{\tiny M}} \\
                                -\mu_{\mbox{\tiny M}} &
                                -\mu_{\mbox{\tiny D}}^{\mbox{\tiny cc}} \\
                              \end{array}
                            \!\!\right)
\ee
and $\,{/\!\!\!\!F}_{\!\!\mbox{\tiny DYM}}\in{\goth Sec}(M,{\rm End}^+(\ep))$ is the (quantized) relative curvature of $\,\DDD_{\!\mbox{\tiny YM}}$. Since the latter is of simple type, it follows that
\bb
F_{\!\mbox{\tiny DYM}}\; =\; F_{\!\!\mbox{\tiny${\cal A}$}} -
(d_{\!\mbox{\tiny${\cal A}$}}\,(i\mu_{\mbox{\tiny YM}})\, +\,
(i\mu_{\mbox{\tiny YM}})^2\Theta)\wedge\Theta\in\Omega^2(M,{\rm End}^+(\ep))\,.
\ee
Here, $d_{\!\mbox{\tiny${\cal A}$}}$ is the exterior covariant derivative with respect to
the (real) Clifford connection $\partial_{\!\!\mbox{\tiny${\cal A}$}}\in
{\cal A}_{\mbox{\tiny Cl}}(\ep)$ and $F_{\!\!\mbox{\tiny${\cal A}$}}\in
\Omega^2(M,{\rm End}_\gamma^+(\ep))$ its twisting curvature. Hence,
\bb
{/\!\!\!\!F}_{\!\!\mbox{\tiny DYM}}\; =\;
\,{/\!\!\!\!F}_{\!\!\!\!\mbox{\tiny${\cal A}$}}\, +\, \mbox{\small$\frac{n - 1}{n}$}
\left(\delta_\gamma(d_{\!\mbox{\tiny${\cal A}$}}\,(i\mu_{\mbox{\tiny YM}})) +
(i\mu_{\mbox{\tiny YM}})^2\right)\,.
\ee

We may then take advantage of Lemma (\ref{lemma lichnerowicz formula}) to obtain:
\bb
{\rm tr}_{\!\mbox{\tiny$\ppp$}}V_{\!\mbox{\tiny D}} =
{\rm tr}_\gamma(curv(\,\ddd_{\!\!\!\mbox{\tiny${\cal A}$}})) -
{\rm tr}_{\!\mbox{\tiny$\ppp$}}\Phi_{\!\mbox{\tiny DYM}}^2 +
\mbox{\small$\frac{\varepsilon}{4}$}\,g_{\mbox{\tiny M}}(e_i,e_j)
\,{\rm tr}_{\!\mbox{\tiny$\ppp$}}\!\left(
\{\gamma_{\mbox{\tiny$\ppp$}}(e^i),\Phi_{\!\mbox{\tiny DYM}}\}
\{\gamma_{\mbox{\tiny$\ppp$}}(e^j),\Phi_{\!\mbox{\tiny DYM}}\}\right)\,,
\ee
where we have neglected an appropriate boundary term and $e_1,\ldots,e_n\in TM$
is any local ($g_{\mbox{\tiny M}}-$orthonormal) basis with dual basis
$e^1,\ldots,e^n\in T^*\!M$.

It follows that
\bb
{\rm tr}_{\!\mbox{\tiny$\ppp$}}\Phi_{\!\mbox{\tiny DYM}}^2
&=&
2{\rm tr}_{\!\mbox{\tiny$\ep$}}\!\left(\mu_{\mbox{\tiny YM}}^2 -
\,{/\!\!\!\!F}_{\!\!\mbox{\tiny DYM}}^2\right)\,,
\nonumber\\[.2cm]
{\rm tr}_{\!\mbox{\tiny$\ppp$}}\!\left(
\{\gamma_{\mbox{\tiny$\ppp$}}(e^i),\Phi_{\!\mbox{\tiny DYM}}\}
\{\gamma_{\mbox{\tiny$\ppp$}}(e^j),\Phi_{\!\mbox{\tiny DYM}}\}\right)
&=&
2{\rm tr}_{\!\mbox{\tiny$\ep$}}\!\left(
\{\gamma_{\!\mbox{\tiny$\ep$}}(e^i),\mu_{\!\mbox{\tiny YM}}\}
\{\gamma_{\!\mbox{\tiny$\ep$}}(e^j),\mu_{\!\mbox{\tiny YM}}\}\right) -
\nonumber\\[.1cm]
&&
\hspace{-.38cm}-\;
2{\rm tr}_{\!\mbox{\tiny$\ep$}}\!\left(
\{\gamma_{\!\mbox{\tiny$\ep$}}(e^i),\,{/\!\!\!\!F}_{\!\!\mbox{\tiny DYM}}\}
\{\gamma_{\!\mbox{\tiny$\ep$}}(e^j),\,{/\!\!\!\!F}_{\!\!\mbox{\tiny DYM}}\}\right)\,.
\ee

Furthermore,
\bb
{\rm tr}_{\!\mbox{\tiny$\ep$}}\,{/\!\!\!\!F}_{\!\!\mbox{\tiny DYM}}^2 =
-\mbox{\small$\frac{1}{2}$}{\rm tr}_g F_{\!\!\!\mbox{\tiny${\cal A}$}}^2 +
\varepsilon\mbox{\small$(\frac{n - 1}{n})^2$}\,
{\rm tr}_g(\partial_{\!\!\mbox{\tiny${\cal A}$}}\,\mu_{\mbox{\tiny YM}})^2 +
\mbox{\small$(\frac{n - 1}{n})^2$}\,
{\rm tr}_{\!\mbox{\tiny$\ep$}}\mu_{\mbox{\tiny YM}}^4
\ee
and
\bb
{\rm tr}_{\!\mbox{\tiny$\ep$}}\!\left(
\{\gamma_{\!\mbox{\tiny$\ep$}}(e^i),\mu_{\!\mbox{\tiny YM}}\}
\{\gamma_{\!\mbox{\tiny$\ep$}}(e^j),\mu_{\!\mbox{\tiny YM}}\}\right) &=& 0\,,
\\[.1cm]
{\rm tr}_{\!\mbox{\tiny$\ep$}}\!\left(
\{\gamma_{\!\mbox{\tiny$\ep$}}(e^i),\,{/\!\!\!\!F}_{\!\!\mbox{\tiny DYM}}\}
\{\gamma_{\!\mbox{\tiny$\ep$}}(e^j),\,{/\!\!\!\!F}_{\!\!\mbox{\tiny DYM}}\}\right)
&=&
{\rm tr}_{\!\mbox{\tiny$\ep$}}\!\left(
\{\gamma_{\!\mbox{\tiny$\ep$}}(e^i),\,{/\!\!\!\!F}_{\!\!\!\!\mbox{\tiny${\cal A}$}}\}
\{\gamma_{\!\mbox{\tiny$\ep$}}(e^j),\,{/\!\!\!\!F}_{\!\!\!\!\mbox{\tiny${\cal A}$}}\}\right)
+\nonumber\\[.1cm]
&&
+\;
\mbox{\small$(\frac{1-n}{n})^2$}\,{\rm tr}_{\!\mbox{\tiny$\ep$}}\!\left(
\{\gamma_{\!\mbox{\tiny$\ep$}}(e^i),\,{/\!\!\!\!\alpha}_{\mbox{\tiny YM}}\}
\{\gamma_{\!\mbox{\tiny$\ep$}}(e^j),\,{/\!\!\!\!\alpha}_{\mbox{\tiny YM}}\}\right)
+\nonumber\\[.1cm]
&&
+\;
\mbox{\small$4(\frac{1-n}{n})^2$}\,
{\rm tr}_{\!\mbox{\tiny$\ep$}}\!\left(\gamma_{\!\mbox{\tiny$\ep$}}(e^i)
\gamma_{\!\mbox{\tiny$\ep$}}(e^i)\mu_{\!\mbox{\tiny YM}}^4\right)\,,
\ee
where we abbreviated $\,{/\!\!\!\!\alpha}_{\mbox{\tiny YM}}\equiv
\delta_\gamma(d_{\!\mbox{\tiny${\cal A}$}}\,(i\mu_{\mbox{\tiny YM}}))$. Also,
\bb
\mbox{\small$\frac{\varepsilon}{4}$}\,g_{\mbox{\tiny M}}(e_i,e_j)\,
{\rm tr}_{\!\mbox{\tiny$\ep$}}\!\left(
\{\gamma_{\!\mbox{\tiny$\ep$}}(e^i),\,{/\!\!\!\!F}_{\!\!\!\!\mbox{\tiny${\cal A}$}}\}
\{\gamma_{\!\mbox{\tiny$\ep$}}(e^j),\,{/\!\!\!\!F}_{\!\!\!\!\mbox{\tiny${\cal A}$}}\}\right)
&=&
\mbox{\small$\frac{2-n}{2}$}\,
{\rm tr}_g(F_{\!\!\!\mbox{\tiny${\cal A}$}}^2)\,,
\\[.1cm]
\mbox{\small$(\frac{1-n}{n})^2$}\,\mbox{\small$\frac{\varepsilon}{4}$}\,
g_{\mbox{\tiny M}}(e_i,e_j)\,
{\rm tr}_{\!\mbox{\tiny$\ep$}}\!\left(
\{\gamma_{\!\mbox{\tiny$\ep$}}(e^i),\,{/\!\!\!\!\alpha}_{\mbox{\tiny YM}}\}
\{\gamma_{\!\mbox{\tiny$\ep$}}(e^j),\,{/\!\!\!\!\alpha}_{\mbox{\tiny YM}}\}\right)
&=&
-\mbox{\small$\varepsilon\frac{(1-n)^3}{n^2}$}
\,{\rm tr}_g(\partial_{\!\!\mbox{\tiny${\cal A}$}}\,\mu_{\mbox{\tiny YM}})^2\,,
\\[.1cm]
\mbox{\small$\varepsilon\,(\frac{1-n}{n})^2$}\,
g_{\mbox{\tiny M}}(e_i,e_j)\,
{\rm tr}_{\!\mbox{\tiny$\ep$}}\!\left(\gamma_{\!\mbox{\tiny$\ep$}}(e^i)
\gamma_{\!\mbox{\tiny$\ep$}}(e^i)\mu_{\!\mbox{\tiny YM}}^4\right)
&=&
\mbox{\small$n(\frac{1-n}{n})^2$}\,{\rm tr}_{\!\mbox{\tiny$\ep$}}
(\mu_{\mbox{\tiny YM}}^4)\,.
\ee

Finally, one ends up with
\bb
-{\rm tr}_{\!\mbox{\tiny$\ppp$}}\Phi_{\!\mbox{\tiny DYM}}^2 +
\mbox{\small$\frac{\varepsilon}{4}$}\,g_{\mbox{\tiny M}}(e_i,e_j)
\,{\rm tr}_{\!\mbox{\tiny$\ppp$}}\!\left(
\{\gamma_{\mbox{\tiny$\ppp$}}(e^i),\Phi_{\!\mbox{\tiny DYM}}\}
\{\gamma_{\mbox{\tiny$\ppp$}}(e^j),\Phi_{\!\mbox{\tiny DYM}}\}\right) =\hspace{3.18cm}
\nonumber\\[.08cm]
-2{\rm tr}_{\!\mbox{\tiny$\ep$}}\mu_{\mbox{\tiny YM}}^2 +
2{\rm tr}_{\!\mbox{\tiny$\ep$}}\,{/\!\!\!\!F}_{\!\!\mbox{\tiny DYM}}^2
-\mbox{\small$\frac{\varepsilon}{2}$}\,g_{\mbox{\tiny M}}(e_i,e_j)\,
{\rm tr}_{\!\mbox{\tiny$\ep$}}
\{\gamma_{\!\mbox{\tiny$\ep$}}(e^i),{/\!\!\!\!F}_{\!\!\mbox{\tiny DYM}}\}
\{\gamma_{\!\mbox{\tiny$\ep$}}(e^j),{/\!\!\!\!F}_{\!\!\mbox{\tiny DYM}}\}=\hspace{2cm}
\nonumber\\[.18cm]
\mbox{\small$(n-3)$}\,
{\rm tr}_g(F_{\!\!\!\mbox{\tiny${\cal A}$}}^2) -
\mbox{\small$2\varepsilon(n-2)(\frac{n-1}{n})^2$}\,
{\rm tr}_g(\partial_{\!\!\mbox{\tiny${\cal A}$}}\,\mu_{\mbox{\tiny YM}})^2 -
\mbox{\small$2\frac{(n-1)^3}{n^2}$}\,
{\rm tr}_{\!\mbox{\tiny$\ep$}}(\mu_{\mbox{\tiny YM}}^4)
-\mbox{\small$2$}\,
{\rm tr}_{\!\mbox{\tiny$\ep$}}(\mu_{\mbox{\tiny YM}}^2)\,,
\ee
which for $\varepsilon := +1$ and anti-Hermitian $\mu_{\mbox{\tiny YM}}$ has the form of the Standard Model Lagrangian. We stress that the ``kinetic term'' of the Higgs,
${\rm tr}_g(\partial_{\!\!\mbox{\tiny${\cal A}$}}\,\mu_{\mbox{\tiny YM}})^2$, drops out, if $\,\DDD_{\!\mbox{\tiny YM}}$ were not of simple type.

The explicit form of the combined Dirac-Majorana mass operator
$\mu_{\mbox{\tiny YM}}\in{\goth Sec}(M,{\rm End}(\ep))$ yields:
\bb
{\rm tr}_g(\partial_{\!\!\mbox{\tiny${\cal A}$}}\,\mu_{\mbox{\tiny YM}})^2 &=&
-\mbox{\small$4$Re}\,{\rm tr}_g(\da\varphi_{\!\mbox{\tiny$e$}})^2\,,\\[.1cm]
a\,{\rm tr}_{\!\mbox{\tiny$\ep$}}\mu_{\mbox{\tiny YM}}^4
+ {\rm tr}_{\!\mbox{\tiny$\ep$}}\mu_{\mbox{\tiny YM}}^2 \!&=&\!
4{\rm Re}\left(a\,{\rm tr}_{\!\mbox{\tiny$\www_{\!e}$}}\varphi_{\!\!\mbox{\tiny$e$}}^4
-\,{\rm tr}_{\!\mbox{\tiny$\www_{\!e}$}}\varphi_{\!\!\mbox{\tiny$e$}}^2
+ \Lambda_{\mbox{\tiny DM,$\nu$}}\right)\,,
\ee
whereby $a \equiv \mbox{\small$2\,\frac{(n-1)^3}{n^2}$}$ and
\bb
\label{true cosmological const.}
\Lambda_{\mbox{\tiny DM,$\nu$}} &\equiv&
a\,{\rm tr}_{\!\mbox{\tiny$\www_{\nu}$}}m_{\mbox{\tiny D,$\nu$}}^4
- {\rm tr}_{\!\mbox{\tiny$\www_{\nu}$}}m_{\mbox{\tiny D,$\nu$}}^2 +
a\,{\rm tr}_{\!\mbox{\tiny$\www_{\nu}$}}m_{\mbox{\tiny M,$\nu$}}^4
- {\rm tr}_{\!\mbox{\tiny$\www_{\nu}$}}m_{\mbox{\tiny M,$\nu$}}^2
\nonumber\\[.1cm]
&&
-\;
2a\,{\rm tr}_{\!\mbox{\tiny$\www_{\nu}$}}
(m_{\mbox{\tiny D,$\nu$}}\circ m_{\mbox{\tiny M,$\nu$}})^2
\ee
is the ``true cosmological constant'', which naturally occurs in the Einstein-Hilbert action
when Majorana masses are taken into account within the geometrical frame of Dirac type
gauge theories. Its possible phenomenological consequences, for instance, with respect to the mass of the Higgs boson and the cosmological issue of ``dark matter'', will be discussed separately in a forthcoming paper. However, because of the significance of the cosmological constant, we summarize the basic steps to obtain the result (\ref{true cosmological const.}). This will also enlighten the subtle interplay between simple type Dirac operators and the peculiar form of $\Lambda_{\mbox{\tiny DM,$\nu$}}$ as the sum of two Higgs potentials and an ``interaction term'' for the Dirac and Majorana masses.

First, it follows that $\mu_{\mbox{\tiny YM}}^2$ structurally reads:
\bb
\mu_{\mbox{\tiny YM}}^2 &=& \left(
                          \begin{array}{cc}
                            u & z \\
                            z^{\mbox{\tiny cc}} & u^{\mbox{\tiny cc}} \\
                          \end{array}
                        \right)\,,\nonumber\\[.1cm]
u &\equiv& \mu_{\mbox{\tiny D}}^2 - \mu_{\mbox{\tiny M}}^2\,,\cr
z &\equiv& \mu_{\mbox{\tiny D}}\circ\mu_{\mbox{\tiny M}} -
\mu_{\mbox{\tiny M}}\circ\mu_{\mbox{\tiny D}}^{\mbox{\tiny cc}}\,.
\ee

Because of the explicit form of the sections
$\varphi_{\mbox{\tiny D}}\in{\goth Sec}(M,{\rm End}_\gamma(\www_{\!\mbox{\tiny$\nu$}}\op
\www_{\!\mbox{\tiny$e$}}))$ and $m_{\mbox{\tiny M}}\in
{\goth Sec}(M,{\rm End}^+_\gamma(\www_{\!\mbox{\tiny$\nu$}}\op\www_{\!\mbox{\tiny$e$}}))$,
one gets
\bb
m_{\mbox{\tiny M}}\circ\varphi_{\mbox{\tiny D}}^{\mbox{\tiny cc}} =
m_{\mbox{\tiny M}}\circ\varphi_{\mbox{\tiny D}}\quad\Rightarrow\quad
\mu_{\mbox{\tiny M}}\circ\mu_{\mbox{\tiny D}}^{\mbox{\tiny cc}} =
-\mu_{\mbox{\tiny M}}\circ \mu_{\mbox{\tiny D}}\,.
\ee
Therefore,
\bb
a\,{\rm tr}_{\!\mbox{\tiny$\ep$}}\mu_{\mbox{\tiny YM}}^4
+ {\rm tr}_{\!\mbox{\tiny$\ep$}}\mu_{\mbox{\tiny YM}}^2
&=&
4{\rm Re}[a\,{\rm tr}_{\!\mbox{\tiny$\ep$}}\varphi_{\mbox{\tiny D}}^4 -
\,{\rm tr}_{\!\mbox{\tiny$\ep$}}\varphi_{\mbox{\tiny D}}^2 +
a\,{\rm tr}_{\!\mbox{\tiny$\ep$}}m_{\mbox{\tiny M}}^4 -
\,{\rm tr}_{\!\mbox{\tiny$\ep$}}m_{\mbox{\tiny M}}^2
\nonumber\\[.1cm]
&&
-\;
2a\,{\rm tr}_{\!\mbox{\tiny$\ep$}}(\varphi_{\mbox{\tiny D}}\circ m_{\mbox{\tiny M}})^2]\,,
\ee
where the occurrence of the Higgs potentials of $\varphi_{\mbox{\tiny D}}$ and
$m_{\mbox{\tiny M}}$ are due to the fact that the Dirac-Yukawa-Majorana operator
$\,\DDD_{\mbox{\tiny YM}} = \,\ddd_{\!\!\!\mbox{\tiny${\cal A}$}} + i\mu_{\mbox{\tiny YM}}\in
{\cal D}_{\!\mbox{\tiny real}}(\ep)$ is of simple type.

Note that also ${\rm tr}_g(F_{\!\!\!\mbox{\tiny${\cal A}$}}^2) =
4{\rm Re}\,{\rm tr}_g(F_{\!\!\mbox{\tiny A}}^2)$. Hence, the total Dirac action with
respect to the real Dirac operator $\,{/\!\!\!\!P}_{\!\!\mbox{\tiny DYM}}\in{\cal D}(\ppp)$
is a real-valued functional, actually. This is independent of whether the section $\varphi_{\!\mbox{\tiny$e$}}\in{\goth Sec}(M,{\rm End}_\gamma(\www_{\!\mbox{\tiny$e$}}))$
and the simple type Dirac operator $\dda\in{\cal D}(\www)$ are supposed to be Hermitian
or anti-Hermitian.

\section{Real Clifford bi-modules and the ``$\pi_{\mbox{\tiny D}}-$map''}
The Pauli map (\ref{pauli-map}) is defined for general real Clifford modules. In the previous
section we demonstrated how the Pauli map of the simple type Dirac operator defined in terms
of a Yang-Mills-Higgs connection on a Majorana module encodes the full STM action functional.

In this section we discuss once again the STM action in view of the Dirac operators (\ref{pauli type dop}), this time, however, in the case where the underlying Majorana module is supposed to have the structure of a Clifford bi-module. The discussion of the previous section exhibited the importance of simple type Dirac operators. However, the Pauli map does not preserve the structure of simple type Dirac operators. Although the Yukawa-coupling term  and the Pauli term are geometrically treated almost in the same manner, there is yet a basic asymmetry between these two terms. Basically, this is because
$i\mu_{\mbox{\tiny YM}}\in{\goth Sec}(M,{\rm End}^-(\ppp))$
anti-commutes with the Clifford action in contrast to
$\iota\,{/\!\!\!\!{\cal F}}_{\!\!\mbox{\tiny D}}\in{\goth Sec}(M,{\rm End}^-(\ppp))$. This apparent asymmetry, however, may be easily overcome in the case where the underlying Majorana module is (embedded into) a Clifford bi-module. This will yield a straightforward geometrical interpretation of the STM action in terms of the Einstein-Hilbert action including a ``cosmological constant'' term.
\begin{definition}
Let $(\ep,\langle\cdot,\cdot\rangle_{\!\mbox{\tiny$\ep$}},\tau_{\!\mbox{\tiny$\ep$}},
\gamma_{\mbox{\tiny$\ep$}},\gamma_{\mbox{\tiny$\ep$,op}},J_{\mbox{\tiny$\ep$}})$ be a real Clifford bi-module over $(M,g_{\mbox{\tiny M}})$. The mapping
\bb
\label{pi-map}
\pi_{\mbox{\tiny D}}:\,{\cal D}_{\!\mbox{\tiny real}}(\ep)&\longrightarrow&
{\cal D}_{\!\mbox{\tiny real}}(\ppp)\nonumber\\[.1cm]
\DDD &\mapsto& {/\!\!\!\!P}_{\!\!\mbox{\tiny D}} := \,\DDD + i{\cal F}_{\!\!\mbox{\tiny D}}
\ee
is called the ``$\pi_{\mbox{\tiny D}}-$map''. Here,
\bb
{\cal F}_{\!\!\mbox{\tiny D}} &:=& \left(
\begin{array}{cc}
0 & -\tau_{\mbox{\tiny$\ep$}}\circ
\,{/\!\!\!\!{\cal F}}_{\!\!\mbox{\tiny D}} \\
\tau_{\mbox{\tiny$\ep$}}\circ
\,{/\!\!\!\!{\cal F}}_{\!\!\mbox{\tiny D}} & 0 \\
\end{array}
\right)\nonumber\\[.1cm]
&\equiv&
-\tau_{\mbox{\tiny$\ppp$}}\circ\,{/\!\!\!\!{\cal F}}_{\!\!\mbox{\tiny D}}\in
{\rm Sec}(M,{\rm End}^-(\ep))\,,\\[.1cm]
{/\!\!\!\!{\cal F}}_{\!\!\mbox{\tiny D}} &:=&
\,{/\!\!\!\!F}_{\!\!\mbox{\tiny D,op}}\ot\varepsilon_{\!\mbox{\tiny 2}}\in
{\rm Sec}(M,{\rm End}^-_\gamma(\ep))
\ee
and
\bb
{/\!\!\!\!F}_{\!\!\mbox{\tiny D,op}}\in{\rm Sec}(M,{\rm End}^+_\gamma(\ep))
\ee
is the relative curvature of $\,\DDD_{\!\!\mbox{\tiny op}}$, quantized with respect to $\gamma_{\mbox{\tiny$\ep$,op}}$.
\end{definition}

Apparently, the real Pauli-like Dirac operator that is defined by (\ref{pi-map})
is most analogous to the real, simple type Dirac operator (\ref{Dirac-Yukawa op}). In particular, if $\,\DDD\in\dep$ is of simple type, then so is its associated Pauli-like operator $\,{/\!\!\!\!P}_{\!\!\mbox{\tiny D}}\in{\cal D}(\ppp = \!\!\!\!\phantom{\ep}^{\mbox{\tiny 2}}\ep)$. In other words: In contrast to the Pauli map
(\ref{pauli-map}), the map (\ref{pi-map}) preserves the distinguished structure of simple type Dirac operators. According to its definition, however, the $\pi_{\mbox{\tiny D}}-$map does not preserve $S-$reducibility, as opposed to the Pauli map ${\cal P}_{\!\mbox{\tiny D}}$.

Starting again with a Yang-Mills-Higgs connection
$\partial_{\mbox{\tiny YMH}}\in{\cal A}(\www)$ on the Majorana module
$\www\twoheadrightarrow M$, we may consider the real Dirac operator of simply type:
\bb
\label{pauli-like DYM op}
{/\!\!\!\!P}_{\!\!\mbox{\tiny DYM}} &:=&
\pi_{\mbox{\tiny D}}(\ddd_{\!\!\!\mbox{\tiny${\cal A}$}} + i\mu_{\mbox{\tiny YM}})\cr
&\equiv&
\ddd_{\!\!\!\mbox{\tiny${\cal A}$}} +
i(\mu_{\mbox{\tiny YM}} + {\cal F}_{\!\!\mbox{\tiny DYM}})\,,
\ee
with ${\cal F}_{\!\!\mbox{\tiny DYM}} := -\tau_{\mbox{\tiny$\ppp$}}\circ
\left(\,{/\!\!\!\!F}_{\!\!\mbox{\tiny DYM,op}}\ot\varepsilon_{\!\mbox{\tiny 2}}\right)$
being defined by the ($\gamma_{\mbox{\tiny$\ep$,op}}-$quantized) relative curvature of $\,\DDD_{\!\mbox{\tiny YM,op}} = \,\ddd_{\!\!\!\mbox{\tiny${\cal A}$,op}} +
i\mu_{\mbox{\tiny YM}}$. Notice that
$\,\DDD_{\!\mbox{\tiny YM,op}}\notin{\cal D}_{\mbox{\tiny S}}(\ep)$, in contrast to $\,\DDD_{\!\mbox{\tiny YM}}.$ However, the $\,\DDD_{\!\mbox{\tiny YM}}-$ induced
Pauli-like curvature term $i{\cal F}_{\!\!\mbox{\tiny DYM}}$ does not contribute to the fermionic part of the total Dirac action. Instead, the latter is fully determined by the Dirac connection
\bb
\partial_{\mbox{\tiny DYM}} &:=& \partial_{\!\!\mbox{\tiny${\cal A}$}} +
iext_\Theta(\mu_{\mbox{\tiny YM}})\cr
&\equiv&
\partial_{\mbox{\tiny${\cal YMH}$}} + iext_\Theta(\mu_{\mbox{\tiny M}})
\ee
of the simple type Dirac operator
$\,\DDD_{\!\mbox{\tiny YM}} = \delta_\gamma\circ\partial_{\mbox{\tiny DYM}}\in
{\cal D}_{\mbox{\tiny S,real}}(\ep)$. Here, $\partial_{\mbox{\tiny${\cal YMH}$}}\in
{\cal A}_{\mbox{\tiny S}}(\ep)$
denotes the real form of the Dirac connection of the simple type Dirac operator (\ref{Dirac-Yukawa op}) on the Dirac module $\sss\twoheadrightarrow M$ that is induced
by the Yang-Mills-Higgs connection $\partial_{\mbox{\tiny YMH}}\in
{\cal A}_{\mbox{\tiny S}}(\www)$ on the underlying Majorana module
$\www\twoheadrightarrow M$.

Since the real Dirac operator $\,{/\!\!\!\!P}_{\!\!\mbox{\tiny DYM}}$ is of simple type
it becomes straightforward to express the Dirac action
\bb
{\cal I}_{\mbox{\tiny DYM}} := \int_M\ast
{\rm tr}_\gamma\!\left(curv(\,{/\!\!\!\!P}_{\!\!\mbox{\tiny DYM}}) -
\varepsilon\,ev_g(\omega^2_{\mbox{\tiny D}})\right)
\ee
in terms of the sections parameterizing
$\,\DDD_{\!\mbox{\tiny YM}}\in{\cal D}_{\mbox{\tiny S,real}}(\ep)$.

First, we mention that the Dirac vector field $\xi_{\mbox{\tiny D}}\in{\goth Sec}(M,TM)$
of any simple type Dirac operator vanishes identically. This holds true for arbitrary Clifford modules $(\ep,\gamma_{\!\mbox{\tiny$\ep$}})\twoheadrightarrow(M,g_{\mbox{\tiny M}})$. Therefore,
\bb
{\rm tr}_{\mbox{\tiny$\ep$}}\!\left(\,\DDD^2 - \triangle_{\mbox{\tiny B}}\right) =
{\rm tr}_\gamma\!\left(curv(\,\DDD) - \varepsilon\,ev_g(\omega^2_{\mbox{\tiny D}})\right)
\ee
does not hold true only up to boundary terms but is an identity for simple type Dirac operators $\,\DDD\in\dep$.

Similar to the last section, we put $\Phi_{\mbox{\tiny DYM}} :=
i(\mu_{\mbox{\tiny YM}} + {\cal F}_{\!\!\mbox{\tiny DYM}})\in
{\goth Sec}(M,{\rm End}^-(\ppp))$ and apply once more Lemma \ref{lemma lichnerowicz formula}. This time, however, we may take advantage of $\{\gamma_{\!\mbox{\tiny$\ppp$}}(\alpha),\Phi_{\mbox{\tiny DYM}}\} \equiv 0$, for all
$\alpha\in T^*\!M$. Consequently, $\Theta\wedge\Phi_{\mbox{\tiny DYM}} = -
\Phi_{\mbox{\tiny DYM}}\wedge\Theta$ and thus
\bb
ev_g(\omega^2_{\mbox{\tiny D}}) &=&
-g_{\mbox{\tiny M}}(e^i,e^j)\,\Theta(e_i)\circ\Theta(e_j)\circ\Phi_{\mbox{\tiny DYM}}^2\cr
&=&
-\mbox{\small$\frac{\varepsilon^2}{n^2}$}\,g_{\mbox{\tiny M}}(e_i,e_j)\,
\gamma_{\!\mbox{\tiny$\ppp$}}(e^i)\circ\gamma_{\!\mbox{\tiny$\ppp$}}(e^j)\circ
\Phi_{\mbox{\tiny DYM}}^2\cr
&=&
-\mbox{\small$\frac{\varepsilon}{n}$}\,\Phi_{\mbox{\tiny DYM}}^2\,.
\ee

Furthermore,
\bb
{\cal I}_{\mbox{\tiny DYM}} &=&
\int_M\left[{\rm tr}_\gamma\!\left(curv(\,\ddd_{\!\!\!\mbox{\tiny${\cal A}$}}) -
(d_{\!\!\mbox{\tiny${\cal A}$}}\Phi_{\mbox{\tiny DYM}} +
\Phi_{\mbox{\tiny DYM}}^2\Theta)\wedge\Theta\right) +
\mbox{\small$\frac{1}{n}$}\,
{\rm tr}_{\mbox{\tiny$\ppp$}}\Phi_{\mbox{\tiny DYM}}^2\right]dvol_{\mbox{\tiny M}}
\nonumber\\[.1cm]
&=&
\int_M\left[{\rm tr}_\gamma curv(\,\ddd_{\!\!\!\mbox{\tiny${\cal A}$}}) +
{\rm tr}_{\mbox{\tiny$\ppp$}}\Phi_{\mbox{\tiny DYM}}^2\right]dvol_{\mbox{\tiny M}}\,.
\ee

The Dirac action with respect to $\,{/\!\!\!\!P}_{\!\!\mbox{\tiny DYM}} =
\pi_{\mbox{\tiny D}}(\ddd_{\!\!\!\mbox{\tiny${\cal A}$}} + i\mu_{\mbox{\tiny YM}})\in
{\cal D}_{\!\mbox{\tiny real}}(\ppp)$ thus dynamically generalizes the
Einstein-Hilbert action (\ref{EH-action plus cosm. const.} --
\ref{cosmological const. vers. Higgs}) with the cosmological constant induced by the Yang-Mills-Higgs connection, whose quantization (together with the Majorana masses)
defines the fermionic action\footnote{It makes no sense to take the Majorana masses into
account directly on the Majorana module $\www\twoheadrightarrow M$. Indeed, for this one has to make use of the induced Dirac module.}. This time, however, the ``cosmological constant''
does depend on the metric as opposed to (\ref{cosmological const. vers. Higgs}). Indeed, according to the explicit form of $\Phi_{\mbox{\tiny DYM}}$, it follows that
\bb
{\rm tr}_{\mbox{\tiny$\ppp$}}\Phi_{\mbox{\tiny DYM}}^2 &=&
2\,{\rm tr}_{\!\mbox{\tiny$\ep$}}\!\left(\,{/\!\!\!\!F}_{\!\!\mbox{\tiny DYM,op}}^2 - \mu_{\mbox{\tiny YM}}^2\right)\nonumber\\[.1cm]
&=&
-{\rm tr}_g F_{\!\!\!\mbox{\tiny${\cal A}$}}^2 +
2\varepsilon\mbox{\small$(\frac{n - 1}{n})^2$}\,
{\rm tr}_g(\partial_{\!\!\mbox{\tiny${\cal A}$}}\mu_{\mbox{\tiny YM}})^2 +
2\mbox{\small$(\frac{n - 1}{n})^2$}\,
{\rm tr}_{\!\mbox{\tiny$\ep$}}\mu_{\mbox{\tiny YM}}^4 -
2{\rm tr}_{\!\mbox{\tiny$\ep$}}\mu_{\mbox{\tiny YM}}^2\,.
\ee
For $\varepsilon := -1$ and Hermitian $\mu_{\mbox{\tiny YM}}$ the ``cosmological constant term'' has the form of the usual Lagrangian of the Standard Model, such that
${\cal I}_{\mbox{\tiny DYM}}$, again, takes the form of the combined Einstein-Hilbert-Yang-Mills-Higgs
action\footnote{For $\varepsilon := +1$ one gets the corresponding Lagrangian with respect to the Euclidean signature of $g_{\mbox{\tiny M}}$.}. This functional basically
coincides with what has been derived from the Pauli-Dirac operator
${\cal P}_{\!\mbox{\tiny D}}(\ddd_{\!\!\!\mbox{\tiny${\cal A}$}} +
i\mu_{\mbox{\tiny YM}})\in{\cal D}_{\mbox{\tiny S,real}}(\ppp)$ in the previous section.
Note, however, that there is a difference concerning the conditions imposed on
$(\varepsilon,\mu_{\mbox{\tiny YM}})$. Also note that there is a significant difference between ${\cal P}_{\!\mbox{\tiny D}}$ and $\pi_{\mbox{\tiny D}}$ in dimension two.

\section{Conclusion}
In this article we discussed the geometrical structure of Pauli-type Dirac operators which encode the STM action
including gravity. This has been done by carefully analyzing the corresponding structure of the Dirac equation and
the Majorana equation in terms of real Clifford (bi-)modules and Dirac operators of simple type. It has been shown
how the geometrical frame presented allows to overcome the issue of ``fermion doubling'' and how the combined
Einstein-Hilbert-Yang-Mills-Higgs (EHYMH-) action can be derived from the distinguished class of real Dirac
operators of simple type. The latter description allows to geometrically recast the EHYMH-action into a form
which formally looks identical to the Einstein-Hilbert action with a cosmological constant. On this basis, we
have demonstrated how Majorana masses are naturally included within the geometrical frame of Dirac type gauge
theories and how they dynamically contribute to the combined EHYMH-action in terms of a peculiar cosmological
constant. This cosmological constant may have interesting phenomenological consequences with respect to dark
matter/energy and the mass of the Higgs boson to be discussed in a forthcoming work.

\vspace{.5cm}

\noindent
{\bf Acknowledgments}\\
The author would like to thank E. Binz and P. Guha for their continuous
interest and stimulating discussions on the presented subject. Especially, the author is
very grateful to J. Jost and W. Spr\"o{\ss}ig for the possibility to perform this work in
an outstandingly stimulating atmosphere within the respective scientific groups.\\

\vspace{.1cm}

\section*{Appendix}
In this appendix, we briefly introduce a specific class of Majorana modules. The
latter will be appropriate to geometrically describe the (minimal) Standard Model in
terms of Dirac type gauge theories when also Majorana masses are taken into account.
The inclusion of massive Dirac neutrinos within the (minimal) Standard Model has been
discussed already in \cite{ToTh:06}.

Let $M$ be an orientable, connected and simply connected four-dimensional spin manifold admitting a Lorentzian structure. For each Lorentz metric
$g_{\mbox{\tiny M}}\in{\goth Sec}(M,\ep_{\mbox{\tiny EH}})$, the corresponding Lorentz manifold $(M,g_{\mbox{\tiny M}})$ is also supposed to be time orientable. For every choice
of a spin-structure let
\bb
\www_{\!\mbox{\tiny$e$}}:=S\ot_\cc E\twoheadrightarrow M
\ee
be a twisted spinor bundle. The Hermitian vector bundle
$E = E_{\mbox{\tiny R}}\op E_{\mbox{\tiny L}}\twoheadrightarrow M$ is assumed to be
associated with a $G-$principal bundle
$G\hookrightarrow{\cal P}_{\!\mbox{\tiny G}}\twoheadrightarrow M$.
In the case of the Standard Model: $G := SU(3)\times SU(2)\times U(1)$. According to Geroch's Theorem, the frame bundle ${\cal F}_{\!\mbox{\tiny M}}\twoheadrightarrow M$ of $M$ is trivial, provided $M$ is open (c.f. \cite{Ger:68} and \cite{Ger:70}). Also, the topological structure of the electroweak gauge sub-bundle:
${\cal P}_{\!\mbox{\tiny${\rm SU(2)}\!\times\! {\rm U(1)}$}}\hookrightarrow
{\cal P}_{\!\mbox{\tiny SU(3)}}\times_M
{\cal P}_{\!\mbox{\tiny${\rm SU(2)}\!\times\! {\rm U(1)}$}}\twoheadrightarrow M$, is fully determined by the moduli space of ground states of the Higgs boson. In particular, the electroweak gauge bundle is trivial, provided the electrically charged weak vector bosons $W^\pm$ are considered as being charge conjugate to each other (c.f. \cite{Tol:05}). In the case of the Standard Model, the Hermitian vector bundle $E\twoheadrightarrow M$ is defined by the fermionic representation of $G$ (c.f., for example, Sec. 3.1 in \cite{ToTh:06}).

We put, respectively, for the uncharged and charged sector of the Majorana module
$\www = \www_{\!\mbox{\tiny$\nu$}}\op\www_{\!\mbox{\tiny$e$}}\twoheadrightarrow M$:
\bb
\www_{\!\mbox{\tiny$\nu$}} :=
\begin{array}{c}
\www_{\!\mbox{\tiny$\nu$,R}} \equiv \www_{\!\mbox{\tiny$\nu$,RR}}\op\www_{\!\mbox{\tiny$\nu$,RL}}\\
\op\\
\www_{\!\mbox{\tiny$\nu$,L}} \equiv  \www_{\!\mbox{\tiny$\nu$,LR}}\op\www_{\!\mbox{\tiny$\nu$,LL}}\,,\\
\end{array}\,,\qquad
\www_{\!\mbox{\tiny$e$}} :=
\begin{array}{c}
\www_{\!\mbox{\tiny$e$,R}} \equiv \www_{\!\mbox{\tiny$e$,RR}}\op\www_{\!\mbox{\tiny$e$,RL}}\\
\op\\
\www_{\!\mbox{\tiny$e$,L}} \equiv  \www_{\!\mbox{\tiny$e$,LR}}\op\www_{\!\mbox{\tiny$e$,LL}}\\
\end{array}\,.
\ee
Here,
\bb
\www_{\!\mbox{\tiny$\nu$,RR}} &:=& S_{\mbox{\tiny R}}\ot V_{\mbox{\tiny R}}\,,\qquad
\www_{\!\mbox{\tiny$\nu$,LL}} := S_{\mbox{\tiny L}}\ot V_{\mbox{\tiny L}}\,,\cr
\www_{\!\mbox{\tiny$\nu$,RL}} &:=& S_{\mbox{\tiny R}}\ot V_{\mbox{\tiny L}}\,,\qquad
\www_{\!\mbox{\tiny$\nu$,LR}} := S_{\mbox{\tiny L}}\ot V_{\mbox{\tiny R}}\,,\\[.1cm]
\www_{\!\mbox{\tiny$e$,RR}} &:=& S_{\mbox{\tiny R}}\ot E_{\mbox{\tiny R}}\,,\qquad
\www_{\!\mbox{\tiny$e$,LL}} := S_{\mbox{\tiny L}}\ot E_{\mbox{\tiny L}}\,,\cr
\www_{\!\mbox{\tiny$e$,RL}} &:=& S_{\mbox{\tiny R}}\ot E_{\mbox{\tiny L}}\,,\qquad
\www_{\!\mbox{\tiny$e$,LR}} := S_{\mbox{\tiny L}}\ot E_{\mbox{\tiny R}}\,.
\ee
The Hermitian vector space $V = V_{\mbox{\tiny R}}\op V_{\mbox{\tiny L}}$ carries the trivial
representation of $G$. Its dimension may be arbitrarily chosen.

Note that
\bb
\www \simeq S\ot_\cc(V\op E)\,.
\ee

Let, respectively, $\tau_{\mbox{\tiny V}}$ and $\tau_{\mbox{\tiny E}}$ be the corresponding grading involutions of $M\times V_{\mbox{\tiny R}}\op V_{\mbox{\tiny L}}\twoheadrightarrow M$ and $E = E_{\mbox{\tiny R}}\op E_{\mbox{\tiny L}}\twoheadrightarrow M$. According to the above decomposition, the grading involution reads:
\bb
\tau_{\!\mbox{\tiny$\www$}} :=
\left(\!\!
  \begin{array}{cc}
    \tau_{\!\mbox{\tiny$\www_{\mbox{\tiny$\nu$}}$}} & 0 \\
    0 & \tau_{\!\mbox{\tiny$\www_{\mbox{\tiny$e$}}$}} \\
  \end{array}
\!\!\right)\,,
\ee
whereby
\bb
\tau_{\!\mbox{\tiny$\www_{\mbox{\tiny$\nu$}}$}} :=
\left(\!\!
  \begin{array}{cc}
    \tau_{\mbox{\tiny V}} & 0 \\
    0 & -\tau_{\mbox{\tiny V}} \\
  \end{array}
\!\!\right)\,,\qquad
\tau_{\!\mbox{\tiny$\www_{\mbox{\tiny$e$}}$}} :=
\left(\!\!
  \begin{array}{cc}
    \tau_{\mbox{\tiny E}} & 0 \\
    0 & -\tau_{\mbox{\tiny E}} \\
  \end{array}
\!\!\right)\,.
\ee

By abuse of notation, we do not distinguish between $\tau_{\mbox{\tiny V}}$ and
${\rm id}_{\mbox{\tiny S}}\!\ot\!\tau_{\mbox{\tiny V}}$. Likewise, $\tau_{\mbox{\tiny E}}$ is
identified with ${\rm id}_{\mbox{\tiny S}}\!\ot\!\tau_{\mbox{\tiny E}}$.

It follows that, for example,
\bb
\www_{\!\mbox{\tiny$\nu$,LR}} &=&
\{\nu\in\www_{\!\mbox{\tiny$\nu$}}\,|\,
\tau_{\mbox{\tiny M}}\nu = -\nu\,,\;\tau_{\mbox{\tiny V}}\nu = +\nu\}
\subset\www_{\!\mbox{\tiny$\nu$}}\,,\\[.1cm]
\www_{\!\mbox{\tiny$e$,RL}} &=&
\{e\in\www_{\!\mbox{\tiny$e$}}\,|\,
\tau_{\mbox{\tiny M}}e = +e\,,\;\tau_{\mbox{\tiny E}}e = -e\}
\subset\www_{\!\mbox{\tiny$e$}}\,,\quad\mbox{etc.}
\ee

In particular, the (total spaces of the) respective eigen bundles of $\tau_{\!\mbox{\tiny$\www_{\mbox{\tiny$\nu$}}$}}$ and $\tau_{\!\mbox{\tiny$\www_{\mbox{\tiny$e$}}$}}$ read:
\bb
\www_{\!\mbox{\tiny$\nu$}} &\simeq& \www_{\!\mbox{\tiny$\nu$}}^+\op\www_{\!\mbox{\tiny$\nu$}}^-\,,
\nonumber\\[.1cm]
\www_{\!\mbox{\tiny$\nu$}}^+ &:=& \www_{\!\mbox{\tiny$\nu$,RR}}\op\www_{\!\mbox{\tiny$\nu$,LL}}\,,\cr
\www_{\!\mbox{\tiny$\nu$}}^- &:=& \www_{\!\mbox{\tiny$\nu$,RL}}\op\www_{\!\mbox{\tiny$\nu$,LR}}\,;
\\[.2cm]
\www_{\!\mbox{\tiny$e$}} &\simeq& \www_{\!\mbox{\tiny$e$}}^+\op\www_{\!\mbox{\tiny$e$}}^-\,,
\nonumber\\[.1cm]
\www_{\!\mbox{\tiny$e$}}^+ &:=& \www_{\!\mbox{\tiny$e$,RR}}\op\www_{\!\mbox{\tiny$e$,LL}}\,,\cr
\www_{\!\mbox{\tiny$e$}}^- &:=& \www_{\!\mbox{\tiny$e$,RL}}\op\www_{\!\mbox{\tiny$e$,LR}}\,.
\ee

The Clifford action is defined in terms of the Clifford mapping:
\bb
\gamma_{\!\mbox{\tiny$\www$}} :=
\left(\!\!
\begin{array}{cc}
 \gamma_{\!\mbox{\tiny$\www_{\mbox{\tiny$\nu$}}$}} & 0 \\
 0 & \gamma_{\!\mbox{\tiny$\www_{\mbox{\tiny$e$}}$}} \\
\end{array}
\!\!\right)\,,\qquad
\gamma_{\!\mbox{\tiny$\www_{\mbox{\tiny$\nu$}}$}} = \gamma_{\!\mbox{\tiny$\www_{\mbox{\tiny$e$}}$}} :=
\left(\!\!
\begin{array}{cc}
 0 & \gamma_{\mbox{\tiny Ch}} \\
 \gamma_{\mbox{\tiny Ch}} & 0 \\
\end{array}
\!\!\right)\,,
\ee
whereby the Clifford mapping $\gamma_{\mbox{\tiny Ch}}:\,T^*\!M\rightarrow{\rm End}_\cc(S)$ acts trivially on the sub-bundles
$E_{\mbox{\tiny R}},\,E_{\mbox{\tiny L}}\subset E\twoheadrightarrow M$ and on the
Hermitian vector spaces $V_{\mbox{\tiny R}},\,V_{\mbox{\tiny L}}$.

The real structure is defined by
\bb
J_{\!\mbox{\tiny$\www$}} :=
\left(\!\!
\begin{array}{cc}
 J_{\!\mbox{\tiny$\www_{\mbox{\tiny$\nu$}}$}} & 0 \\
 0 & J_{\!\mbox{\tiny$\www_{\mbox{\tiny$e$}}$}} \\
\end{array}
\!\!\right)\,,\qquad
J_{\!\mbox{\tiny$\www_{\mbox{\tiny$\nu$}}$}} =
J_{\!\mbox{\tiny$\www_{\mbox{\tiny$e$}}$}} :=
\left(\!\!
\begin{array}{cc}
 0 & J \\
 J & 0 \\
\end{array}
\!\!\right)\,,
\ee
whereby $J\circ\gamma_{\mbox{\tiny Ch}}(\alpha) = -\gamma_{\mbox{\tiny Ch}}(\alpha)\circ J$ and $J\circ\tau_{\mbox{\tiny M}} = -\tau_{\mbox{\tiny M}}\circ J$ is assumed to hold for all $\alpha\in T^*\!M$.

The Dirac type operator $\dda + i\varphi_{\mbox{\tiny D}}$ decomposes as follows:
\begin{itemize}
\item On the uncharged sector: $m_{\mbox{\tiny D,$\nu$}}\in
{\goth Sec}(M,{\rm End}_\gamma(\www_{\!\mbox{\tiny$\nu$}}))\;
(\mbox{real and constant})$
\bb
\ddd + im_{\mbox{\tiny D,$\nu$}} \equiv
\left(\!\!
\begin{array}{cc}
im_{\mbox{\tiny D,$\nu$}} & \ddd \\
\ddd & im_{\mbox{\tiny D,$\nu$}} \\
\end{array}
\!\!\right)\,;
\ee
\item On the charged sector:
\bb
\dda + i\varphi_{\!\mbox{\tiny$e$}} \equiv
\left(\!\!
\begin{array}{cc}
i\varphi_{\!\mbox{\tiny$e$}} & \dda \\
\dda & i\varphi_{\!\mbox{\tiny$e$}} \\
\end{array}
\!\!\right) \,,
\ee
where either $\varphi_{\!\mbox{\tiny$e$}}\in
{\goth Sec}(M,{\rm End}^-_\gamma(\www_{\!\mbox{\tiny $e$}}))$, or $\varphi_{\!\mbox{\tiny$e$}}\in
{\goth Sec}(M,{\rm End}^+_\gamma(\www_{\!\mbox{\tiny $e$}}))$.
\end{itemize}
As mentioned already, the latter case holds true only for left-right symmetric gauge theories. It does not hold true in the usual (minimal) Standard Model where parity is maximally violated.

The block matrix notation used for the Dirac type operators refers to the embedding
\bb
\www_{\!\mbox{\tiny$e$,R}}&\hookrightarrow&\www_{\!\mbox{\tiny$e$}}\nonumber\\[.1cm]
e_{\mbox{\tiny R}}&\mapsto&
\mbox{\small$\left(\!\!
\begin{array}{c}
e_{\mbox{\tiny R}} \\
0 \\
\end{array}
\!\!\right)$}\,,
\ee
where, for example, on the left-hand side
$e_{\mbox{\tiny RR}}\in\{e\in\www_{\!\mbox{\tiny$e$}}\,|\,
\tau_{\mbox{\tiny M}}e = +e\,,\;\tau_{\mbox{\tiny E}}e = +e\} = \www_{\!\mbox{\tiny$e$,RR}}\subset\www_{\!\mbox{\tiny$e$}}$, etc.

Accordingly, the Majorana mass matrix $m_{\mbox{\tiny M,$\nu$}}\in
{\goth Sec}(M,{\rm End}^+_\gamma(\www_{\!\mbox{\tiny$\nu$}}))$ (real and constant) reads:
\bb
m_{\mbox{\tiny M,$\nu$}}\equiv
\left(\!\!
\begin{array}{cc}
 m_{\mbox{\tiny M,$\nu$}} & 0  \\
 0  & m_{\mbox{\tiny M,$\nu$}} \\
\end{array}
\!\!\right) \,,
\ee
such that the combined Dirac-Yukawa-Majorana equations (\ref{diracmajomasseq}--\ref{diracmasseq}) explicitly takes the form:
\bb
i\ddd\nu &=& m_{\mbox{\tiny D,$\nu$}}\nu + m_{\mbox{\tiny M,$\nu$}}\nu^{\mbox{\tiny cc}}
\quad\Leftrightarrow\qquad\left\{
\begin{array}{ccc}
  i\ddd\nu_{\mbox{\tiny R}} & = & m_{\mbox{\tiny D,$\nu$}}\nu_{\mbox{\tiny L}} +
    m_{\mbox{\tiny M,$\nu$}}\nu_{\mbox{\tiny R}}^{\mbox{\tiny cc}}\,, \\[.1cm]
  i\ddd\nu_{\mbox{\tiny L}} & = & m_{\mbox{\tiny D,$\nu$}}\nu_{\mbox{\tiny R}} +
    m_{\mbox{\tiny M,$\nu$}}\nu_{\mbox{\tiny L}}^{\mbox{\tiny cc}} \,,
\end{array}
\right.\\[.5cm]
i\dda e & = & \varphi_{\!\mbox{\tiny$e$}}e
\hspace{2.6cm}\Leftrightarrow\qquad\left\{
\begin{array}{ccc}
  i\dda e_{\mbox{\tiny R}} & = & \varphi_{\!\mbox{\tiny$e$}}e_{\mbox{\tiny L}}\,, \\[.1cm]
  i\dda e_{\mbox{\tiny L}} & = & \varphi_{\!\mbox{\tiny$e$}}e_{\mbox{\tiny R}}\,.
\end{array}
\right.
\ee

In the case of the Standard Model a ``charged state'' is geometrically represented by a section of $\www_{\!\mbox{\tiny$e$}}^+\subset\www_{\!\mbox{\tiny$e$}}\twoheadrightarrow M$:
\bb
e = \left(\!\!
      \begin{array}{c}
        e_{\mbox{\tiny RR}} \\
        e_{\mbox{\tiny LL}} \\
      \end{array}
    \!\!\right)\in{\goth Sec}(M,\www^+_{\!\mbox{\tiny$e$}})\,.
\ee
Whence,
\bb
i\dda e & = & \varphi_{\!\mbox{\tiny$e$}}e
\qquad\Leftrightarrow\qquad
\left\{
\begin{array}{ccc}
  i\dda e_{\mbox{\tiny RR}} & = & \varphi_{\!\mbox{\tiny$e$,RL}}e_{\mbox{\tiny LL}}\,, \\[.1cm]
  i\dda e_{\mbox{\tiny LL}} & = & \varphi_{\!\mbox{\tiny$e$,LR}}e_{\mbox{\tiny RR}}\,,
\end{array}
\right.
\ee
where
\bb
\varphi_{\!\mbox{\tiny$e$,RL}}&\equiv&
\pi_{\mbox{\tiny E,R}}\circ\varphi_{\!\mbox{\tiny$e$}}\circ\pi_{\mbox{\tiny E,L}}\in
{\goth Sec}(M,{\rm Hom}_\gamma(\www_{\!\mbox{\tiny$e$,L}},\www_{\!\mbox{\tiny$e$,R}}))\,,\cr
\varphi_{\!\mbox{\tiny$e$,LR}}&\equiv&
\pi_{\mbox{\tiny E,L}}\circ\varphi_{\!\mbox{\tiny$e$}}\circ\pi_{\mbox{\tiny E,R}}\in
{\goth Sec}(M,{\rm Hom}_\gamma(\www_{\!\mbox{\tiny$e$,R}},\www_{\!\mbox{\tiny$e$,L}}))\,,
\ee
and $\pi_{\mbox{\tiny E,R/L}} := ({\rm id}_{\mbox{\tiny E}} \pm \tau_{\mbox{\tiny E}})/2$ are
the complementary idempotents with respect to the $\zz_2-$grading
$E = E_{\mbox{\tiny R}}\!\op\! E_{\mbox{\tiny L}}\twoheadrightarrow M$.

Finally, in the case of the (minimal) Standard Model, the section
$\varphi_{\!\mbox{\tiny$e$}}\in{\goth Sec}(M,{\rm End}^-_\gamma(\www_{\!\mbox{\tiny e}}))$ is related to the usual Higgs field via the ``Yukawa mapping'':
\bb
\varphi_{\!\mbox{\tiny$e$,LR }}&:=& G_{\!\mbox{\tiny Y}}(\varphi)\equiv
\left(\!\!
  \begin{array}{cc}
    ({\mbox{\bf g}'}^{\mbox{\tiny q}}\ot\varphi,-{\mbox{\bf g}}^{\mbox{\tiny q}}
    \ot{\rm I}_{\mbox{\tiny 2}}\ot\varphi^{\mbox{\tiny cc}}) & 0 \\
    0 &  {\mbox{\bf g}}^{\mbox{\tiny l}}\ot\varphi\\
  \end{array}
\!\!\right)\,,\\[.1cm]
\varphi_{\!\mbox{\tiny$e$,RL }} &:=& \varphi_{\!\mbox{\tiny$e$,LR }}^\dagger\,.
\ee
Here, respectively, ${\mbox{\bf g}'}^{\mbox{\tiny q}},\,
{\mbox{\bf g}}^{\mbox{\tiny q}}\in\cc(N)$ and ${\mbox{\bf g}}^{\mbox{\tiny l}}\in\cc(N)$ are the matrices of the ``Yukawa coupling constants'' of the quarks of electrical charge $-1/3$ and $+2/3$ and the leptons of electrical charge equal to $-1$. The section
$\varphi\in{\goth Sec}(M,E_{\mbox{\tiny H}})$
geometrically describes the (semi-classical state of the) Higgs field. According to the minimal Standard Model the Higgs boson carries a rank two sub-representation
$E_{\mbox{\tiny H}}\hookrightarrow E\twoheadrightarrow M$ of the fermionic representation $E\twoheadrightarrow M$. This sub-representation is fixed by the ``hyper-charge relations'' between the hyper-charges carried by the quarks and leptons (see again, for example, Sec. 3.1 in loc. site; for a geometrical discussion of the Yukawa mapping:
$G_{\!\mbox{\tiny Y}}:\,E_{\mbox{\tiny H}}\hookrightarrow E$, see also \cite{ToTh:05}). When these relations are known, the hyper-charge of the Higgs boson is fixed by the demand that the Dirac type operators
$\dda \pm i\varphi_{\!\mbox{\tiny e}}\in{\cal D}(\www_{\!\mbox{\tiny$e$}})$ transform with respect to the adjoint representation of the Yang-Mills gauge sub-group of
${\cal G}_{\mbox{\tiny D}}$.

\end{document}